\documentclass{article}

\usepackage{epubtk}    
\usepackage{booktabs}  
\usepackage{graphicx}  

\usepackage{amsmath,amssymb,amscd,amsfonts,mathrsfs}

\showlistoftablesfalse

\def\nn{\nonumber}
\def\eps{{\epsilon}}
\def\th{{\theta}}
\def\t{{\theta}}
\def\p{{\partial}}
\def\s{\sigma}
\newcommand{\half}{\frac{1}{2}}
\newcommand{\cL}{\mathcal{L}}
\newcommand{\cJ}{\mathcal{J}}
\newcommand{\cQ}{\mathcal{Q}}
\newcommand{\cS}{\mathcal{S}}
\newcommand{\cM}{\mathcal{M}}
\newcommand{\cN}{\mathcal{N}}
\newcommand{\bigO}{\mathcal{O}}

\newcommand{\be}{\begin{equation}}
\newcommand{\ee}{\end{equation}}
\newcommand{\bea}{\begin{eqnarray}}
\newcommand{\eea}{\end{eqnarray}}

\newcommand{\abs}{\mathrm{abs}}
\newcommand{\ext}{\mathrm{ext}}
\newcommand{\AdS}{\mathrm{AdS}}
\newcommand{\CFT}{\mathrm{CFT}}

\newcommand{\df}{\textrm{d}}
\newcommand{\fr}{\frac}
\newcommand{\sr}{\sqrt}

\newcommand{\pd}{\partial}
\newcommand{\nnr}{\nonumber \\}
\newcommand{\expe}[1]{\mathrm{e}^{#1}}
\newcommand{\im}{\mathrm{i}}

\begin{document}

\title{The Kerr/CFT Correspondence and its Extensions}

\author{\epubtkAuthorData{Geoffrey Comp\`ere}{%
Physique Th\'eorique et Math\'ematique\\
Universit\'e Libre de Bruxelles, CP 231,\\
1050 Bruxelles, Belgium}{%
gcompere@ulb.ac.be}{%
http://www.ulb.ac.be/sciences/ptm/pmif/gcompere/}%
}

\date{}
\maketitle

\begin{abstract}
We present a first-principles derivation of the main results of the Kerr/CFT correspondence and its extensions using only tools from gravity and quantum field theory. Firstly, we review properties of extremal black holes with in particular the construction of an asymptotic Virasoro symmetry in the near-horizon limit. The entropy of extremal spinning or charged black holes is shown to match with a chiral half of Cardy's  formula.  Secondly, we show how a thermal 2-dimensional conformal field theory (CFT) is relevant to reproduce the dynamics of near-superradiant probes around near-extremal black holes in the semi-classical limit. Thirdly, we review the hidden conformal symmetries of asymptotically-flat black holes away from extremality and present how the non-extremal entropy can be matched with Cardy's formula. We follow an effective field theory approach and consider the Kerr-Newman black hole and its generalizations in various supergravity theories. The interpretation of these results by deformed dual conformal field theories is discussed and contrasted with properties of standard 2-dimensional CFTs. We conclude with a list of open problems. \vspace{2cm}
\end{abstract}

\epubtkKeywords{Black holes, Holography, AdS/CFT, Conformal field theory}

\epubtkUpdate
    [Id=A,
     ApprovedBy=subjecteditor,
     AcceptDate={10 June 2016},
     PublishDate={10 June 2016},
     Type=major]{%
Major revision, updated and expanded. The main changes are as follows. I now emphasize the conceptual limitations of the Kerr/CFT correspondence with respect to the AdS/CFT correspondence. The Introduction~(Section \ref{sec:introduction}) and Summary~(Section \ref{ccl}) have been totally rewritten. Sections~\ref{sec:ext}, \ref{sec:2dCFT}, \ref{sec:hidden} have been merged/remodeled. I added content to Sections~\ref{sec:the}, \ref{sec:nearext1}, \ref{sec:nobulkdof}, \ref{sec:DLCQ}, \ref{sec:deformations}, \ref{sec:hidden}, and \ref{ccl}. The list of references has increased from 271 to 364.
}


 \newpage

 \tableofcontents

\newpage

\section{Introduction}
\label{sec:introduction}

It is known since the work of Bardeen, Bekenstein, Carter and Hawking~\cite{Bekenstein:1972tm,Bardeen:1973gs,Hawking:1974sw} that black holes are thermodynamical systems equipped with a temperature and an entropy. In analogy to Bolzmann's statistical theory of gases, one expects that the entropy of black holes counts microscopic degrees of freedom. Understanding what these degrees of freedom actually are is one of the main challenges that a theory of quantum gravity should address. 

Since the advent of string theory, several black holes enjoying supersymmetry have been understood microscopically. In such cases, supersymmetry and its non-renormalization theorems allow to map the black hole states to dual states in a weakly-coupled description in terms of elementary strings and D-branes, which also provides a method to microscopically reproduce Hawking radiation slightly away from extremality~\cite{Strominger:1996sh,Callan:1996dv}, see \cite{Gubser:1998ex,David:2002wn,Dabholkar:2012zz} for reviews. 

These results can be contrasted with the challenge of describing astrophysical black holes that are non-supersymmetric and non-extremal, for which these methods cannot be directly applied. Astrophysical black holes are generically rotating and have approximately zero electromagnetic charge. Therefore, the main physical focus should be to understand the microstates of the Kerr black hole and to a smaller extent the microstates of the Schwarzschild, the Kerr--Newman and the Reissner--Nordstr\"om black hole. 

All black holes in Einstein gravity coupled to matter admit an entropy equal to their area in Planck units divided by $4$. This universality deserves an explanation which is missing so far. One temptative explanation comes from the holographic principle proposed in \cite{'tHooft:1993gx,Susskind:1994vu} which states that gravity can be described equivalently by a theory with a lower number of dimensions. In particular, black holes would be holographic in the sense that their microscopic degrees of freedom are encoded on a holographic plate on their horizon.

One of the greatest achievements of modern theoretical physics to have provided explicit realizations of the holographic principle: the exact AdS/CFT correspondences. They provide a dual description of specific systems in Type IIB supergravity or M-theory in terms of specific conformal field theories \cite{Maldacena:1997de,Maldacena:1998re,Witten:1998qj}. Some supersymmetric black holes are described in such correspondences. They contain in their near-horizon limit a factor of three-dimensional anti-de~Sitter spacetime AdS\sub{3}~\cite{Maldacena:1997ih,Cvetic:1997xv}  or more precisely a quotient thereof known as the BTZ black hole \cite{Banados:1992wn,Banados:1992gq}. The existence of a dual $2d$ CFT description is enough to account for the black hole entropy thanks to universality of the asymptotic growth of states, namely Cardy's formula \cite{Cardy:1986ie}. Cardy's formula only depends upon the Virasoro zero modes and the CFT central charges which can be evaluated in classical gravity using asymptotic symmetry methods \cite{Brown:1986nw}. Therefore, the exact microscopic description in terms of elementary strings and branes becomes unnecessary details and the black hole entropy follows from a universal relation whose ingredients can be computed in classical gravity \cite{Strominger:1997eq}. 

A special limit of $\AdS_3/\CFT_2$ correspondences is relevant for our purposes. When the BTZ black hole that appears in the near-horizon limit is taken extremal, it admits itself a near-horizon limit, the so-called self-dual orbifold \cite{Coussaert:1994tu} which consists of $\AdS_2$ with a twisted $U(1)$ fiber. The self-dual orbifold is sometimes called the ``very near-horizon limit'' of the original extremal black hole \cite{Azeyanagi:2008dk}. It turns out that a chiral half of the conformal structure of the $2d$ CFT extends to the very near-horizon limit \cite{Balasubramanian:2009bg,Compere:2015knw}. The very near-horizon limit admits one copy of the Virasoro algebra as asymptotic symmetry algebra which extends the $U(1)$ rotational Killing symmetry. The entropy in the very near-horizon geometry is then reproduced by a chiral half of Cardy's formula, which is inherited from the $2d$ CFT.

Motivated by the universality of growth of states in a $2d$ CFT, the authors of \cite{Guica:2008mu} formulated the original version of the Kerr/CFT correspondence which conjectures that ``Quantum gravity near the extreme Kerr horizon is dual to a two-dimensional CFT''. The Kerr/CFT correspondence can be viewed as a concrete proposal for realizing the holographic principle in a physically realistic gravitational setting. The starting point of the Kerr/CFT correspondence is the observation that the extremal Kerr geometry admits a decoupled near-horizon limit \cite{Bardeen:1999px}. This decoupled geometry contains an $\AdS_2$ factor and has a $SL(2,\mathbb R) \times U(1)$ symmetry which extends the 2 Killing symmetries of Kerr. This near-horizon geometry differs in two important ways with respect to the decoupled geometries appearing in exact $\AdS_3/\CFT_2$ correspondences. First, there is no $\AdS_3$ factor but a deformation thereof known as warped $\AdS_3$ \cite{Bengtsson:2005zj} which can be understood in string inspired models as an irrelevant deformation of the CFT dual to $AdS_3$ \cite{Compere:2010uk}. Second and most importantly, as realized soon after the original conjecture was made \cite{Azeyanagi:2008dk,Amsel:2009ev,Dias:2009ex}, the near-horizon geometry does not contain a black hole with arbitrary energy in contrast to the BTZ black hole in the $\AdS_3/\CFT_2$ case. Instead, it contains a warped deformation of the self-dual orbifold. The near-horizon region of the extremal Kerr black hole is therefore a ``very near-horizon limit''.  The asymptotic symmetry algebra of the near-horizon geometry was found to admit one copy of the Virasoro algebra with central charge $12 J$ where $J$ is the angular momentum \cite{Guica:2008mu}\footnote{A smooth covariant phase space could not be constructed using the original ansatz for the Virasoro generator \cite{Amsel:2009pu} but a modification of the ansatz allows for an explicit construction \cite{Compere:2015mza,Compere:2015bca} which as a by-product also realizes the Virasoro symmetry in the entire near-horizon region. This provides an example of asymptotic symmetries which extend into the bulk, also known as symplectic symmetries \cite{Compere:2015knw} (see also \cite{Barnich:2010eb,Compere:2014cna}).}. A thermal version of a chiral sector of Cardy's formula then exactly equates the black hole entropy. Here, such a relationship cannot be qualified as a microscopic counting since there is no definition of a dual field theory in terms of elementary fields; there is no justification for the validity of Cardy's formula and there are even strong reasons to think that such a dual theory does not exist. 

The main obstruction towards the existence of a dual $2d$ CFT is simply that away from extremality the near-horizon region couples to the asymptotic region which is not scale invariant. Turning on finite energy in the near-horizon description amounts to quitting exact extremality \cite{Amsel:2009ev,Dias:2009ex}. Turning on finite energy should therefore also turn on irrelevant couplings (with respect to scale invariance) which encode the couplings to the asymptotic region \cite{Baggio:2012db}. On the contrary, in the $\AdS_3/\CFT_2$ correspondences, finite energy perturbations (e.g., the BTZ black hole) exist in the near-horizon region described by a $2d$ dual CFT. Such intermediate region does not exist for the extremal Kerr black hole. At best, one might conjecture a $2d$ dual field theory obtained from a CFT with irrelevant deformations in both sectors, so nothing like a standard CFT\epubtkFootnote{Conjectured properties of such a field theory dual are detailed, e.g., in \cite{ElShowk:2011cm,Song:2011sr,Bena:2012wc,Baggio:2012db}. Also note that such a theory will not be unitary since it couples to the asymptotic region. A signature of non-unitarity are the complex conformal weights, as explained in Section~\ref{sec:CFTmatch}.}. 

Given this state of affairs, it might come as a surprize that the chiral thermal version of Cardy's formula universally reproduces the entropy of all known extremal black holes with a $U(1)$ axial symmetry. The matching extends to the Kerr--Newman black hole but also to large classes of black holes in gravity coupled to matter, with anti-de Sitter asymptotic regions, with higher curvature corrections or in higher dimensions as extensively detailed later on. Moreover, a non-trivial matching can be done with a standard thermal $2d$ CFT correlation function slightly away from extremality for a limited number of gravitational observables which explore the near-horizon region \cite{Bredberg:2009pv}. Finally, a Cardy formula also applies away from extremality and $SL(2,\mathbb R) \times SL(2,\mathbb R)$ scale invariance can be identified for certain probes away from extremality. This points to the relevance of the concepts of $2d$ CFTs away from extremality \cite{Cvetic:1997xv,Castro:2010fd,Cvetic:2011dn}  even though it should not be considered as a duality with a standard CFT \cite{Baggio:2012db,Castro:2013kea}\epubtkFootnote{Other approaches arguing for the presence of conformal symmetry around arbitrary black holes can be found in~\cite{Carlip:1998wz,Solodukhin:1998tc,Kang:2004js}. Such approaches will not be discussed here.}. These results shall be collectively referred to as the ``Kerr/CFT correspondence''.  The main scope of this review is to present the first-principle arguments for the Kerr/CFT correspondence with the hope that a deeper explanation for their origin and meaning could be achieved in the future. 

\subsection{Extension to gauge fields}
\label{KKup}

Another notable extension of the thermal chiral Cardy relation exists for the Reissner--Nordstr\"om black hole \cite{Hartman:2008pb}. This extension is valid only after one assumes that the $U(1)$ electromagnetic field can be promoted to be a Kaluza--Klein vector of a higher-dimensional spacetime. The $U(1)$ electric charge is then uplifted as a $U(1)$ axial angular momentum in the higher dimensional spacetime. Both angular momenta are then treated on an equal footing by the higher-dimensional version of the Kerr/CFT correspondence. This construction strengthen the strong parallel between the physics of static charged black holes and rotating black holes. Our point of view is that a proper understanding of the concepts behind the Kerr/CFT correspondence is facilitated by studying in parallel static charged black holes and rotating black holes. The relevance of the Kaluza-Klein construction also motivates to consider the Kerr/CFT correspondence in higher dimensions. 

In more detail, the  Reissner--Nordstr\"om/CFT arguments apply if our spacetime $\mathcal M_4$ can be uplifted to $\mathcal M_4 \times X$ where $X$ is compact and contains at least a $U(1)$ cycle (the total manifold might not necessarily be a direct product). Experimental constraints on such scenarios can be set from bounds on the deviation of Newton's law at small scales~\cite{Long:2003ta,Adelberger:2003zx}. A toy model for such a construction consists in adding a fifth compact dimension $\chi \sim \chi+2\pi R_\chi$, where $2\pi R_\chi$ is the length of the $U(1)$ Kaluza--Klein circle. We then define
\bea
ds^2 = ds^2_{(4)}+(d\chi+A)^2\, .\label{geohigher}\label{KKlift}
\eea
The metric~\eqref{geohigher} does not obey five-dimensional Einstein's equations unless the metric is complemented by matter fields. One simple choice consists of adding a $U(1)$ gauge field $A_{(5)}$, whose field strength is defined as
\be
dA_{(5)} = \frac{\sqrt{3}}{2}\star_{(4)} F\, ,\label{KKliftA}
\ee
where $\star_{(4)}$ is the four-dimensional Hodge dual. The five-dimensional metric and gauge field are then solutions to the five-dimensional Einstein--Maxwell--Chern--Simons theory, as reviewed, e.g., in~\cite{Kim:2010bf}.\epubtkFootnote{These considerations can also be applied to black holes in anti-de~Sitter spacetimes. However, the situation is more intricate because no consistent Kaluza--Klein reduction from five dimensions can give rise to the four-dimensional Einstein--Maxwell theory with cosmological constant~\cite{Lu:2009gj}. As a consequence, the four-dimensional Kerr--Newman--AdS black hole cannot be lifted to a solution of any five-dimensional theory. Rather, embeddings in eleven-dimensional supergravity exist, which are obtained by adding a compact seven-sphere~\cite{Chamblin:1999tk,Cvetic:1999xp}.}
Therefore, in order to review the arguments for the Reissner--Nordstr\"om/CFT correspondence and its generalizations, it is necessary to discuss \emph{five}-dimensional gravity coupled to matter fields.

\subsection{Classes of effective field theories}
\label{sec:eff}

We already motivated the study of the Kerr/CFT correspondence in the Einstein-matter system in 4 and higher dimensions.  Embedding this correspondence in string theory has the potential to give microscopic realizations of these correspondences. Efforts in that direction include~\cite{Nakayama:2008kg,Azeyanagi:2008dk,Guica:2010ej,Compere:2010uk,Azeyanagi:2010pw,deBoer:2010ac,Balasubramanian:2010ys,Song:2011sr,SheikhJabbaria:2011gc,Song:2011ii,deBoer:2011zt,ElShowk:2011cm,Bena:2012wc,Compere:2014bia,Bena:2015pua}.\epubtkFootnote{Some classes of black holes admit a vanishing horizon area $A_h$ and zero temperature $T$ limit such that the ratio $A_h/T$ is finite. Such extremal vanishing horizon (EVH) black holes admit near-horizon limits, which contain (singular) identifications of AdS\sub{3} that can be used for string model building~\cite{Guica:2010ej,Compere:2010uk,deBoer:2010ac,ElShowk:2011cm,deBoer:2011zt}. Most of the ideas developed for the Kerr/CFT correspondence can be developed similarly for EVH black holes~\cite{SheikhJabbaria:2011gc}.}
 Such constructions are only theoretical since there is no reasonable control on how the standard model of particle physics and cosmology fits in string theory despite active research in this area, see, e.g.,~\cite{Grana:2005jc,Denef:2008wq,McAllister:2007bg,Maharana:2012tu}.

Instead we will focus in this review on the effective field theory approach. Since all relevant physics is well below the Planck scale, all arguments of the Kerr/CFT correspondence can be formulated using semi-classical gravity. We will limit our arguments to the action
\bea
S &=& \frac{1}{16\pi G}\int d^dx\sqrt{-g}\bigg( R - \frac{1}{2}f_{AB}(\chi)
\partial_\mu \chi^A \partial^\mu \chi^B - V(\chi) - k_{IJ}(\chi)F^I_{\mu\nu}F^{J\mu\nu} 
 +L^d \bigg)\, ,
\label{generalaction}
\eea
where the dimension-dependent Lagrangian piece $L^d$ is given by
\bea
L^3 &=& -\frac{1}{2}b_{IJ}(\chi)A_\mu^I A^{J \mu} + \frac{1}{2}c_{IJ}(\chi)\eps^{\mu\nu\rho}A_\mu^I F_{\nu\rho}^J , \\
L^4 &=& h_{IJ}(\chi)\epsilon^{\mu\nu\rho\sigma}F^I_{\mu\nu}F^J_{\rho\sigma} , \\
L^5 &=& \frac{1}{2}C_{IJK}(\chi)\eps^{\mu\nu\rho\sigma\tau}A_\mu^I F_{\nu\rho}^J F_{\sigma\tau}^K , \; \dots
\eea
where $C_{IJK} = C_{(IJK)}$. The action is possibly supplemented with highly suppressed higher-derivative corrections. This theory allows to discuss in detail the embedding \eqref{geohigher}\,--\,\eqref{KKliftA} since the five-dimensional Einstein--Maxwell--Chern--Simons theory falls into that class of theories. The $4d$ theory also contains the bosonic sector of $\mathcal N$ = 8 supergravity.  A class of $5d$ theory contained in the general class \eqref{generalaction} is $\cN = 4$ supergravity obtained by compactification of $11D$ supergravity on $T^5$. Note that the action~(\ref{generalaction}) does not contain charged scalars, non-abelian gauge fields nor fermions.\epubtkFootnote{The most general stationary axisymmetric single-center spinning--black-hole solution of the theory~\eqref{generalaction} is not known (see however~\cite{Youm:1997rz,Mei:2010wm,Chow:2014cca} for general ans\"atze). The general 4-dimensional non-extremal rotating dyonic black hole in $\mathcal N=8$ supergravity has been found recently in a specific U-duality frame \cite{Chow:2013tia,Chow:2014cca}. The general charged rotating black hole of 5 dimensional $\cN = 4$ supergravity was found in \cite{Cvetic:1996xz}.
}

One additional motivation for studying this general class of theories comes from the $AdS/CFT$ correspondence~\cite{Maldacena:1998re,Witten:1998qj}. While asymptotically flat black holes are the most physically relevant black holes, gauge/gravity correspondences are mostly understood with the $AdS$ asymptotics. Studying the possible relationship between the Kerr/CFT correspondence and $AdS/CFT$ correspondences therefore naturally leads to considering such actions. We focus on the case where $f_{AB}(\chi)$, $k_{IJ}(\chi)$ and $b_{IJ}(\chi)$ are positive definite and the scalar potential $V(\chi)$ is non-positive in (\ref{generalaction}). This ensures that matter obeys the usual energy conditions and it covers the case of zero and negative cosmological constant. However, we will not discuss the supergravities required to embed AdS--Einstein--Maxwell theory. Three-dimensional models are also relevant with regards to the $\AdS_3/\CFT_2$ correspondences. In three dimensions we allow for massive vector fields which naturally arise in string theory compactifications \cite{Colgain:2010rg,Detournay:2012dz,Karndumri:2013dca}. 

The final motivation for studying this class of theories is simply that the near-horizon limits of extremal solutions take a universal form for any theory in the class~\eqref{generalaction}. We will discuss this point in Section \ref{sec:spin}, see also the review \cite{Kunduri:2013gce}. It is therefore convenient to discuss the theory~\eqref{generalaction} in one swoop.

\subsection{Extremal black holes and astrophysics}

Another motivation for the study of extremal black holes comes from astrophysics. Astrophysical black holes are usually assumed to have approximately zero electric charge. They are usually embedded in magnetic fields and surrounded by an accretion disk. In the first approximation they are described by the Kerr geometry \cite{Middleton:2015osa}.  The bound on the Kerr angular momentum derived from the cosmic-censorship hypothesis is $J \leq G M^2$. No physical process exists that would turn a non-extremal black hole into an extremal one. This is the third law of black hole thermodynamics \cite{Bardeen:1973gs}.  Using detailed models of accretion disks around the Kerr black hole, Thorne derived the bound $J \leq 0.998 \, G M^2$~\cite{Thorne:1974ve}. 
 
Quite surprisingly, it has been claimed that several known astrophysical black holes, such as the black holes in the X-ray binary GRS~1905+105~\cite{McClintock:2006xd} and Cygnus~X-1~\cite{Gou:2011nq}, are more than 95\% close to the extremality bound. More recent observations push the bound even further around 98\% for both the GRS 105+1915 \cite{Blum:2009ez,2013ApJ...775L..45M} and Cygnus X-1 \cite{Gou:2013dna} black hole. Also, the spin-to-mass--square ratio of the supermassive black holes in the active galactic nuclei MCG-6-30-15~\cite{Brenneman:2006hw} and 1H~0707-495~\cite{2009Natur.459..540F} have been claimed to be around 98\%. However, these measurements are subject to controversy since independent data analyses based on different assumptions led to opposite results as reviewed in~\cite{2010MNRAS.406.1425F}: e.g., the spin-to-mass--square ratio of the black hole in Cygnus~X-1 has been evaluated as $J/(GM^2) = 0.05$~\cite{2009astro2010S.197M}. If the measurements of high angular momenta are confirmed and generally accepted, it would promote near-extremal spinning black holes as physical objects of nature.

Due to enhanced $SL(2,\mathbb R)$ symmetry, extremal black holes can be considered as critical conformal systems in the sense of condensed matter theory. Even though such systems are never reached, their symmetries control the physics in the vicinity of the horizon of near-extremal black holes. More precisely, the $SL(2,\mathbb R)$ symmetry controls the near-horizon physics of all fields which can be written in an asymptotically matched expansion between the asymptotically flat region and the near-horizon region.\epubtkFootnote{Note that even though there is an important redshift in the vicinity of the horizon, particles orbit faster and the boost exactly compensates for the redshift \cite{Piran:1977dm,Banados:2009pr,Banados:2010kn,Gralla:2016jfc}. Therefore, signatures of the near-horizon region can be observable by the asymptotic observer.} This study was initiated in \cite{Bredberg:2009pv} which we shall review and which points to a symmetry enhancement beyond the $SL(2,\mathbb R)$ isometry. 

This approach led to further developments which we shall only mention here. $SL(2,\mathbb R)$ symmetry allows to analytically solve for gravitational wave emission on plunge orbits in the near-horizon region by relating this emission to the one on more easily computable circular orbits \cite{Porfyriadis:2014fja,Hadar:2014dpa,Hadar:2015xpa,Gralla:2015rpa}. The profile of gravitational waves arising from probes falling in the near-horizon geometry carries signatures of scale invariance for nearly extreme spins which differ from the otherwise characteristic signature \cite{Gralla:2016qfw}. The presence of $SL(2,\mathbb R)$ symmetry also allows to deduce analytic solutions for force-free electromagnetic fields in the near-horizon region \cite{Lupsasca:2014pfa,Li:2014bta,Zhang:2014pla,Lupsasca:2014hua,Compere:2015pja} but only few such solutions arise as a limit from the asymptotic region \cite{Gralla:2016jfc}.

\subsection{Organization of the review}

Since extremal black holes are the main objects of study, we will spend a large amount of time describing their properties in Section~\ref{sec:ext}. We will contrast the properties of static extremal black holes and of rotating extremal black holes. We will discuss how one can decouple the near-horizon region from the exterior region. We will then show that one can associate thermodynamical properties with any extremal black hole and we will argue that near-horizon geometries contain no local bulk dynamics. Since we aim at drawing parallels between black holes and two-dimensional CFTs, we will quickly review some relevant properties of standard $2d$ CFTs in Section~\ref{sec:2dCFT}.

After this introductory material, we will discuss the core of the Kerr/CFT correspondence starting from the Cardy matching of the entropy of extremal black holes in Section~\ref{sec:KerrCFT1}. There, we will review how the near-horizon region admits a set of asymptotic symmetries at its boundary, which form a Virasoro algebra. Several choices of boundary conditions exist, where the algebra extends a different compact $U(1)$ symmetry of the black hole. Following semi-classical quantization rules, the operators, which define quantum gravity in the near-horizon region, form a representation of the Virasoro algebra. We will also review the arguments that the thermodynamical potential associated with the $U(1)$ symmetry could be interpreted as a limiting temperature of the density matrix dual to the black hole. This leads to considering matching the black hole entropy with the thermal chiral Cardy formula. In Section~\ref{sec:KerrCFT2} we will move to the description of non-extremal black holes, and we will concentrate our analysis on asymptotically-flat black holes for simplicity. We will describe how part of the dynamics of probe fields in the near-extremal Kerr--Newman black hole matches with the thermal 2-point functions of CFTs  with both a left and a right-moving sector. The left-moving sector of the CFTs will match with the corresponding chiral limit of the CFTs derived at extremality. In Section~\ref{sec:hidden} we will review the hidden local conformal symmetry that is present in some probes around the generic Kerr--Newman black hole. Finally, we will summarize the key results of the Kerr/CFT correspondence in Section~\ref{ccl} and provide a list of open problems.

This review complements the lectures on the Kerr black hole presented in~\cite{Bredberg:2011hp} by providing an overview of the Kerr/CFT correspondence and its extensions for general rotating or charged black holes in gravity coupled to matter fields in a larger and updated context. Since we follow an effective field-theory approach, we will cover string-theory models of black holes only marginally. We refer the interested reader to the complementary string theory-oriented review of extremal black holes~\cite{Simon:2011zza}.

\newpage

\section{Extremal Black Holes}
\label{sec:ext}

In this section, we review some key properties of extremal black holes in the context of four-dimensional theories of gravity coupled to matter. We first contrast how to decouple from the asymptotic region the near-horizon region of static and rotating black holes.  We then derive the thermodynamic properties of black holes at extremality. 
We then discuss near-horizon geometries close to extremality and emphasize their lack of local bulk dynamics.

\subsection{General properties}
\label{propBH}

For simplicity, we will strictly concentrate our analysis on stationary black holes. Since we are concerned with the region close to the horizon, one could only require that the near-horizon region is stationary,  while radiation would be allowed far enough from the horizon. Such a situation could be treated in the framework of isolated horizons~\cite{Ashtekar:1998sp,Ashtekar:2000sz} (see~\cite{Ashtekar:2004cn} for a review). However, for our purposes, it will be sufficient and much simpler to assume stationarity everywhere. We expect that all results derived in this review could be generalized for isolated horizons (see~\cite{Wu:2009di} for results along these lines).

Many theorems have been derived that characterize the generic properties of four-dimensional stationary black holes that admit an asymptotically-timelike Killing vector. First, they have one additional axial Killing vector -- they are axisymmetric\epubtkFootnote{``Stationarity implies axisymmetry'' has been proven for any non-extremal black hole in $d=4$ Einstein gravity coupled to any matter obeying the weak energy condition with hyperbolic equations of motion and asymptotically-flat boundary conditions~\cite{Hawking:1971vc,HawkingEllis,Sudarsky:1992ty,Chrusciel:1993cv,Friedrich:1998wq}. The proof has been extended to extremal black holes, to higher dimensions and to anti-de~Sitter asymptotics in~\cite{Hollands:2006rj,Hollands:2008wn,Chrusciel:2008js}.} -- and their event horizon is a Killing horizon\epubtkFootnote{The original proofs were limited to non-extremal black holes, which have a bifurcation surface~\cite{Carter1973,HawkingEllis}. The proof for extremal black holes can now be found in~\cite{Hollands:2008wn}.}. In asymptotically-flat spacetimes, black holes have spherical topology~\cite{HawkingEllis}.

Extremal black holes are defined as stationary black holes with vanishing Hawking temperature,
\be
T_H = 0.
\ee
Equivalently, extremal black holes are defined as stationary black holes whose inner and outer horizons coincide. No physical process is known that would make an extremal black hole out of a non-extremal black hole.\epubtkFootnote{Nevertheless, one can describe the process of spontaneous creation of extremal black holes in an electromagnetic field as an analogue to the Schwinger process of particle creation~\cite{Dowker:1994up}.} 
If one attempts to send finely-tuned particles or waves into a near-extremal black hole in order to further approach extremality, one realizes that there is a smaller and smaller window of parameters that allows one to do so when approaching extremality. In effect, a near-extremal black hole has a potential barrier close to the horizon, which prevents it from reaching extremality. Note that in the other way around, if one starts with an extremal black hole, one can simply throw in a massive particle to make the black hole non-extremal. Therefore, extremal black holes are finely tuned black holes. Nevertheless, studying the extremal limit is very interesting because many simplifications occur and powerful specialized methods can be used.

We will recall the most relevant properties of spinning or charged rotating black holes (some of them also outside extremality) which we will use in the following. We refer the reader to the excellent lecture notes~\cite{Townsend:1997ku} for the derivation of some of these properties.

\begin{itemize}
\item \textit{Angular velocity.} Spinning black holes are characterized by a chemical potential -- the angular velocity $\Omega_J$ -- conjugate to the angular momentum. The angular velocity can be defined in geometrical terms as the coefficient of the black-hole--horizon generator proportional to the axial Killing vector
\be
\xi = \p_t + \Omega_J \p_\phi\, .\label{defxi}
\ee
The net effect of the angular velocity is a frame-dragging effect around the black hole. This gravitational kinematics might be the clue of an underlying microscopic dynamics. Part of the intuition behind the Kerr/CFT correspondence is that the degrees of freedom responsible for the black hole entropy are rotating at the speed of light at the horizon.

\item \textit{Electrostatic potential.} Electrically-charged black holes are characterized by a chemical potential -- the electrostatic potential $\Phi_e$ -- conjugated to the electric charge. It is defined on the horizon $r=r_+$ as
\be
\Phi^I_e = -\xi^\mu A^I_\mu|_{r=r_+},\label{defPhi}
\ee
where $\xi$ is the horizon generator defined in \eqref{defxi}. Similarly, one can associate a magnetic potential $\Phi^I_m$ to the magnetic monopole charge. The form of the magnetic potential can be obtained by electromagnetic duality, or reads as the explicit formula derived in~\cite{Copsey:2005se} (see also~\cite{Compere:2009zh} for a covariant expression). Part of the intuition behind the Reissner--Nordstr\"om/CFT correspondence is that this kinematics is the sign of microscopic degrees of freedom ``moving along the gauge direction''. We will make that statement more precise in Section~\ref{sec:BC}.

\item \textit{Ergoregion.} Although the Killing generator associated with the mass of the black hole, $\p_t$, is timelike at infinity, it does not need to be timelike everywhere outside the horizon. The region where $\p_t$ is spacelike is called the ergoregion and the boundary of that region where $\p_t$ is lightlike is the ergosphere. If there is no ergoregion, $\p_t$ is a global timelike Killing vector outside the horizon. However, it should be noted that 
the presence of an ergoregion does not preclude the existence of a global timelike Killing vector. For example, the extremal spinning Kerr--AdS black hole has an ergoregion. When the horizon radius is smaller than the AdS length, the horizon generator becomes spacelike at large enough distances and there is no global timelike Killing vector, as for the Kerr black hole. On the contrary, when the horizon radius is larger than the AdS length, the horizon generator is timelike everywhere outside the horizon.

\item \textit{Superradiance.} One of the most fascinating properties of some rotating black holes is that neutral particles or waves sent towards the black hole with a frequency $\omega$ and angular momentum $m$ inside a specific band 
\be
0 < \omega < m \Omega_J \label{superradrangeJ}
\ee
come back to the exterior region with a higher amplitude. This amplification effect or Penrose effect allows the extraction of energy very efficiently from the black hole. Superradiance occurs for the Kerr and Kerr--Newman black hole and is related to the presence of the ergoregion and the lack of a global timelike Killing vector. Because of the presence of a global timelike Killing vector, there is no superradiance for large Kerr--AdS black holes (when reflective boundary conditions for incident massless waves are imposed)~\cite{Hawking:1999dp,Winstanley:2001nx}.

\item \textit{Electromagnetic analogue to superradiance.} Charged black holes contain electrostatic energy that can also be extracted by sending charged particles or waves with frequency $\omega$ and charge $q_e$ inside a specific band~\cite{Christodoulou:1972kt} (see~\cite{Jacobson:1996aa} for a review)
\be
0 < \omega < q_e \Phi_e\, . \label{superradrangeQ}
\ee
There is no ergoregion in the four-dimensional spacetime. However, for asymptotically-flat black holes, there is a five-dimensional ergoregion when considering the uplift~\eqref{KKlift}. For the Reissner--Nordstr\"om black hole, the five-dimensional ergoregion lies in the range $r_+ < r < 2M$, where $M$ is the mass and $r$ the standard Boyer--Lindquist radius.

The combined effect of rotation and charge allows one to extract energy in the range 
\be
0 < \omega < m \Omega_J +q_e \Phi_e \, .\label{superradrange}
\ee
When considering a wave scattering off a black hole, one can define the absorption probability $\s_{\abs}$ or macroscopic greybody factor as the ratio between the absorbed flux of energy at the horizon and the incoming flux of energy from infinity,
\be
\s_{\abs} = \frac{dE_{\abs} / dt}{dE_{\text{in}}/dt}\, .
\ee
In the superradiant range~\eqref{superradrange}, the absorption probability is negative because the outgoing flux of energy is higher than the incoming flux. 

\item \textit{No thermal radiation but spontaneous emission.} Taking quantum mechanical effects into account, non-extremal black holes radiate with a perfect black-body spectrum at the horizon at the Hawking temperature $T_H$~\cite{Hawking:1974sw}. The decay rate of a black hole as observed from the asymptotic region is the product of the black-body spectrum decay rate with the greybody factor $\s_{\abs}$,
\be
\Gamma = \frac{1}{e^{\frac{\omega - m\Omega_J-q_e \Phi_e}{T_H} }- 1}\s_{\abs}\, .
\ee
The greybody factor accounts for the fact that waves (of frequency $\omega$, angular momentum $m$ and electric charge $q_e$) need to travel from the horizon to the asymptotic region in the curved geometry. In the extremal limit, the thermal factor becomes a step function. The decay rate then becomes
\be
\Gamma_{\ext} = -\Theta(-\omega+m\Omega_J+q_e\Phi_e)\s_{\abs}\, .
\ee
As a consequence, ordinary Hawking emission with $\s_{\abs}>0$ and $\omega > m \Omega_J+q_e\Phi_e$ vanishes while quantum superradiant emission persists. Therefore, extremal black holes that exhibit superradiance, spontaneously decay to non-extremal black holes by emitting superradiant waves.

\item \textit{Innermost stable orbit approaching the horizon in the extremal limit.} Near-extremal black holes have an innermost stable circular orbit (ISCO) very close to the horizon. (In Boyer--Lindquist coordinates, the radius of such an orbit coincides with the radius of the horizon. However, since the horizon is a null surface, while the ISCO is timelike, the orbit necessarily lies outside the horizon, which can be seen explicitly in more appropriate coordinates. See Figure~2 of~\cite{Bardeen:1972cc}\epubtkFootnote{We thank the anonymous referee for pointing out this reference.}). As a consequence, the region of the black hole close to the horizon can support accretion disks of matter and, therefore, measurements of electromagnetic waves originating from the accretion disk of near-extremal rotating black holes contain (at least some marginal) information from the near-horizon region. For a careful analysis of the physical processes around rotating black holes, see~\cite{Bardeen:1972cc}. 

\item \textit{Classical singularities approaching the horizon in the extremal limit.} Stationary axisymmetric non-extremal black holes admit a smooth inner and outer horizon, where curvatures are small. However, numerical results~\cite{Brady:1995ni,Brady:1995un,Brady:1998ht,Dafermos:2003wr} and the identification of unstable linear modes using perturbation theory~\cite{McNamara:1978aa,Dotti:2008yr,Dotti:2011ix} showed that the inner horizon is unstable and develops a curvature singularity when the black hole is slightly perturbed. The instability is triggered by tiny bits of gravitational radiation that are blueshifted at the inner Cauchy horizon and which create a null singularity. In the near-extremality limit, the inner horizon approaches the outer horizon and it can be argued that test particles encounter a curvature singularity immediately after they enter the horizon of a near-extremal black hole~\cite{Marolf:2010nd}. In fact, there is an instability at the horizon of both extreme Reissner--Nordstr\"om and Kerr as subsequently proven \cite{Aretakis:2011ha,Aretakis:2011hc,Aretakis:2011gz,Aretakis:2012ei}. 

\end{itemize}

\subsection{Near-horizon geometries of extremal black holes}
\label{sec:NH}

We will define the near-horizon limit of static and rotating black holes at extremality and describe in detail their properties. We will also present some explicit examples of general interest. The extension of the near-horizon limit for near-extremal geometries will be described in Section \ref{sec:nearext1} after first presenting the thermodynamic properties of extremal horizons in Section \ref{sec:the}.

\subsubsection{Static near-horizon geometries}
\label{sec:stat}

As a warm-up, let us first review the near-horizon limit of static $4d$ extremal black holes. In that case, the generator of the horizon (located at $r = r_+$) is the generator of time translations $\p_t$ and the geometry has $SO(3)$ rotational symmetry. Since the horizon generator is null at the horizon, the coordinate $t$ diverges there. The near-horizon limit is then defined as 
\bea
t \rightarrow \frac{r_0\, t}{\lambda} \, , \qquad r \rightarrow r_+ + \lambda r_0\, r \, ,\label{nearcoord00}
\eea
with $\lambda \rightarrow 0$. The scale $r_0$ is introduced for convenience in order to factor out the overall scale of the near-horizon geometry. In the presence of electrostatic potentials, a change of gauge is required when taking the near-horizon limit~\eqref{nearcoord00}. Indeed, in the near-horizon coordinates~\eqref{nearcoord00} the gauge fields take the following form, 
\be
A^I = - \frac{\Phi^I_e}{\lambda} r_0\, dt + A^I_r dr + A^I_\th d\th + A^I_\phi d\phi \, ,
\ee
where $\Phi^I_e$ is the static electric potential of the gauge field $A^I$. Upon taking the near-horizon limit one should, therefore, perform a gauge transformation $A^I \rightarrow A^I + d\Lambda^I$ of parameter
\be
\Lambda^I =\frac{\Phi_e^{I,\ext}}{\lambda} r_0\, t\, ,\label{valLambda}
\ee
where $\Phi_e^{I,\ext}$ is the static electric potential at extremality. 

It is important to note that one is free to redefine the near-horizon limit parameter $\lambda$ as $\lambda \rightarrow \alpha \lambda$ for any $\alpha >0$. This transformation scales $r$ inversely proportionally to $t$. Therefore, the near-horizon geometry admits the enhanced symmetry generator
\be
\zeta_0 = r \p_r - t \p_t
\ee
in addition to $\zeta_{-1} = \p_t$ and the $SO(3)$ symmetry generators. Furthermore, using the kinematical properties of the near-horizon limit, one can deduce the existence of either an $\AdS_2$, $dS_2$ or $\mathbb R^{1,1}$ geometry in the near-horizon limit, with either $SL(2,\mathbb R)$ or Poincar\'e $iso(1,1)$ symmetry which extends to the entire near-horizon geometry. Assuming the strong energy condition, the geometry $\AdS_2$ is singled out, see~\cite{Kunduri:2007vf} and the review \cite{Kunduri:2013gce} for a detailed derivation. The general near-horizon solution is then given by
\bea
ds^2 = v_1 (-r^2\, dt^2 + \frac{dr^2}{r^2})+v_2 (d\theta^2 + \sin^2\theta d\phi^2)\, , \nn \\
\chi^A = \chi^A_\star \, , \qquad A^I = e_I r\, dt - \frac{p^I}{4 \pi} \cos\theta d\phi\, ,\label{nearstatic}
\eea
where $v_1,v_2,\chi^A_\star,e_I,p^I$ are parameters, which are constrained by the equations of motion. The geometry consists of the direct product $\\AdS_2 \times S^2$.

For some supersymmetric theories, the values $v_1,v_2,\chi^A_\star,e_I$ are generically completely fixed by the electric ($q^I$) and magnetic ($p^I$) charges of the black hole and do not depend continuously on the asymptotic value of the scalar fields in the asymptotic region -- the \emph{scalar moduli}. This is the \emph{attractor mechanism}~\cite{Ferrara:1995ih,Strominger:1996kf,Ferrara:1996dd}. It was then realized that it still applies in the presence of certain higher-derivative corrections~\cite{LopesCardoso:1998wt,LopesCardoso:1999ur,LopesCardoso:2000qm}. The attractor mechanism was also extended to non-supersymmetric extremal static black holes~\cite{Ferrara:1997tw,Sen:2005wa,Goldstein:2005hq,Kallosh:2005ax}. As a consequence of this mechanism, the entropy of these extremal black hole does not depend continuously on any moduli of the theory.\epubtkFootnote{In some special cases, there may be some continuous dependence of the near-horizon parameters on the scalar moduli, but the entropy is constant under such continuous changes~\cite{Astefanesei:2006dd}.
} The entropy can however still have discrete jumps when crossing \emph{walls of marginal stability} in the scalar moduli space since the index which captures it has jumps~\cite{Ooguri:2004zv,Denef:2007vg}. The attractor mechanism generally allows to account for the black hole entropy by varying the moduli to a weakly-coupled description of the system without gravity, where states with fixed conserved charges can be counted. Therefore, the attractor mechanism led to an explanation~\cite{Astefanesei:2006sy,Dabholkar:2006tb} of the success of previous string theory calculations of the entropy of certain nonsupersymmetric extremal black holes~\cite{Kaplan:1996ev,Horowitz:1996ac,Dabholkar:1997rk,Tripathy:2005qp,Emparan:2006it,Emparan:2007en}.

As will turn out to be useful in the development of the Reissner--Nordstr\"om/CFT correspondence, let us discuss some features of the near-horizon geometry~\eqref{nearstatic} under the assumption that one gauge field $A$ can be lifted as a Kaluza--Klein vector to a higher-dimensional spacetime, as discussed in Section~\ref{KKup}. In the simple model~\eqref{geohigher}, the change of gauge $A \rightarrow A +d\Lambda$ is implemented as the change of coordinates $\chi \rightarrow \chi+\Lambda$. Using the definition of the electrostatic potential $\Phi^{\ext}_e$ \eqref{defPhi} at extremality, it is straightforward to obtain that in the geometry~\eqref{geohigher} the horizon is generated by the vector field $\xi_{\mathrm{tot}} = \p_t+\Phi^{\ext}_e \p_{\chi}$. The change of coordinates~\eqref{nearcoord00} combined with $\chi \rightarrow \chi+\Lambda$ with $\Lambda$ defined in \eqref{valLambda} then maps this vector to 
\be
\xi_{\mathrm{tot}}\rightarrow \frac{\lambda}{r_0}\p_t\, .\label{scalexi}
\ee

\subsubsection{Spinning near-horizon geometries}
\label{sec:spin}

Let us now consider extremal $4d$ spinning black holes. Let us denote the axis of rotation to be $\p_\phi$, where $\phi \sim \phi + 2\pi$ and let $r= r_+$ be the black-hole horizon. The generator of the horizon is $\xi \equiv \p_t + \Omega^{\ext}_J \p_\phi$ where $\Omega^{\ext}_J$ is the extremal angular velocity. We choose a coordinate system such that the coordinate $t$ diverges at the horizon, which is equivalent to the fact that $g^{tt}$ diverges at the horizon. As in the static case, one needs to perform a gauge transformation of parameter~\eqref{valLambda}, when electrostatic fields are present. One can again interpret this change of gauge parameter as a change of coordinates in a higher-dimensional auxiliary spacetime \eqref{geohigher}. The near-horizon limit is then defined as 
\bea
t \rightarrow r_0 \frac{t}{\lambda} \, , \nn\\
r \rightarrow r_+ + \lambda r_0\, r \, ,\label{nearcoord}\\
\phi \rightarrow \phi + \Omega^{\ext}_J \frac{r_0 \,t}{\lambda}\, ,\nn \\
A \rightarrow A + \frac{\Phi^{\ext}_e }{\lambda}r_0\, dt\, ,\nn
\eea
with $\lambda \rightarrow 0$. The scale $r_0$ is again introduced in order to factor out the overall scale of the near-horizon geometry. The additional effect with respect to the static near-horizon limit is the shift in the angle $\phi$ in order to reach the frame co-moving with the horizon. The horizon generator becomes $\xi = \lambda/r_0 \p_t$ in the new coordinates. Including the gauge field, one has precisely the relation~\eqref{scalexi}. As in the static case, any finite energy excitation of the near-horizon geometry is confined and amounts to no net charges in the original (asymptotically flat of AdS) geometry. 

One is free to redefine $\lambda$ as $\lambda \rightarrow \alpha \lambda$ for any $\alpha >0$ and, therefore, the near-horizon geometry admits the enhanced symmetry generator
\be
\zeta_0 = r \p_r - t \p_t \, , \label{zeta0}
\ee
in addition to $\zeta_{-1} = \p_t$ and $L_0 = \p_\phi$. Together $\zeta_0$ and $\zeta_{-1}$ form a non-commutative algebra under the Lie bracket. 

Now in $d \geq 4$, contrary to the static case, the existence of a third Killing vector is not guaranteed by geometric considerations. Nevertheless, it turns out that Einstein's equations derived from the action~\eqref{generalaction} imply that there is an additional Killing vector $\zeta_{1}$ in the near-horizon geometry~\cite{Kunduri:2007vf,Astefanesei:2007bf} (see also~\cite{daCunha:2010jj} for a geometrical derivation). The vectors $\zeta_{-1},\zeta_0,\zeta_1$ turn out to obey the $SL(2,\mathbb R) \sim SO(2,1)$ algebra when the strong energy condition holds. This dynamical enhancement is at the origin of many simplifications in the near-horizon limit. More precisely, one can prove~\cite{Kunduri:2007vf} that any stationary and axisymmetric asymptotically-flat or anti-de~Sitter extremal black-hole solution of the theory described by the Lagrangian~\eqref{generalaction} admits a near-horizon geometry with $SL(2,\mathbb R) \times U(1)$ isometry. The result also holds in the presence of higher-derivative corrections in the Lagrangian provided that the black hole is big, in the technical sense that the curvature at the horizon remains finite in the limit where the higher-derivative corrections vanish\epubtkFootnote{In $d=3$, the kinematics is sufficiently constrained and the existence of the fourth Killing vector $\zeta_1$ is guaranteed by the near-horizon limit.  It is natural to assume that the strong energy condition will again implies the existence of an $\AdS_2$ factor. The review \cite{Kunduri:2013gce} gives the proof for electrovacuum geometries in $d=3$ but the generalization to the action \eqref{generalaction} is straightforward.}. The general near-horizon geometry of $4d$ extremal spinning black holes consistent with these symmetries is given by
\begin{eqnarray}
 ds^2 &=& \Gamma(\theta)\left[-r^2 d t^2+\frac{d r^2}{r^2}
+ \alpha (\t)^2 d \theta^2+
 \gamma(\theta)^2(d \phi + k\, rd t)^2 \right] , \nn\\
&& 
\chi^A = \chi^A (\t) \, , \qquad A^I = f^I(\t) (d \phi + k\, rdt) - \frac{e_I}{k}d\phi \, , 
\label{GenExt}
\end{eqnarray}
where $\Gamma(\theta) > 0$, $\gamma(\theta) \geq 0$, $\chi^A(\t), f^I(\th)$ and $k,\, e_I \in \mathbb R $ are fixed by the equations of motion. By inverting $t$ and redefining $A^I \rightarrow -A^I$, we can always set $k \geq 0 $, $e_I \geq 0$. The function $\alpha (\t) \geq 0$ can be removed by redefining $\theta$ but it is left for convenience because some near-horizon geometries are then more easily described.\epubtkFootnote{We fix the range of $\theta$ as $\theta \in [0,\pi]$. Since the original black hole has $S^2$ topology and no conical singularities, the functions $\gamma(\th)$, $\alpha(\th)$ also obey regularity conditions at the north and south poles
\be
\frac{\gamma(\th)^2}{\alpha(\th)^2}\sim \th^2 +O(\th^3) \sim (\pi -\th)^2 +O((\pi-\th)^3) \, .
\ee
Similar regularity requirements apply for the scalar and gauge fields.} 

The term $-\frac{e_I}{k}d\phi$ in \eqref{GenExt} is physical since it cannot be gauged away by an allowed gauge transformation. For example, one can check that the near-horizon energy $\mathcal Q_{\p_t}$ would be infinite in the Kerr--Newman near-horizon geometry if this term would be omitted. One can alternatively redefine $f^I(\t) = b^I(\t)+e_I/k$ and the gauge field takes the form 
\be
A^I = b^I(\t) (d \phi + k\, rdt) + e_I rdt \, .
\ee
 
The static near-horizon geometry~\eqref{nearstatic} is recovered upon choosing only $\text{SO}(3)$ covariant quantities with a well-defined static limit. This requires $k \rightarrow 0$ and it requires the form
\be
b^I(\theta)=-\frac{p^I}{4\pi}\cos\theta \, ,
\ee
where $p^I$ are some pure numbers, which are the magnetic charges. 

Going back to the spinning case, the $SL(2,\mathbb R)\times U(1)$ symmetry is generated by 
\begin{align}
&\zeta_{-1}=\partial_t,\qquad \zeta_0=t\partial_t-r\partial_r \, ,\nn \\
&\zeta_1=\bigg(\frac{1}{2r^2}+\frac{t^2}{2}\bigg)\partial_t-tr\partial_r -\frac kr\partial_\phi \, , \qquad 
L_0=\partial_\phi \, .
\label{killing}
\end{align}
In addition, the generator $\zeta_1$ should be accompanied by the gauge transformation of parameter $\Lambda^I = -e_I/r$ so that $\mathcal L_{\zeta_1}A^I_\mu+\p_\mu \Lambda^I=0$. Note that all of these symmetries act within a three-dimensional slice of fixed polar angle $\theta$. The metric is also invariant under discrete symmetry, which maps
\begin{equation}
(t,\phi) \to (-t,-\phi) \, . \label{PT}
\end{equation} 
This is often called the $t$-$\phi$ reflection symmetry in black-hole literature. The parity/time reversal transformation~\eqref{PT} reverses the electromagnetic charges of the solution.

The geometry~\eqref{GenExt} is a warped and twisted product of $\\AdS_2 \times S^2$. The $(r,t)$ coordinates are analogous to Poincar\'e coordinates on $\AdS_2$ with an horizon at $r=0$. One can find global coordinates in the same way that the global coordinates of $\AdS_2$ are related to the Poincar\'e coordinates~\cite{Bardeen:1999px}. Let 
\be
r = (1+y^2)^{1/2}\cos\tau + y \, , \qquad t\, r = (1+y^2)^{1/2}\sin\tau \, .
\ee
The new axial angle coordinate $\varphi$ is chosen so that $d\phi+k rdt = d\varphi + k y d\tau $, with the result
\be
\phi = \varphi+k \log \left| \frac{\cos\tau + y \sin\tau}{1+(1+y^2)^{1/2}\sin\tau} \right| \, .
\ee
In these new coordinates, the near-horizon geometry becomes
\begin{equation}
\begin{split}
&ds^2 = \Gamma(\theta)\left[-(1+y^2) d \tau^2+\frac{d y^2}{1+y^2}
+ \alpha (\t)^2 d \theta^2+ \gamma(\theta)^2(d \varphi + k\, y d \tau)^2 \right] , \\
&\hspace{1cm}
\chi^A = \chi^A (\t) \, , \qquad A^I = f^I(\t) (d \varphi + k\, yd\tau ) - \frac{e_I}{k}d\varphi \, , 
\end{split}
\label{GenExtGlobal}
\end{equation}
after performing an allowed gauge transformation (as the change of gauge falls into the boundary conditions \eqref{extcond2} derived in Section~\ref{sec:BC}). Note that the $\tau = 0$ hypersurface coincides with the $t=0$ hypersurface, and that $\phi = \varphi$ on this hypersurface. The geometry has two boundaries at $y=-\infty$ and $y=+\infty$. 

Geodesic completeness of these geometries has not been shown in general, even though it is expected that they are geodesically complete. For the case of the near-horizon geometry of Kerr, geodesic completeness has been proven explicitly in~\cite{Bardeen:1999px} after working out the geodesic equations.

At fixed polar angle $\theta$, the geometry can be described in terms of \emph{$3d$ warped anti-de~Sitter geometries}; see~\cite{Anninos:2008fx} for a relevant description and~\cite{Nutku:1993eb,Gurses:1994aa,Rooman:1998xf,Duff:1998cr,Moussa:2003fc,Israel:2003ry,Israel:2004vv,Andrade:2005ur,
Detournay:2005fz,Bengtsson:2005zj,Banados:2005da,Compere:2007in,Moussa:2008sj} for earlier work on these three-dimensional geometries. Warped anti-de~Sitter spacetimes are deformations of AdS\sub{3}, where the $S^1$ fiber is twisted around the $\AdS_2$ base. Because of the identification $\phi \sim \phi +2 \pi$, the geometries at fixed $\theta$ are quotients of the warped AdS geometries, which are characterized by the presence of a Killing vector of constant norm (namely $\p_\phi$). These quotients are often called self-dual orbifolds by analogy to similar quotients in AdS\sub{3}~\cite{Coussaert:1994tu}.\epubtkFootnote{In singular limits where both the temperature and horizon area of black holes can be tuned to zero, while keeping the area-over-temperature--ratio fixed, singular near-horizon geometries can be constructed. Such singular near-horizon geometries contain a local AdS\sub{3} factor, which can be either a null self-dual orbifold or a \emph{pinching orbifold}, as noted in~\cite{Bardeen:1999px,Balasubramanian:2007bs,Fareghbal:2008ar,Azeyanagi:2010pw} (see~\cite{deBoer:2010ac} for a comprehensive study of the simplest three-dimensional model and~\cite{SheikhJabbaria:2011gc} for a partial classification of four-dimensional vanishing area near-horizon solutions of \eqref{generalaction}).} 

The geometries enjoy a global timelike Killing vector (which can be identified as $\p_\tau$) if and only if 
\be
k \gamma (\th) < 1 \, , \qquad \forall \theta \in [0,\pi] \, .\label{globalKillingtime}
\ee
If there is no global timelike Killing vector, there is at least one special value of the polar angle $\theta_\star$, where $k \gamma (\th_\star) = 1$. At that special value, the slice $\theta = \theta_\star$ is locally an ordinary AdS\sub{3} spacetime and acquires a local $SL(2,\mathbb R) \times SL(2,\mathbb R)$ isometry. At all other values of $\theta$, one $SL(2,\mathbb R)$ is broken to $U(1)$. Note that there is still a global \emph{time function} for each near-horizon geometry. Constant global time $\tau$ in the global coordinates~\eqref{GenExtGlobal} are spacelike surfaces because their normal is timelike,
\be
g^{ab}\p_a \tau \p_b \tau = g^{\tau\tau} = -(1+y^2)^{-1}\Gamma^{-1}(\th) < 0 \, .
\ee
Hence, there are no closed timelike curves.

One can show the existence of an attractor mechanism for extremal spinning black holes, which are solutions of the action~\eqref{generalaction}~\cite{Astefanesei:2006dd}. According to~\cite{Astefanesei:2006dd}, the complete near-horizon solution is generically independent of the asymptotic data and depends only on the electric charges $\cQ^I_{e}$, magnetic charges $\cQ^I_{m}$ and angular momentum $\cJ$ carried by the black hole, but in special cases there may be some dependence of the near horizon background on this asymptotic data.
In all cases, the entropy only depends on the conserved electromagnetic charges and the angular momentum of the black hole and might only jump discontinuously upon changing the asymptotic values of the scalar fields, as it does for static charged black holes~\cite{Ooguri:2004zv,Denef:2007vg}.

One can generalize the construction of near-horizon extremal geometries to higher dimensions. In five dimensions, there are two independent planes of rotation since the rotation group is a direct product $SO(4) \sim SO(3) \times SO(3)$. Assuming the presence of two axial $U(1)$ symmetries $\p_{\phi_i}$, $i=1,2$ (with fixed points at the poles), one can prove~\cite{Kunduri:2007vf} that the near-horizon geometry of a stationary, extremal black-hole solution of the five-dimensional action~\eqref{generalaction} is given by
\bea
ds^2 &=& \Gamma (\theta) \left[ -r^2\, dt^2 + \frac{dr^2}{r^2}+\alpha(\theta)^2 d\theta^2 + \sum_{i,j=1}^2\gamma_{ij}(\theta)^2 (d\phi^i + k_i r\, dt) (d\phi^j + k_j r\, dt) \right], \label{near5d}\\
\chi^A &=& \chi^A(\theta) \, , \qquad A^I = \sum_{i=1}^2 f^I_i(\theta) (d\phi^i + k_i r\, dt)-\frac{e_I}{k^i}d\phi^i \, .\nn
\eea
In particular, the solutions obtained from the uplift~\eqref{KKlift}\,--\,\eqref{KKliftA} fall into this class. In general, these solutions can be obtained starting from both black holes (with $S^3$ horizon topology) and black rings (with $S^2 \times S^2$ horizon topology)~\cite{Emparan:2001wn}. We refer the reader to the review \cite{Kunduri:2013gce} for further information. Additional properties of near-horizon geometries are also reviewed in \cite{Compere:2015bca}.

\subsubsection{Explicit near-horizon geometries}
\label{sec:explicitnear}

Let us now present explicit examples of near-horizon geometries of interest. We will discuss the cases of the $4d$ extremal Kerr and Reissner--Nordstr\"om black holes as well as the $4d$ extremal Kerr--Newman and Kerr--Newman--AdS black holes. We will also present the $3d$ extremal BTZ black hole since it is quite universal. Other near-horizon geometries of interest can be found, e.g., in~\cite{Clement:2001ny,Dias:2007nj,Lu:2008jk}.

\subsubsection{Extremal Kerr}

The near-horizon geometry of extremal Kerr with angular momentum $\cJ = J$ can be obtained by the above procedure, starting from the extremal Kerr metric written in usual Boyer--Lindquist coordinates; see the original derivation in~\cite{Bardeen:1999px} as well as in~\cite{Guica:2008mu,Bredberg:2011hp}. The result is the so-called ``NHEK geometry'', which is written as \eqref{GenExt} without matter fields and with 
\bea
\alpha(\theta) &=& 1, \qquad \Gamma(\th) = J (1+\cos^2\theta) \, , \nn\\ 
\gamma(\th) &=& \frac{2 \sin\theta}{1+\cos^2\theta} \, , \qquad k=1 \, . \label{NHEKvals}
\eea
The angular momentum only affects the overall scale of the geometry. There is a value $\theta_\star = \arcsin(\sqrt{3}-1)\sim 47$ degrees for which $\p_t$ becomes null. For $\theta_\star < \theta < \pi - \theta_\star$ (a finite range around the equator $\theta = \frac{\pi}{2}$), $\p_t$ is spacelike. This feature is a consequence of the presence of the ergoregion in the original Kerr geometry. Near the equator we have a ``stretched'' AdS\sub{3} self-dual orbifold (as the $S^1$ fiber is streched), while near the poles we have a ``squashed'' AdS\sub{3} self-dual orbifold (as the $S^1$ fiber is squashed). 

\subsubsection{Extremal Reissner--Nordstr\"om}

The extremal Reissner--Nordstr\"om black hole is determined by only one parameter: the electric charge $Q$. We use the normalization of the gauge field such that the Lagrangian is proportional to $R-F_{ab}F^{ab}$. The mass is $\cM =Q$ and the horizon radius is $r_+=r_-=Q$. This black hole is static and, therefore, its near-horizon geometry takes the form~\eqref{nearstatic}. We have explicitly
\be
\nu_1=Q^2,\quad \nu_2=Q^2,\quad e=Q,\quad p = 0.\label{KNvals}
\ee

\subsubsection{Extremal Kerr--Newman}

It is useful to collect the different functions characterizing the near-horizon limit of the extremal Kerr--Newman black hole. The black hole has mass $\cM = \sqrt{a^2 + Q^2}$. The horizon radius is given by $r_+ = r_- =\sqrt{a^2 + Q^2}$. One finds
\bea
\alpha(\theta) &=& 1,\qquad \Gamma(\th) = r_+^2+a^2 \cos^2\theta,\nn\\ 
\gamma(\th) &=& \frac{(r_+^2+a^2) \sin\theta}{r_+^2+a^2\cos^2\theta},\qquad k=\frac{2a r_+}{r_+^2+a^2}\, ,\label{fctsKN} \\
f(\theta) &=& Q \left(\frac{r_+^2+a^2}{2ar_+}\right)\frac{r_+^2-a^2\cos^2\theta}{r_+^2+a^2\cos^2\theta},\qquad e=\frac{Q^3}{r_+^2+a^2}.\nn
\eea
In the limit $Q \rightarrow 0$, the NHEK functions~\eqref{NHEKvals} are recovered. The near-horizon geometry of extremal Kerr--Newman is therefore smoothly connected to the near-horizon geometry of Kerr. In the limit $a \rightarrow 0$ one finds the near-horizon geometry of the Reissner--Nordstr\"om black hole~\eqref{KNvals}. The limiting procedure is again smooth.

\subsubsection{Extremal Kerr--Newman--AdS}
\label{sec:defAdSKerrN}

As a last example of near-horizon geometry, let us discuss the extremal spinning charged black hole in AdS or Kerr--Newman--AdS black hole in short. The Lagrangian is given by $L \sim R+6/l^2-F^2$ where $l^2 > 0 $. It is useful for the following to start by describing a few properties of the non-extremal Kerr--Newman--AdS black hole. The physical mass, angular momentum, electric and magnetic charges at extremality are expressed in terms of the parameters $(M,a,Q_e,Q_m)$ of the solution as
\bea
\cM &=& \frac{M}{\Xi^2},\qquad \cJ = \frac{a M}{\Xi^2}, \\
\cQ_e &=& \frac{Q_e}{\Xi}, \qquad \cQ_m = \frac{Q_m}{\Xi} ,
\eea
where $\Xi= 1 - a^2/l^2$ and $Q^2 = Q_e^2 +Q_m^2$. The horizon radius $r_+\; (r_-)$ is defined as the largest (smallest) root, respectively, of
\be
\Delta_r = (r^2+a^2)(1+r^2/l^2)-2M r + Q^2. \label{defDeltar}
\ee
Hence, one can trade the parameter $M$ for $r_+$. If one expands $\Delta_r$ up to quadratic order around $r_+$, one finds 
\be
\Delta_r = \Delta_0(r_+-r_\star)(r-r_+)+\Delta_0(r-r_+)^2+O(r-r_+)^3\, ,\label{Deltaexpanded}
\ee
where $\Delta_0$ and $r_\star$ are defined by 
\bea
\Delta_0 &=& 1+a^2/l^2+6r_+^2/l^2, \nn \\ 
\Delta_0(r_+-r_\star) &=& r_+ \left( 1+\frac{a^2}{l^2}+\frac{3r_+^2}{l^2}-\frac{a^2+Q^2}{r_+^2}\right).\label{def2spAdSKN}
\eea
In AdS, the parameter $r_\star$ obeys $r_- \leq r_\star \leq r_+$, and coincides with $r_-$ and $r_+$ only at extremality. In the flat limit $l \rightarrow \infty$, we have $\Delta_0 \rightarrow 1$ and $r_\star \rightarrow r_-$. The Hawking temperature is given by 
\be
T_H = \frac{\Delta_0(r_+-r_\star)}{4\pi (r_+^2+a^2)}.
\ee
The extremality condition is then $r_+ = r_\star=r_-$ or, more explicitly, the following constraint on the four parameters $(r_+,a,Q_e,Q_m)$,
\be
1+\frac{a^2}{l^2}+\frac{3r_+^2}{l^2}-\frac{a^2+Q^2}{r_+^2} = 0.\label{extconstraintAdSKN}
\ee

The near-horizon geometry was obtained in~\cite{Hartman:2008pb,Chen:2010jj} (except the coefficient $e$ given here). The result is
\bea
\Gamma(\theta) &=& \frac{\rho_+^2}{\Delta_0},\qquad \alpha(\theta) = \frac{\Delta^{1/2}_0}{\Delta^{1/2}_\theta},\qquad k = \frac{2ar_+ \Xi}{\Delta_0 (r_+^2+a^2)},\nn \\
\gamma(\theta) &=& \frac{\Delta^{1/2}_\theta \Delta^{1/2}_0 (r_+^2+a^2) \sin\theta}{\rho_+^2 \Xi}, \quad e = \frac{Q_e}{\Delta_0} \frac{r_+^2-a^2}{r_+^2+a^2},\\
f(\th)&=& \frac{(r_+^2+a^2)[Q_e (r_+^2-a^2 \cos^2\theta)+2Q_m a r_+ \cos\theta ]}{2\rho^2_+ \Xi a r_+}\, ,\nn 
\eea
where we defined
\bea
\Delta_\theta &=& 1- \frac{a^2}{l^2}\cos^2\theta,\qquad \rho_+^2 = r_+^2 + a^2 \cos^2\theta,\nn \\
r_0^2 &=& \frac{r_+^2+a^2}{\Delta_0}.\label{defspAdSKN}
\eea
The near-horizon geometry of the extremal Kerr--Newman black hole is recovered in the limit $l \rightarrow \infty$.

\subsubsection{Extremal BTZ}

Next we consider the simplest 3-dimensional example: the BTZ black hole solution of Einstein gravity with negative cosmological constant \cite{Banados:1992wn,Banados:1992gq}. Our convention for the Lagrangian is $L \sim R+2/l^2$. The black hole is characterized by its mass $\mathcal M$ and angular momentum $\mathcal J$. There are two extremal limits defined as $\mathcal M l = \pm \mathcal J$. In either case, the near-horizon geometry has $SL(2,\mathbb R) \times U(1)$ isometry and is given by
\be
ds^2 = \frac{l^2}{4} \left(-r^2\, dt^2 + \frac{dr^2}{r^2}+2{|\mathcal J|}(d\phi+\frac{r}{\sqrt{2{|\mathcal J|}}} dt)^2 \right),\label{selfdualorb}
\ee
with $\phi\sim\phi+2\pi$, which is known as the self-dual AdS\sub{3} orbifold~\cite{Coussaert:1994tu}.

\subsection{Thermodynamics at extremality}
\label{sec:the}

Black holes are characterized by thermodynamic variables which obey an analogue of the standard four laws of thermodynamics \cite{Bardeen:1973gs}. In this section we review various definitions of thermodynamic variables at extremality, discuss their equivalence and scope and deduce the consequences of the four laws of thermodynamics at extremality. We also mention two dynamical properties of the entropy which go beyond the thermodynamic limit: the entropy as extremum of the entropy function and the form of quantum logarithmic corrections. 

\subsubsection{Entropy}
\label{sec:entropy}

The classical entropy of any black hole in Einstein gravity coupled to matter fields such as~\eqref{generalaction} is given by 
\be
\cS = \frac{1}{4 G_N \hbar}\int_\Sigma vol(\Sigma),\label{entropy1}
\ee
where $\Sigma$ is a cross-section of the black-hole horizon and $G_N$ is the four-dimensional Newton's constant. In the near-horizon geometry, the horizon is formally located at any value of $r$ as a consequence of the definition~\eqref{nearcoord}. Nevertheless, we can move the surface $\Sigma$ to any finite value of $r$ without changing the integral, thanks to the scaling symmetry $\zeta_0$ of \eqref{killing}. Evaluating the expression~\eqref{entropy1}, we obtain
\be
\cS = \frac{\pi}{2 G_N \hbar}\int_0^\pi d\theta \;\alpha(\th)\Gamma(\th)\gamma(\th)\, .\label{SEinstein}
\ee
In particular, the entropy of the extremal Kerr black hole is given by
\be
\cS = 2\pi \cJ\, .\label{SKerr}
\ee
In units of $\hbar$ the angular momentum $\cJ$ is a dimensionless half-integer. The main result~\cite{Guica:2008mu,Lu:2008jk,Azeyanagi:2008kb,Hartman:2008pb,Nakayama:2008kg,Chow:2008dp,Isono:2008kx,Azeyanagi:2008dk,Lu:2009gj,Compere:2009dp} of the Kerr/CFT correspondence that we will review below 
is the derivation of the entropy~\eqref{SEinstein} using Cardy's formula~\eqref{Cardy}. 

When higher derivative corrections are considered, the entropy does not scale any more like the horizon area. The black-hole entropy at equilibrium can still be defined as the quantity that obeys the first law of black-hole mechanics, where the mass, angular momenta and other extensive quantities are defined with all higher-derivative corrections included. More precisely, the entropy is first defined for non-extremal black holes by integrating the first law, and using properties of non-extremal black holes, such as the existence of a \emph{bifurcation surface}~\cite{Wald:1993nt,Iyer:1994ys}. The resulting entropy formula is unique and given by 
\be
\cS = -\frac{2\pi }{\hbar}\int_{\Sigma} \frac{\delta^{\text{cov}} L }{\delta R_{abcd}}\eps_{ab}\eps_{cd} vol(\Sigma)\, ,\label{entropy3}
\ee
where $\eps_{ab}$ is the \emph{binormal} to the horizon, i.e., the volume element of the normal bundle to $\Sigma$. One can define it simply as $\eps_{ab} = n_a \xi_b - \xi_a n_b$, where $\xi$ is the generator of the horizon and $n$ is an outgoing null normal to the horizon defined by $n^2 = 0$ and $n^a \xi_a = -1$. Since the Lagrangian is diffeomorphism invariant (possibly up to a boundary term), it can be expressed in terms of the metric, the matter fields and their covariant derivatives, and the Riemann tensor and its derivatives. This operator $\delta^{\text{cov}}/\delta R_{abcd}$ acts on the Lagrangian while treating the Riemann tensor as if it were an independent field. It is defined as a covariant Euler--Lagrange derivative as 
\be
\frac{\delta^{\text{cov}} }{\delta R_{abcd}} = \sum_{i=0} (-1)^i \nabla_{(e_1}\dots \nabla_{e_i)}\frac{\p}{\p \nabla_{(e_1} \dots \nabla_{e_i)}R_{abcd}}\,.\label{Riemcov}
\ee
Moreover, the entropy formula is conserved away from the bifurcation surface along the future horizon as a consequence of the zeroth law of black-hole mechanics~\cite{Jacobson:1993vj}. Therefore, one can take the extremal limit of the entropy formula evaluated on the future horizon in order to define entropy at extremality. Quite remarkably, the Iyer--Wald entropy~\eqref{entropy3} can also be reproduced~\cite{Azeyanagi:2009wf} using Cardy's formula as we will detail below. 

The non-extremal entropy formula \eqref{entropy3} is the covariant phase space conserved charge associated with the Killing generator of the horizon, normalized with the inverse temperature, $\xi \equiv \frac{2\pi}{\kappa} (\p_t + \Omega_J \p_\phi)$. In the extremal limit, $\kappa \rightarrow 0$ and the generator is not well-defined. One can then ask whether the extremal entropy \eqref{entropy3} is still a covariant phase space conserved charge, this time  associated with a Killing vector defined in the near-horizon region. The answer is positive \cite{Hajian:2013lna,Ali2014}. One defines an arbitrary surface $\Sigma$ as $r=r_H$, $t= t_H$ in the near-horizon region. This surface is a bifurcation surface because it exists a Killing vector in the near-horizon geometry which vanishes on $\Sigma$. It is given by
\bea
\xi \equiv  n_\Sigma^{(a)} \zeta_{(a)} - k_J \p_\phi
\eea
where the $SL(2,\mathbb R)$ generators $\zeta_{(a)}$, $(a)=-1,0,1$ are given in \eqref{killing} and $n_\Sigma^{(a)}$ are chosen such that $\xi =0$ on $\Sigma$ and $g_{(a)(b)}n_\Sigma^{(a)}n_\Sigma^{(b)}=-1$ where $g_{(a)(b)}$ is the $SL(2,\mathbb R)$ Killing metric. Using $\nabla_{[a} \xi_{b]} = \eps_{ab}$ one then find the associated conserved charge \eqref{entropy3} \cite{Hajian:2013lna}.

In five-dimensional Einstein gravity coupled to matter, the entropy of extremal black holes can be expressed as 
\be
\cS = \frac{\pi^2}{\hbar G_N} \int_0^\pi d\theta \, \alpha(\theta)\Gamma(\theta) \gamma(\theta)\, ,\label{entropy5d}
\ee
where $\Gamma(\th)$ and $\alpha(\th)$ have been defined in \eqref{near5d} and $\gamma(\th)^2 = \det(\gamma_{ij}(\theta)^2)$.

From the attractor mechanism for four-dimensional extremal spinning black holes~\cite{Astefanesei:2006dd}, the entropy at extremality can be expressed as an extremum of the functional
\bea
\mathcal E(\Gamma(\th),\alpha(\th),\gamma(\th),f^I(\theta),\chi^A(\theta),k,e_I) &=& 2\pi(k \cJ+e_I \cQ^I - \int_\Sigma d\theta d\phi \sqrt{-g}\mathcal L),
\eea
where $\mathcal L$ is the Lagrangian. More precisely, the extremum is defined from the equations
\bea
\frac{\delta \mathcal E}{\delta \Gamma(\th)} =\frac{\delta \mathcal E}{\delta \alpha(\th)} =\frac{\delta \mathcal E}{\delta \gamma(\th)} =\frac{\delta \mathcal E}{\delta f^I(\th)} = \frac{\delta \mathcal E}{\delta \chi^A(\th)} = \frac{\delta \mathcal E}{\delta k} =  \frac{\delta \mathcal E}{\delta e_I}  = 0
\eea
which are equivalent to the restriction of the equations of motion on the near-horizon ansatz \eqref{GenExt}. The entropy can be shown to depend only on the angular momentum $\cJ$ and the conserved charges $\cQ^I_{e,m}$, 
\be
\cS=\cS_{\ext}(\cJ,\cQ^I_e,\cQ^I_m),
\ee
and depend in a discontinuous fashion on the scalar moduli~\cite{Sen:2005wa}. The result holds for any Lagrangian in the class~\eqref{generalaction}, including higher-derivative corrections, and the result can be generalized straightforwardly to five dimensions.

When quantum effects are taken into account, the entropy formula also gets modified in a non-universal way, which depends on the matter present in quantum loops. In Einstein gravity, the main correction to the area law is a logarithmic correction term. The logarithmic corrections to the entropy of extremal rotating black holes of area $A_+$ can be obtained using the quantum entropy function formalism~\cite{Sen:2012cj} as 
\bea
\cS  = \frac{A_+}{4} +\frac{1}{180}\log A_+ (64+2 n_S -26 n_V+7 n_F - \frac{233}{2}n_{3/2}) +\dots
\eea
where besides gravity, it is assumed that the low energy action contains $n_S$ minimally coupled massless scalars, $n_V$ minimally coupled massless vector fields, $n_F$ minimally coupled massless Dirac fermions and $n_{3/2}$ minimally coupled massless Majorana Rarita-Schwinger fields.

\subsubsection{Temperature and chemical potentials}
\label{sec:temp}

Even though the Hawking temperature is zero at extremality, quantum states just outside the horizon are not pure states when one defines the vacuum using the generator of the horizon. Let us review these arguments following~\cite{Guica:2008mu,Hartman:2008pb,Chow:2008dp}. We assume that all thermodynamical quantities are analytic as function of the parameters defining the near-horizon geometries. We will drop the index $I$ distinguishing different gauge fields since this detail is irrelevant to the present arguments.

From the expression of the entropy in terms of the charges $\cS_{\ext}(\cJ,\cQ_e,\cQ_m)$, one can define the chemical potentials
\be
\frac{1}{T_{\phi}} = \left( \frac{\p \cS_{\ext}}{\p \cJ}\right)_{\cQ_{e,m}},\quad \frac{1}{T_{e}} = \left( \frac{\p \cS_{\ext}}{\p \cQ_e}\right)_{\cJ,\cQ_{m}},\quad
\frac{1}{T_{m}} = \left( \frac{\p \cS_{\ext}}{\p \cQ_m}\right)_{\cJ,\cQ_{e}} .\nn
\ee
Note that electromagnetic charges are quantized, but when the charges are large one can use the continuous thermodynamic limit. These potentials obey the balance equation
\be
\delta \cS_{\ext} = \frac{1}{T_{\phi}}\delta \cJ + \frac{1}{T_e}\delta \cQ_e+ \frac{1}{T_m}\delta \cQ_m\, .\label{limitfirstlaw}
\ee

Another way to obtain these potentials is as follows. At extremality, any fluctuation obeys
\be
0=T_H \delta \cS = \delta \cM - (\Omega^{\ext}_J \delta \cJ + \Phi^{\ext}_e \delta \cQ_e + \Phi^{\ext}_m \delta \cQ_m),\label{eq:1stlawext}
\ee
where $\Omega^{\ext}_J$ is the angular potential at extremality and $\Phi^{\ext}_{e,m}$ are electric and magnetic potentials at extremality; see Section~\ref{propBH} for a review of these concepts.

One can express the first law at extremality~\eqref{eq:1stlawext} as follows: any variation in $\cJ$ or $\cQ_{m,e}$ is accompanied by an energy variation. One can then solve for $\cM=\cM_{\ext}(\cJ,\cQ_e,\cQ_m)$. The first law for a non-extremal black hole can be written as
\be
\delta \cS = \frac{1}{T_H}\left( \delta \cM - (\Omega_J \delta \cJ + \Phi_e \delta \cQ_e + \Phi_m \delta \cQ_m) \right)\, .\label{eq:89}
\ee
Let us now take the extremal limit using the following ordering. We first take extremal variations with $\delta \cM = \delta \cM_{\ext}(\cJ,\cQ_e,\cQ_m)$. Then, we take the extremal limit of the background configuration. We obtain \eqref{limitfirstlaw} with
\bea
T_{\phi} &=& \lim_{T_H \rightarrow 0} \frac{T_H}{\Omega_J^{\ext} - \Omega_J} = -\left. \frac{\p T_H / \p r_+}{ \p \Omega_J/ \p r_+} \right|_{r_+ = r_{\ext}}\, ,\label{formTF1} \\
T_{e,m} &=& \lim_{T_H \rightarrow 0} \frac{T_H}{\Phi_{e,m}^{\ext} - \Phi_{e,m}} = -\left. \frac{\p T_H / \p r_+}{ \p \Phi_{e,m}/ \p r_+} \right|_{r_+ = r_{\ext}}\, ,\label{formTem}
\eea
where the extremal limit can be practically implemented by taking the limit of the horizon radius $r_+$ to the extremal horizon radius $r_{\ext}$. 

The interpretation of these chemical potentials can be made in the context of quantum field theories in curved spacetimes; see~\cite{Birrell:1982ix} for an introduction. The Hartle--Hawking vacuum for a Schwarzschild black hole, restricted to the region outside the horizon, is a density matrix $\rho = e^{-\omega/T_{H}}$ at the Hawking temperature $T_H$. For spacetimes that do not admit a global timelike Killing vector, such as the Kerr geometry, the Hartle--Hawking vacuum does not exist, but one can use the generator of the horizon to define positive frequency modes and, therefore, define the vacuum in the region where the generator is timelike (close enough to the horizon). This is known as the Frolov--Thorne vacuum~\cite{Frolov:1989jh} (see also~\cite{Duffy:2005mz}). One can take a suitable limit of the definition of the Frolov--Thorne vacuum to provide a definition of the vacuum state for any spinning or charged extremal black hole.

Quantum fields for non-extremal black holes can be expanded in eigenstates with asymptotic energy $\hat \omega$ and angular momentum $\hat m$ with $\hat t$ and $\hat \phi$ dependence as $e^{-i\hat \omega \hat t + i\hat m \hat \phi}$. When approaching extremality, one can perform the change of coordinates~\eqref{nearcoord} in order to zoom close to the horizon. By definition, the scalar field $\phi$ in the new coordinate system $x^a=(t,\phi,\theta,r)$ reads in terms of the scalar field $\hat \phi$ in the asymptotic coordinate system $\hat x^a=(\hat t,\hat \phi,\theta,\hat r)$ as $\phi(x^a)=\hat \phi(\hat x^a)$. We can then express
\be
e^{- i \omega t + i m \phi} = e^{-i \hat \omega \hat t + i \hat m \phi},
\ee
and the near-horizon parameters are
\be
m = \hat m,\qquad \omega = \frac{\hat\omega - m \Omega_J}{\lambda} \, .\label{modes1}
\ee
When no electromagnetic field is present, any finite energy $\omega$ in the near-horizon limit at extremality $\lambda \rightarrow 0$ corresponds to eigenstates with $\hat \omega = \hat m \Omega^{\ext}_J$. When electric fields are present, zooming in on the near-horizon geometry from a near-extremal solution requires one to perform the gauge transformation $A(x) \rightarrow A(x) + d\Lambda(x) $ with gauge parameter given in \eqref{valLambda}, which will transform the minimally-coupled charged scalar wavefunction by multiplying it by $e^{i q_e \Lambda(x)}$. Finite energy excitations in the near-horizon region then require $\hat \omega = m \Omega_J^{\ext}+q_e \Phi_e^{\ext}$. Invoking (classical) electromagnetic duality, the magnetic contribution has the same form as the electric contribution. In summary, the general finite-energy extremal excitation has the form
\be
\hat \omega = m \Omega_J^{\ext}+q_e \Phi_e^{\ext}+q_m \Phi_m^{\ext}\, .\label{hatomega}
\ee

Following Frolov and Thorne, we assume that quantum fields in the non-extremal geometry are populated with the Boltzmann factor
\be
\exp\left(\hbar\frac{\hat \omega - \hat m \Omega_J - \hat q_e \Phi_e - \hat q_m \Phi_m}{T_H}\right),
\ee
where $\hat q_{e,m}$ are the electric and magnetic charge operators. We also assume that modes obey \eqref{hatomega} at extremality. Using the definitions~\eqref{formTF1}\,--\,\eqref{formTem}, we obtain the non-trivial extremal Boltzmann factor in the extremal and near-horizon limit
\be
\exp\left(-\hbar\frac{m}{T_{\phi}} - \hbar\frac{q_e}{T_e}-\hbar\frac{q_m}{T_m}\right),\label{bol1}
\ee
where the mode number $m$ and charges $q_{e,m}$ in the near-horizon region are equal to the original mode number and charges $\hat m,\hat q_{e,m}$. This completes the argument that the Frolov--Thorne vacuum is non-trivially populated in the extremal limit.

Now, as noted in~\cite{Amsel:2009ev}, there is a caveat in the previous argument for the Kerr black hole and, as a trivial generalization, for all black holes that do not possess a global timelike Killing vector. For any non-extremal black hole, the horizon-generating Killing field is timelike just outside the horizon. If there is no global timelike Killing vector, this vector field should become null on some surface at some distance away from the horizon. This surface is called the velocity of light surface. For positive-energy matter, this timelike Killing field defines a positive conserved quantity for excitations in the near-horizon region, ruling out instabilities. However, when approaching extremality, it might turn out that the velocity of light surface approaches asymptotically the horizon. In that case, the horizon-generating Killing field of the extreme black hole may not be everywhere timelike. This causes serious difficulties in defining quantum fields directly in the near-horizon geometry~\cite{Kay:1988mu,Ottewill:2000qh,Ottewill:2000yr}. However, (at least classically) dynamical instabilities might appear only if there are actual bulk degrees of freedom in the near-horizon geometries. We will argue that this is not the case in Section~\ref{sec:nobulkdof}. As a conclusion, extremal Frolov--Thorne temperatures can be formally and uniquely defined as the extremal limit of non-extremal temperatures and chemical potentials. However, the physical interpretation of these quantities is better understood finitely away from extremality.

The condition for having a global timelike Killing vector was spelled out in \eqref{globalKillingtime}. This condition is violated for the extremal Kerr black hole or for any extremal Kerr--Newman black hole with $a \geq Q/\sqrt{3}$, as can be shown by using the explicit values defined in \eqref{sec:explicitnear}. (The extremal Kerr--Newman near-horizon geometry does possess a global timelike Killing vector when $a < Q/\sqrt{3}$ and the Kerr--Newman--AdS black holes do as well when $4a^2/(\Delta_0 r_+^2)<1$, which is true for large black holes with $r_+ \gg l$. Nevertheless, there might be other instabilities due to the electric superradiant effect.) 

The extremal Frolov--Thorne temperatures should also be directly encoded in the metric~\eqref{GenExt}. More precisely, these quantities should only depend on the metric and matter fields and not on their equations of motion. Indeed, from the derivation~\eqref{formTF1}\,--\,\eqref{formTem}, one can derive these quantities from the angular velocity, electromagnetic potentials and surface gravity, which are kinematical quantities. More physically, the Hawking temperature arises from the analysis of free fields on the curved background, and thus depends on the metric but not on the equations of motion that the metric solves. It should also be the case for the extremal Frolov--Thorne temperatures. Using a reasonable ansatz for the general black-hole solution of~\eqref{generalaction}, including possible higher-order corrections, and assuming analyticity at the horizon, one can derive~\cite{Chow:2008dp,Azeyanagi:2009wf,Hajian:2013lna} the very simple formula
\be
T_{\phi} = \frac{1}{2\pi k}\, .\label{TL}
\ee
From similar considerations, it should also be possible to derive a formula for $T_e$ in terms of the functions appearing in \eqref{GenExt} as
\be
T_e = \frac{1}{2\pi e}\, .\label{Teansatz}
\ee
and prove the equivalence between \eqref{Teansatz} and \eqref{formTem}. The formula \eqref{Teansatz} is consistent with the thermodynamics of (AdS)--Kerr--Newman black holes as one can check from the formulae in Section~\ref{sec:explicitnear}. 

Similarly, one can work out the thermodynamics of five-dimensional rotating black holes. Since there are two independent angular momenta $\cJ_1$, $\cJ_2$, there are also two independent chemical potentials $T_{\phi_1}$, $T_{\phi_2}$ associated with the angular momenta. The same arguments lead to 
\be
T_{\phi_1} = \frac{1}{2\pi k_1},\qquad T_{\phi_2} = \frac{1}{2\pi k_2}\, ,\label{T125d}
\ee
where $k_1$ and $k_2$ are defined in the near-horizon solution~\eqref{near5d}. 

When considering the uplift~\eqref{KKlift} of the gauge field along a compact direction of length $2\pi R_\chi$, one can use the definition \eqref{T125d} to define the chemical potential associated with the direction $\p_\chi$. Since the circle has a length $2\pi R_\chi$, the extremal Frolov--Thorne temperature is expressed in units of $R_\chi$, 
\be
T_\chi \equiv T_e R_\chi = \frac{R_\chi}{2\pi e}\, ,\label{defTchi}
\ee
where $T_e$ is defined in \eqref{Teansatz}. 

\subsubsection{The three laws of near-horizon geometries}

In summary, near-horizon geometries of 4-dimensional extremal black holes obey the following three laws:

\begin{description}
\item[Balance equation] The entropy at extremality obeys the balance equation
\be
\delta \cS_{\ext} = \frac{1}{T_{\phi}}\delta \cJ + \frac{1}{T_e}\delta \cQ_e+ \frac{1}{T_m}\delta \cQ_m\, .\label{limitfirstlaw2}
\ee

\item[Zero law] The angular chemical potential $T_\phi$ and electromagnetic potential $T_e$ are given by
\bea
T_\phi = \frac{1}{2 \pi k},\qquad T_e = \frac{1}{2\pi e}
\eea
where $k,\, e$ are constants over the near-horizon geometry.

\item[Entropy function law] The entropy at extremality is the extremum of the entropy function
\bea
\mathcal E \equiv \frac{\cJ}{T_\phi}+\frac{\cQ_e}{T_e}+\frac{\cQ_m}{T_m} - 2\pi \int_\Sigma d\theta d\phi \sqrt{-g}\mathcal L,
\eea
among near-horizon geometries of the form \eqref{GenExt} where $\mathcal L$ is the Lagrangian and $\Sigma$ is a sphere at fixed time and radius.
\end{description}

Let us comment on these laws. The balance equation relates the variation of charges at extremality. It can be deduced from the first law of thermodynamics at leading order in the extremal limit, assuming analyticity and smoothness of thermodynamical quantities in the extremal limit. The balance equation can be generalized in two ways. First, instead of assuming extremal variations, one can take arbitrary parametric variations of the black hole, including non-extremal ones. It was shown in \cite{Johnstone:2013ioa} that analyticity at the horizon implies $\delta \cM = \delta \cM_{\ext}(\cJ,\cQ_e,\cQ_m)+O(\eps^2)$ where $\eps$ is an extremality parameter with $T_H \sim \eps$. Therefore, the equation \eqref{limitfirstlaw2} follows from the non-extremal first law \eqref{eq:89} and from \eqref{formTF1}-\eqref{formTem}. One can also deduce \eqref{limitfirstlaw2} for generic linear perturbations around the near-horizon geometry using the Wald covariant phase space approach, under the condition that these perturbations have vanishing $SL(2,\mathbb R)$ charges \cite{Hajian:2013lna}.

The zero law is a consequence of both the zero law of thermodynamics for non-extremal black holes and the presence of $SL(2,\mathbb R)$ invariance in the extremal limit, which ensures that $k$ and $e$ are constant. Since no dynamical processes outside equilibrium are  allowed in the near-horizon geometries as we will discuss in Section \ref{sec:nobulkdof}, there is no analogue of the second law of thermodynamics $\delta \mathcal S \geq 0$ at extremality. 

The entropy function law is a statement about the dynamics of gravity among the class of near-horizon geometries and it has no obvious analogue away from extremality since it depends upon the existence of the near-horizon limit. It provides with the ground state entropy at extremality as a function of the other dynamical quantities in the system. 
The generalization to lower and higher dimensions is straightforward.

\subsubsection{Temperatures and entropies of specific extremal black holes}

The entropy of the extremal Kerr black hole is $\cS_{\ext} = 2\pi J$. Integrating \eqref{limitfirstlaw} or using the explicit near-horizon geometry and using \eqref{TL}, we find
\be
T_\phi = \frac{1}{2\pi}\, ,
\ee
and $T_e$ is not defined. 

The entropy of the extremal Reissner--Nordstr\"om black hole is $\cS_{\ext} = \pi Q^2$. Integrating \eqref{limitfirstlaw}, we obtain 
\be
T_e = \frac{1}{2\pi Q}\, ,\label{TeRN}
\ee
while $T_\phi$ is not defined.

For the electrically-charged Kerr--Newman black hole, the extremal entropy reads as $\cS_{\ext} = \pi (a^2+r_+^2)$. Expressing the entropy in terms of the physical charges $Q = \sqrt{r_+^2-a^2}$ and $J = a r_+$, we obtain 
\be
\cS_{\ext} = \frac{\pi}{2}\left( \frac{4J^2}{\sqrt{Q^4+4J^2}-Q^2}+\sqrt{Q^4+4 J^2}-Q^2 \right)\, .
\ee
Using the definition \eqref{limitfirstlaw} and re-expressing the result in terms of the parameters $(a,r_+)$ we find
\be
T_\phi = \frac{a^2+r_+^2}{4\pi a r_+},\qquad T_e = \frac{a^2+r_+^2}{2\pi (r_+^2-a^2)^{3/2}}.\label{TKN}
\ee
We can also derive $T_\phi$ from \eqref{TL} and the explicit near-horizon geometry~\eqref{fctsKN}. $T_e$ is consistent with \eqref{Teansatz}.

For the extremal Kerr--Newman--AdS black hole, the simplest way to obtain the thermodynamics at extremality is to compute \eqref{formTF1}\,--\,\eqref{formTem}. Using the extremality constraint~\eqref{extconstraintAdSKN}, we obtain 
\be
T_\phi = \frac{(a^2+r_+^2)\Delta_0}{4\pi a r_+ \Xi}, \qquad 
T_e = \frac{(a^2+r_+^2)\Delta_0}{2\pi Q_e (r_+^2-a^2)}\, ,
\ee
where we used the definitions~\eqref{defspAdSKN}. The magnetic potential $T_m$ can then be obtained by electromagnetic duality. The expressions coincide with \eqref{TL}\,--\,\eqref{Teansatz}. These quantities reduce to \eqref{TKN} in the limit of no cosmological constant when there is no magnetic charge, $q_m = 0$. The extremal entropy is given by $\cS_{\ext} = \pi(r_+^2+a^2)/\Xi$.

\subsection{Near-extremal near-horizon geometries} 
\label{sec:nearext1}

An important question about near-horizon geometries is the following: how much dynamics of gravity coupled to matter fields is left in a near-horizon limit such as~\eqref{nearcoord}? 
We already discussed in Section~\ref{sec:spin} the absence of non-perturbative solutions in near-horizon geometries, such as black holes. In this section, we will discuss the existence of near-extremal solutions obtained from a combined near-horizon limit and zero temperature limit. We will show that these solutions are related to extremal near-horizon geometries via a non-trivial diffeomorphism and we will point out that their temperature has to be fixed in order to be able to define the energy (which is then fixed as well). There is therefore no black hole of arbitrary energy in near-horizon geometries. In Section~\ref{sec:nobulkdof}, we will argue for the absence of local bulk degrees of freedom. We will discuss later in Section~\ref{sec:microS} the remaining non-trivial boundary dynamics generated by large diffeomorphisms.

Let us first study infinitesimal perturbations of the near-horizon geometry~\eqref{GenExt}. As a consequence of the change of coordinates and the necessary shift of the gauge field~\eqref{nearcoord}, the near-horizon energy $ \delta \mathcal Q_{\p_t}$ of an infinitesimal perturbation is related to the charge associated with the generator of the horizon $\xi_{\mathrm{tot}} \equiv (\xi,\Lambda) = (\p_t + \Omega^{\ext}_J \p_\phi,\Phi_e^{\ext})$ as follows, 
\be
\delta \mathcal Q_{\xi_{\mathrm{tot}}} = \frac{\lambda}{r_0} \delta \mathcal Q_{\p_t} \, ,\qquad \lambda \rightarrow 0, \label{extcond3}
\ee
as derived in Sections~\ref{sec:stat} and \ref{sec:spin}. Assuming no magnetic charges for simplicity, the conserved charge $\delta \mathcal Q_{\xi_{\mathrm{tot}}} $ is given by $\delta \mathcal M-\Omega^{\ext}_J \delta\cJ -\Phi^{\ext}_e \delta \mathcal Q_e$.\epubtkFootnote{Our conventions for the infinitesimal charges associated with symmetries is as follows: the energy is $\delta\mathcal M = \delta\mathcal Q_{\p_t}$, the angular momentum is $\delta\mathcal J = \delta\mathcal Q_{-\p_\phi}$ and the electric charge is $\delta\mathcal Q_e =\delta \mathcal Q_{-\p_{\chi}}$. In other words, the electric charge is associated with the gauge parameter $\Lambda=-1$. The first law then reads $T_H\delta \cS=\delta \mathcal M-\Omega_J \delta \mathcal J-\Phi_e \delta \mathcal Q_e$.} Using the first law of thermodynamics valid for arbitrary (not necessarily stationary) perturbations, the left-hand side of \eqref{extcond3} can be expressed as
\be
T_H \delta \cS_{\ext} = \delta \mathcal Q_{\xi_{\mathrm{tot}}} \, . \label{extcond4}
\ee
Any geometry that asymptotes to \eqref{GenExt} will have finite near-horizon energy $ \mathcal Q_{\p_t}$. Indeed, an infinite near-horizon energy would be the sign of infrared divergences in the near-horizon geometry and it would destabilize the geometry. It then follows from \eqref{extcond3}\,--\,\eqref{extcond4} that any infinitesimal perturbation of the near-horizon geometry~\eqref{GenExt} will correspond to an extremal black-hole solution with vanishing Hawking temperature, at least such that $T_H = O(\lambda)$. Common usage refers to black-hole solutions, where $T_H \sim \lambda$ as near-extremal black holes. Nevertheless, it should be emphasized that after the exact limit $\lambda \rightarrow 0$ is taken the Hawking temperature of such a solution is exactly zero.

We can obtain a near-extremal near-horizon geometry as follows. Starting from a stationary non-extremal black hole of mass $M$ in Boyer--Lindquist coordinates, we perform the near-horizon scaling limit~\eqref{nearcoord} together with the scaling of the temperature
\be
 T_H \rightarrow \frac{\lambda}{r_0} T^{\mathrm{near-ext}}\, .\label{defThl}
\ee
While the form of the general non-extremal solution would be required to perform that limit in detail, all examples so far in the class of theories~\eqref{generalaction}, such as the Kerr--Newman--AdS black hole, lead to the following metric
\bea 
ds^2 &=& \Gamma(\theta)\left[-r(r+4\pi T^{\mathrm{near-ext}}) d t^2+\frac{d r^2}{r(r+4\pi T^{\mathrm{near-ext}})}
+ \alpha (\t) d \theta^2+ \gamma(\theta)(d \phi + k\, rd t)^2 \right] ,\nn\\
\chi^A &=&\chi^A(\th),\qquad A^I= f^I(\t) (d \phi + k\, rdt) -\frac{e_I}{k}d\phi\, .\label{GenNearExt} 
\eea 
The near-extremal near-horizon solution~\eqref{GenNearExt} is diffeomorphic to the near-horizon geometry in Poincar\'e coordinates~\eqref{GenExt}. Denoting the finite temperature coordinates by a subscript $T$ and the Poincar\'e coordinates by a subscript $P$, and defining 
\bea
\kappa_F \equiv 2 \pi T^{\mathrm{near-ext}},
\eea
the change of coordinates reads as~\cite{Maldacena:1998uz,Spradlin:1999bn,Amsel:2009ev,Bredberg:2009pv}
\bea
t_P &=& -e^{-\kappa_F t_F} \frac{r_F + \kappa_F}{\sqrt{r_F (r_F + 2\kappa_F)}},\\
r_P &=& \frac{1}{\kappa_F}e^{\kappa_F t_F}\sqrt{r_F(r_F+2\kappa_F)},\\
\phi_P &=& \phi_F -\frac{1}{2}\log \frac{r_F}{r_F+2\kappa_F}.
\eea
or, conversely, 
\bea
t_F &=& \frac{1}{\kappa_F}\log \frac{r_P}{\sqrt{t_P^2 r_P^2 - 1}}, \\
r_F &=& -\kappa_F (1+ t_P r_P),\\
\phi_F &=& \phi_P + \frac{1}{2}\log \frac{t_P r_P + 1}{t_P r_P - 1}.
\eea
Therefore, the classical geometries are equivalent. However, since the diffeomorphism is non-trivial, there is a distinction at the quantum level. At large radius, the transformation of coordinates reads as $t_P = - e^{-\kappa_F t_F}$, $\phi_P = \phi_F$ which is the transformation between Minkowski space and Rindler space at temperature $T^{\mathrm{near-ext}}$ (compactified on a cylinder of radius $2\pi$). Fields will be therefore be quantized in a different manner in the two geometries.

Let us now compute the energy. Multiplying Eq.~\eqref{extcond3} by $r_0/\lambda$ and using \eqref{extcond4} and \eqref{defThl}, we get that the energy variation around the near-extremal geometry is given by 
\be
\slash\hspace{-5pt}\delta \mathcal Q_{\p_t} = T^{\mathrm{near-ext}} \delta \cS_{\ext}\, ,\label{heat}
\ee
where the extremal entropy $\cS_{\ext}$ can be expressed in terms of the near-horizon quantities as~\eqref{SEinstein}\epubtkFootnote{Since the derivation of the formula~\eqref{heat} was rather indirect, we checked that it is correct for the Kerr--Newman--AdS family of black holes by computing the energy variation directly using the Lagrangian charges defined in~\cite{Barnich:2007bf,Compere:2007az,Compere:2009dp}.}. Since $T^{\mathrm{near-ext}}$ is independent from the near-horizon quantities at extremality, the energy is not an exact quantity as long as $T^{\mathrm{near-ext}}$ is allowed to be varied, which we emphasize by using the notation $\slash\hspace{-5pt}\delta$. In other words, for general variations, the charge $\slash\hspace{-5pt}\delta \mathcal Q_{\p_t}$ is a heat term, which does not define a conserved energy. Requiring the energy to be defined, we need to fix $T^{\mathrm{near-ext}}$ and the energy is then given by $T^{\mathrm{near-ext}} \cS_{ext}$ which is fixed. This implies that different temperatures or equivalently different energies define distinct boundary conditions in the near-horizon region.

\subsection{Absence of bulk dynamics in near-horizon geometries} 
\label{sec:nobulkdof}

In this section, we will review arguments pointing to the absence of local degrees of freedom in the near-horizon geometries~\eqref{GenExt} or \eqref{GenNearExt}, following the arguments of~\cite{Amsel:2009ev,Dias:2009ex} for Einstein gravity in the NHEK geometry. The only non-trivial dynamics will be argued to occur at the boundary of the near-horizon geometries due to non-trivial diffeomorphisms. The analysis of these diffeomorphisms will be deferred until Section~\ref{sec:BC}. This lack of dynamics is familiar from the $\\AdS_2 \times S^2$ geometry~\cite{Maldacena:1998uz}, which, as we have seen in Sections \ref{sec:stat}-\ref{sec:spin}, is the static limit of the spinning near-horizon geometries\epubtkFootnote{In the arguments of \cite{Maldacena:1998uz}, the presence of the compact $S^2$ is crucial. Conversely, in the case of non-compact horizons, such as the extremal planar AdS--Reissner--Nordstr\"om black hole, flux can leak out the $\mathbb R^2$ boundary and the arguments do not generalize straightforwardly. There are indeed interesting quantum critical dynamics around $\\AdS_2 \times \mathbb R^2$ near-horizon geometries~\cite{Faulkner:2009wj}, but we will not touch upon this topic here since we concentrate exclusively on compact black holes.}.

Propagating degrees of freedom have finite energy. If near-horizon geometries contain propagating modes, one expects that a highly-symmetric solution would exist which has a non-trivial spectrum of energy. Such solution would then approximate the late-time thermalized state after dissipation has occurred. However, in the case of $4d$ Einstein gravity, one can prove that the NHEK (near-horizon extremal Kerr geometry) is the unique (up to diffeomorphisms) regular stationary and axisymmetric solution asymptotic to the NHEK geometry with a smooth horizon~\cite{Amsel:2009ev}. This can be understood as a Birkoff theorem for the NHEK geometry. This can be paraphrased by the statement that there are no black holes ``inside'' of the NHEK geometry. The near-extremal near-horizon geometries are not candidates for thermalized states since they do not have a non-trivial spectrum of energy, as we showed in Section \ref{sec:nearext1}.

One can also prove that there is a near-horizon geometry in the class~\eqref{GenExt}, which is the unique (up to diffeomorphisms) near-horizon stationary and axisymmetric solution of $4d$ AdS--Einstein--Maxwell theory~\cite{Kunduri:2008rs,Kunduri:2008tk,Kunduri:2011ii}. The assumption of axisymmetry can be further relaxed since stationarity implies axisymmetry~\cite{Hollands:2008wn}. Additional results can be obtained for various theories of the class \eqref{generalaction} in lower and higher dimensions, see \cite{Kunduri:2013gce}. 

It is then natural to conjecture that any stationary solution of the more general action~\eqref{generalaction}, which asymptotes to a near-horizon geometry of the form~\eqref{GenExt} is diffeomorphic to it. This conjecture remains to be proven but if correct, it would imply, together with the result in Section \ref{sec:nearext1}, that there is no non-trivial candidate stationary near-horizon solution with arbitrary finite energy in any theory of the form~\eqref{generalaction}. One can then argue that there will be no solution asymptotic to \eqref{GenExt} -- even non-stationary -- with a non-zero  energy above the background near-horizon geometry, except solutions related via a diffeomorphism.

In order to test directly whether or not there exist any local bulk dynamics in the class of geometries, which asymptote to the near-horizon geometries~\eqref{GenExt}, one can perform a linear analysis and study which modes survive at the non-linear level after backreaction is taken into account. This analysis has been performed with care for the massless spin 0 and spin 2 field around the NHEK geometry in~\cite{Amsel:2009ev,Dias:2009ex} under the assumption that all non-linear solutions have vanishing $SL(2,\mathbb R) \times U(1)$ charges (which is justified by the existence of a Birkoff theorem as mentioned earlier). The conclusion is that there is no linear mode that is the linearization of a non-linear solution. In other words, there is no local massless spin 0 or spin 2 bulk degree of freedom around the NHEK solution. 
The result could very likely be extended to massive scalars, gauge fields and gravitons propagating on the general class of near-horizon solutions~\eqref{GenExt} of the action~\eqref{generalaction}, but such an analysis has not been done at that level of generality.

\newpage

\section{Two-Dimensional Conformal Field Theories}
\label{sec:2dCFT}

Since we aim at drawing parallels between black holes and two-dimensional CFTs ($2d$~CFTs), it is useful to describe some of their key properties. An important caveat is that, as discussed in the Introduction~(\ref{sec:introduction}), there is no standard $2d$ CFT dual to the Kerr black hole, nor a chiral part of a standard $2d$ CFT dual to the extremal Kerr black hole. The language of $2d$ CFTs is however relevant to describe the properties of gravity and its probes \cite{Bredberg:2009pv,Hartman:2009nz,Porfyriadis:2014fja}.

A $2d$~CFT is defined as a local quantum field theory with local conformal invariance. Background material can be found, e.g., in~\cite{DiFrancesco:1997nk,Ginsparg:1988ui,Polchinski:1998rq}. In two-dimensions, the local conformal group is an infinite-dimensional extension of the globally-defined conformal group $SL(2,\mathbb R)\times SL(2,\mathbb R)$ on the plane or on the cylinder. It is generated by two sets of vector fields $L_n,\bar L_n$, $n\in \mathbb Z$ obeying the Lie bracket algebra known as the Witt algebra
\bea
\,[L_m,L_n] &=& (m-n)L_{m+n} \, , \nn\\
\,[L_m,\bar L_n] &=& 0 \, ,\\
\,[\bar L_m,\bar L_n] &=& (m-n)\bar L_{m+n} \, . \nn
\eea
From Noether's theorem, each symmetry is associated to a quantum operator. The local conformal symmetry is associated with the conserved and traceless stress-energy tensor operator, which can be decomposed into left and right moving modes $\cL_n$ and $\bar \cL_n$, $n\in \mathbb Z$. The operators $\cL_n,\bar \cL_n$ form two copies of the Virasoro algebra
\bea
\left[\cL_m,\cL_n \right] &=& (m-n)\cL_{m+n}+\frac{c_L}{12}m(m^2-A_L)\delta_{m+n,0} \, , \nn\\ 
\left[\cL_m,\bar \cL_n \right] &=& 0 \, , \label{Virc1} \\
\left[\bar \cL_m,\bar \cL_n \right] &=& (m-n)\bar \cL_{m+n}+\frac{c_R}{12}m(m^2-A_R)\delta_{m+n,0} \, , \nn
\eea
where $\cL_{-1},\cL_0,\cL_1$ (and $\bar \cL_{-1},\bar \cL_0,\bar \cL_1$) span a $SL(2,\mathbb R)$ subalgebra. The pure numbers $c_L$ and $c_R$ are the left and right-moving central charges of the CFT. It is generally accepted that the central charges need to be taken large in order to possibly admit a gravitationnal dual \cite{ElShowk:2011ag}. The auxiliary parameters $A_L,A_R$ depend if the CFT is defined on the plane or on the cylinder. They correspond to shifts of the background value of the zero eigenmodes $\cL_0,\bar \cL_0$. In many examples of CFTs, additional symmetries are present in addition to the two sets of Virasoro algebras.

A $2d$~CFT can be uniquely characterized by a list of (primary) operators $\mathcal O$, the conformal dimensions of these operators (their eigenvalues under $\cL_0$ and $\bar \cL_0$) and the \emph{operator product expansions} between all operators. Since we will only be concerned with universal properties of CFTs here, such detailed data of individual CFTs will not be important for our considerations. 

We will describe in the next short sections some properties of CFTs that are most relevant to the Kerr/CFT correspondence and its extensions: the Cardy formula and its range of validity, some properties of the discrete light-cone quantization (DLCQ), another closely related class of conformally invariant theories namely the warped conformal field theories, and some classes of irrelevant deformations. The material in this section is far too preliminary to formulate concrete proposals for dual theories to black holes but it contains some ingredients which are expected to play a role in such a holographic correspondence.

\subsection{Cardy's formula and its extended range}
\label{sec:Cardy}

In any unitary and modular invariant CFT, the asymptotic growth of states in the microcanonical ensemble is determined only by the left and right central charge and the left and right eigenvalues $E_L$, $E_R$ of $\mathcal L_0$, $\bar{\mathcal L}_0$ as
\be
\cS_{\text{CFT}}(E_L,E_R) = 2\pi \left( \sqrt{\frac{c_L E_L}{6}}+\sqrt{\frac{c_R E_R}{6}} \right) ,
\ee
when $E_L \gg c_L$, $E_R \gg c_R$. This is known as Cardy's formula derived originally in~\cite{Cardy:1986ie,Bloete:1986qm} using modular invariance of the CFT. A review can be found, e.g., in~\cite{Carlip:1998qw}. Transforming to the canonical ensemble using the definition of the left and right temperatures,
\be
\left( \frac{\p \cS_{\text{CFT}}}{\p E_L}\right)_{E_R} = \frac{1}{T_L},\qquad \left( \frac{\p \cS_{\text{CFT}}}{\p E_R }\right)_{E_L} = \frac{1}{T_R} \, ,
\ee
we get
\be
E_L = \frac{\pi^2}{6}c_L T_L^2,\qquad E_R = \frac{\pi^2}{6}c_R T_R^2 \, ,
\ee
and, therefore, we obtain an equivalent form of Cardy's formula, 
\be
\cS_{\text{CFT}} = \frac{\pi^2}{3} ( c_L T_L +c_R T_R ) \, , \label{Cardy}
\ee
valid when $T_L \gg 1$, $T_R \gg 1$. It will be referred to as the thermal Cardy formula.

The origin of Cardy's formula lies in an IR/UV connection implied by modular invariance. The spectrum of states at high energies is dictated by the spectrum of states at small energies.  As we will discuss in Section \ref{sec:microS}, the matching between the entropy of extremal black holes and Cardy's formula will not be performed in its range of validity. It is therefore crucial to investigate whether or not its range of validity can be extended for classes of CFTs that might be relevant for holography. Two such extensions have been derived which we review herebelow. 

\subsubsection{Extended validity for large central charge and sparse light spectrum} 
\label{extC}

We only consider CFTs which have identical left and right central charges $c \equiv c_L=c_R$ which are very large, $c \gg 1$ as required for holography \cite{ElShowk:2011ag}. Light states are defined as states with total energy in the range
\bea
-\frac{c}{12} \leq E_L + E_R \leq \eps,
\eea
where $\eps$ is a small positive number which asymptotes to 0 in the large $c$ limit. The energy of the vacuum is $E_L+E_R = -\frac{c}{12}$ on the cylinder and zero in the plane. We assume a sparse light spectrum in the sense that the density of states is bounded as
\bea
\rho(E) = \text{exp} [S(E)] \lesssim \text{exp} \left[ 2\pi \big( E + \frac{c}{12} \big) \right],\qquad E \leq \eps.\label{bound1}
\eea
This feature is expected to be consistent with the $\AdS_3$ holographic duality where there is a gap without black hole solutions between the global $\AdS_3$ vacuum with energy $-\frac{c}{12}$ and massless BTZ black hole with zero energy, and it might be also relevant for holography related to deformations of $\AdS_3$. 

With these assumptions, it was shown in \cite{Hartman:2014oaa} assuming modular invariance and unitarity that Cardy's formula  
\be
\cS_{\text{CFT}} = \frac{\pi^2 c}{3} (  T_L + T_R ) \, , \label{HartmanKS}
\ee
holds in the range
\bea
T_L > \frac{1}{2\pi},\qquad T_R > \frac{1}{2\pi}. 
\eea
If one further restricts the mixed density of states as
\bea
\rho(E_L,E_R) \lesssim \text{exp} \left[  4\pi\sqrt{(E_L+\frac{c}{24})(E_R+\frac{c}{24})}\right], \qquad (E_L < 0 \,\, \text{or}\, \, E_R < 0),\label{bound2}
\eea
then Cardy's formula holds for all $E_L E_R > \left( \frac{c}{24}\right)^2$ (outside a small sliver which is conjectured to vanish) or, equivalently, for all 
\bea
T_L T_R > \frac{1}{(2\pi)^2}
\eea
(outside a small sliver which is conjectured to vanish). 

An important class of CFTs which obey the bounds \eqref{bound1}-\eqref{bound2} are the symmetric product orbifold CFTs. Let us briefly review their construction~\cite{Klemm:1990df}. Given a conformally-invariant sigma-model with target space manifold $\mathcal M$, one can construct the symmetric product orbifold by considering the sigma-model with $N$ identical copies of the target space $\mathcal M$, identified up to permutations,
\be
\text{Sym}^N(\mathcal M) \equiv \left( \otimes^N \mathcal M \right)/S_N\,,
\ee
where $S_N$ is the permutation group on $N$ objects. The symmetric orbifold CFT has central charge $c_{Sym} = N c$ if the original CFT has central charge $c$. The symmetric product orbifold with target space $\mathcal M = K3$ or $T^4$ is holographically dual to IIB string theory on $\AdS_3 \times S^3 \times \mathcal M$~\cite{Maldacena:1997re,deBoer:1998ip,Dijkgraaf:1998gf} (see also~\cite{Pakman:2009zz} and references therein). This particular CFT was instrumental in the first microscopic counting of black holes in string theory \cite{Strominger:1996sh}.

It was shown in \cite{Hartman:2014oaa} that symmetric orbifold theories at the orbifold point (where the theory is free and its spectrum known) not only obey the bounds \eqref{bound1}-\eqref{bound2} but also saturate them. This shows that the bounds are optimal and that these theories are the most dense theories that are still compatible with Cardy's growth of states.

\subsubsection{Extended validity for long strings} 
\label{sec:longshort}

An older argument for the extension of Cardy's formula goes as follows. Given a set of Virasoro generators $\mathcal L_n$ and a non-zero integer $N \in \mathbb Z_0$, one can always redefine a subset or an extension of the generators, which results in a different central charge (see, e.g.~\cite{Banados:1998wy}). One can easily check that the generators
\be
\cL_n^{\text{short}} = \frac{1}{N}\cL_{N\, n} \label{Virshort}
\ee
obey the Virasoro algebra with a larger central charge $c^{\text{short}} = N\, c$. Conversely, one might define
\be
\cL_n^{\text{long}} = N\cL_{n/N}. \label{Virlong}
\ee
In general, the generators $\cL_n^{\text{long}}$ with $n \neq N k$, $k\in \mathbb Z$ do not make sense because there are no fractionalized Virasoro generators in the CFT. Such generators would be associated with multivalued modes $e^{in \phi/N}$ on the cylinder $(t,\phi)\sim (t,\phi+2\pi)$. However, in some cases, as we review below, the Virasoro algebra~\eqref{Virlong} can be defined. The resulting central charge is smaller and given by $c^{\text{long}} = c/N$.

If a CFT with generators~\eqref{Virlong} can be defined such that it still captures the entropy of the original CFT, the Cardy formula~\eqref{Cardy} applied in the original CFT could then be used outside of the usual Cardy regime $T_L \gg 1$. Indeed, using the CFT with left-moving generators~\eqref{Virlong} and their right-moving analogue, one has 
\be
\cS_{\text{CFT}} = \frac{\pi^2}{3} ( \frac{c_L}{N} (N T_L) +\frac{c_R}{N} (N T_R)),\label{Cardy3}
\ee
which is valid when $N T_L \gg 1$, $N T_R \gg 1$. If $N$ is very large, Cardy's formula~\eqref{Cardy} would then always apply.


The ``long string CFT'' can be made more explicit in the context of symmetric product orbifold CFTs. The Virasoro generators of the resulting infrared CFT can then be formally constructed from the generators $\cL_m$ of the original infrared CFT as~\eqref{Virshort}. Conversely, if one starts with a symmetric product orbifold, one can isolate the ``long string'' sector, which contains the ``long'' twisted operators. One can argue that such a sector can be effectively described in the infrared by a CFT, which has a Virasoro algebra expressed as~\eqref{Virlong} in terms of the Virasoro algebra of the low energy CFT of the symmetric product orbifold~\cite{Maldacena:1996ds}. The derivation of Section \ref{extC} makes a more precise statement on the range of validity of Cardy's formula for orbifold CFTs.

\subsection{DLCQ as a chiral limit}
\label{sec:DLCQ}

The role of the DLCQ of CFTs in the context of the Kerr/CFT correspondence was emphasized in~\cite{Balasubramanian:2009bg} (for closely related work see~\cite{Strominger:1998yg,Azeyanagi:2008dk}). Here, we will review how a DLCQ is performed and how it leads to a chiral half of a CFT. A chiral half of a CFT is here defined as a sector of a $2d$~CFT defined on the cylinder, where the right-movers are set to the Ramond-Ramond ground state after the limiting DLCQ procedure. We will use these considerations in Section~\ref{sec:microS}.

Let us start with a CFT defined on a cylinder of radius $R$, 
\be
ds^2 = - dt^2 +d\phi^2 = -du\, dv,\quad u=t-\phi,\quad v=t+\phi\, .
\ee
Here the coordinates are identified as $(t,\phi) \sim (t,\phi+2\pi R)$, which amounts to
\be
(u,v)\sim (u-2\pi R,v+2\pi R)\, .\label{id2}
\ee
The momentum operators $P^v$ and $P^u$ along the $u$ and $v$ directions are $L_0$ and $\bar L_0$, respectively. They have a spectrum 
\bea
P^v |O \rangle = L_0 |O \rangle = \left( h - \frac{c}{24}+ n \right) \frac{1}{R} |O\rangle, \\
P^u |O \rangle = \bar L_0 |O \rangle =\left( \bar h- \frac{c}{24}+ \bar n \right) \frac{1}{R} |O\rangle, 
\eea
where the conformal dimensions obey $h,\bar h\geq 0$ and $n,\bar n \neq 0$ are quantized left and right momenta. 

Following Seiberg~\cite{Seiberg:1997ad}, consider a boost with rapidity $\gamma$
\be
u^\prime = e^\gamma u,\qquad v^\prime = e^{-\gamma}v\, .
\ee
The boost leaves the flat metric invariant. The discrete light-cone quantization of the CFT is then defined as the limit $\gamma \rightarrow \infty$ with $R^\prime \equiv Re^\gamma$ fixed. In that limit, the identification~\eqref{id2} becomes 
\be
(u^\prime,v^\prime)\sim (u^\prime -2\pi R^\prime, v^\prime)\, .
\ee
Therefore, the resulting theory is defined on a null cylinder. Because of the boosted kinematics, we have 
\bea
P^{v^\prime} |O \rangle &=& \left( h- \frac{c}{24}+ n \right) \frac{1}{R e^\gamma} |O\rangle, \\
P^{u^\prime} |O \rangle &=& \left( \bar h- \frac{c}{24}+ \bar n \right) \frac{e^\gamma}{R } |O\rangle\, . 
\eea
Keeping $P^{u^\prime}$ (the momentum along $v^\prime$) finite in the $\gamma \rightarrow \infty$ limit requires $\bar h=\frac{c}{24}$ and $\bar n=0$. 

Therefore, the DLCQ limit requires one to freeze the right-moving sector to the vacuum state. The resulting theory admits an infinite energy gap in that sector. The left-moving sector still admits non-trivial states. All physical finite-energy states in this limit only carry momentum along the compact null direction $u^\prime$. Therefore, the DLCQ limit defines a Hilbert space $\mathcal H$,
\be
\mathcal H = \{ | \text{anything} \rangle_L \otimes |  \frac{c}{24} \rangle_R \} \label{spec1}
\ee
with left chiral excitations around the Ramond-Ramond vacuum of the CFT $| \frac{c}{24} \rangle_L \otimes |  \frac{c}{24} \rangle_R$. As a consequence, the right-moving Virasoro algebra does not act on that Hilbert space. This is by definition a chiral half of a CFT. 

In summary, the DLCQ of a $2d$~CFT leads to a chiral half of the CFT with central charge $c$. The limiting procedure removes most of the dynamics of the original CFT. Conversely, given a spectrum such as \eqref{spec1} one possible completion of the theory which is modular invariant is a $2d$ CFT. 

\subsection{Warped conformal field theories as chiral irrelevant deformations}
\label{sec:deformations}

A unitary relativistic field theory with time and space translation invariance and scale invariance is necessarily a $2d$ CFT under some additional technical assumptions on its spectrum \cite{Polchinski:1987dy}. Relaxing the assumptions of Lorentz invariance and scale invariance and imposing instead chiral scale invariance leads to another possibility: a warped conformal field theory whose symmetry group is the direct product of a Virasoro algebra with a $U(1)$ Ka\v{c}-Moody algebra \cite{Hofman:2011zj}. Only a few such quantum theories are known at present \cite{Hofman:2014loa,Castro:2015uaa}. One might think of such theories as arising from specific irrelevant deformations of a $2d$ CFT which preserve chiral conformal invariance. Such deformations can be described perturbatively as a deformation of the action by an irrelevant operator of conformal weights $(1,\bar h)$ with $\bar h \geq 2$, 
\bea
S_{deformed} = S_{CFT} + \lambda \int dx^+ dx^- \mathcal O_{1,\bar h}(x^+,x^-) + O(\lambda^2)
\eea
where $\lambda$ is the deformation parameter. It was shown in \cite{Guica:2010sw} that the resulting theory is exactly marginal with respect to left-moving conformal symmetry. It was also shown that the right-moving conformal weight $\bar \Delta$  of generic operators in the CFT gets modified as
\bea
\bar\Delta_{deformed} = \bar\Delta + \sum_{n > 0} c_n (\lambda k_+)^n
\eea
where $k_+$ is the right-moving momentum and $c_n$ are coefficients that can be computed in principle in perturbation theory. An important property of such theories is that they admit an analogue of Cardy's formula \cite{Detournay:2012pc}. 

As a field theory with universal properties, it is of interest in order to formulate possible holographic correspondences. Nevertheless, the role of warped conformal field theories in the description of extremal (not to say non-extremal) black holes is far from clear. A DLCQ is necessary in order to match the extremal limit on the gravitational side. The occurence of complex weights in near-horizon geometries without global timelike Killing vector such as NHEK also indicates that the putative dual theory will not be standard. The common properties of warped conformal field theories and near-horizon extremal geometries are the exact $SL(2,\mathbb R) \times U(1)$ symmetries and the occurrence of a Virasoro algebra. A precise relationship was attempted for several classes of extremal black holes in supergravity whose near-horizon geometry contains a warped deformation of $\AdS_3$ both perturbatively \cite{Compere:2010uk,ElShowk:2011cm} (see also \cite{Maldacena:2008wh}) and non-perturbatively \cite{Song:2011ii,Bena:2012wc}. However, such efforts did not lead to precise correspondences for realistic extremal black holes. Another issue is that warped $\AdS_3$ geometries (as well as $\AdS_3$ geometries themselves) admit boundary conditions which are described by either the conformal algebra \cite{Guica:2013jza,Guica:2011ia,Compere:2013bya} or the warped conformal algebra \cite{Compere:2007in,Compere:2008cv,Compere:2009zj,Blagojevic:2009ek,Anninos:2010pm,Henneaux:2011hv,Castro:2014ima,Compere:2014bia} and it is not clear which one is realized in a consistent quantum gravity, see \cite{Compere:2010uk,ElShowk:2011cm,Song:2011sr,Song:2011ii,Bena:2012wc} for proposals.\epubtkFootnote{Deformations of CFTs for near-horizon geometries with vanishing horizon are discussed in \cite{Guica:2010ej,Azeyanagi:2010pw, deBoer:2010ac,SheikhJabbaria:2011gc,deBoer:2011zt}.}

\subsection{Irrelevant deformations on both sectors}

The description of non-extremal black holes with an asymptotically flat or $AdS$ region requires to consider $2d$ CFTs with irrelevant deformations in both left and right sectors. One issue with such a description is that the IR cutoff set by the temperature will in general be of the same order of magnitude as the UV cutoff set by the mass scale associated with irrelevant perturbations \cite{Baggio:2012db}. Therefore the CFT description will have no range of validity. Exceptions are special black holes in string theory where the $\AdS_3/\CFT_2$ correspondence precisely applies in the extremal limit. The CFT then controls part of the physics away from extremality. Given such considerations, it might then come as a surprise that there are some CFT features of black holes away from extremality (an effective string description \cite{Cvetic:1997xv}, a Cardy-type formula and the $SL(2,\mathbb R) \times SL(2,\mathbb R)$ invariance of some probes \cite{Castro:2010fd}) as we will describe in Section \ref{sec:hidden}. 

\newpage

\section{Matching the Entropy of Extremal Black Holes}
\label{sec:KerrCFT1}

We discussed that near-horizon geometries of compact extremal black holes are isolated systems with universal properties and we reviewed that they have no local bulk dynamics. Given the non-trivial thermodynamic properties of these systems even at extremality, one can suspect that some dynamics is left over. It turns out that there is one remaining dynamical sector: non-trivial diffeomorphisms which are associated with non-vanishing conserved surface charges. We will discuss that four-dimensional spinning extremal black holes belong to a phase space which represent one copy of the Virasoro algebra with a specific central charge. The chiral Cardy formula reproduces the black hole entropy which points to the relevance of a CFT description. We will discuss the generalization to charged extremal black holes and to higher dimensions.

\subsection{Boundary conditions and asymptotic symmetry algebra}
\label{sec:BC}

The theory of non-trivial diffeomorphisms in gravity goes back to the work of ADM on the definition of asymptotically conserved quantities in asymptotically flat spacetimes \cite{ADM}. The framework to define asymptotic conserved charges and their algebra was then generalized in several respects in Hamiltonian \cite{Regge:1974zd,Brown:1986ed,Troessaert:2015nia} and Lagrangian formalisms \cite{Lee:1990nz,Iyer:1994ys,Barnich:2001jy,Barnich:2007bf,Compere:2007az}. In gravity, most diffeomorphisms are pure gauge because they are associated with trivial canonical surface charges. Some diffeomorphisms are however too large at the boundary: they are associated with infinite charges and should be discarded. In the intermediate case, some diffeomorphisms are associated with finite surface charges. The quotient of allowed diffeomorphism by trivial diffeomorphisms constitutes a Lie algebra known as the \emph{asymptotic symmetry algebra} or by extension to the group, the \emph{asymptotic symmetry group}.  A given set of boundary conditions comes equipped with an asymptotic symmetry group which preserves the boundary conditions. Boundary conditions are restricted by the condition that all surface charges are finite and integrable. There is no universal method or uniqueness in the construction of boundary conditions but once boundary conditions are proposed, their consistency can be checked. 

After these general consideration, let us discuss the existence and the construction of a consistent set of boundary conditions that would define ``the set of solutions in the near-horizon region of extremal black holes'' and their associated asymptotic symmetry algebra. One has to propose boundary conditions from first principles. We will use the Lagrangian methods~\cite{Barnich:2001jy,Barnich:2007bf}. Restricting our discussion to the fields appearing in \eqref{generalaction}, the boundary conditions should be preserved by a set of allowed diffeomorphisms and $U(1)$ gauge transformations $(\zeta^\mu,\Lambda)$, which act on the fields as 
\bea
\delta_{(\zeta,\Lambda)} g_{\mu\nu} &=& \mathcal L_{\zeta} g_{\mu\nu},\qquad \delta_{(\zeta,\Lambda)}A_\mu = \mathcal L_\zeta A_\mu + \p_{\mu} \Lambda\, ,\nn \\
\delta_{(\zeta,\Lambda)}\chi^A &=& \mathcal L_\zeta \chi^A .\label{actfields}
\eea
Asymptotic symmetries are the set of all these allowed transformations that are associated with non-trivial surface charges. The set of allowed transformations that are associated with zero charges are pure gauge/trivial transformations. The set of asymptotic symmetries inherits a Lie algebra structure from the Lie commutator of diffeomorphisms and $U(1)$ gauge transformations. Therefore, the asymptotic symmetries form an algebra,
\be
\left[ (\zeta_m,\Lambda_m), (\zeta_n,\Lambda_n)\right]\equiv ( [\zeta_m,\zeta_n],[\Lambda_m,\Lambda_n]_\zeta)\,,\label{bragen}
\ee
where $[\zeta_m,\zeta_n]$ is the Lie commutator and 
\be
[\Lambda_m,\Lambda_n]_\zeta \equiv \zeta^\mu_m \p_\mu \Lambda_n - \zeta^\mu_n \p_\mu \Lambda_m \, .
\ee
Consistency requires that the charge associated with each element of the asymptotic symmetry algebra be finite and well defined. Moreover, as we are dealing with a spatial boundary, the charges are required to be conserved in time. By construction, one always first defines the ``infinitesimal variation of the charge'' $\delta \mathcal Q$ from infinitesimal variations of the fields around a solution. If $\delta \mathcal Q$ is the exact variation of a quantity $ \mathcal Q$, the quantity $ \mathcal Q$ is the \emph{well-defined} charge and the charges are said to be \emph{integrable}. 

Imposing consistent boundary conditions and obtaining the associated asymptotic symmetry algebra requires a careful analysis of the asymptotic dynamics of the theory. If the boundary conditions are too strong, all interesting excitations are ruled out and the asymptotic symmetry algebra is trivial. If they are too weak, the boundary conditions are inconsistent because transformations preserving the boundary conditions are associated to infinite or ill-defined charges. In general, there is a narrow window of consistent and interesting boundary conditions. There is not necessarily a unique set of consistent boundary conditions.

As an illustration, asymptotically anti-de~Sitter spacetimes in spacetime dimensions $d+1$ admit the $SO(2,d)$ asymptotic symmetry algebra for $d \geq 3$~\cite{Abbott:1981ff,Ashtekar:1984,Henneaux:1985tv,Henneaux:1985ey} and two copies of the Virasoro algebra for $d=2$~\cite{Brown:1986nw}. However, other boundary conditions are also possible \cite{Compere:2008us,Compere:2013bya,Troessaert:2013fma,Avery:2013dja,Donnay:2015abr,Afshar:2016wfy,Perez:2016vqo}. Asymptotically-flat spacetimes admit as asymptotic symmetry algebra the Poincar\'e algebra at spatial infinity \cite{ADM,Geroch:1972aa,Regge:1974zd,Ashtekar:1978aa,Ashtekar:1990gc,Ashtekar:1991vb,Compere:2011db,Compere:2011ve} and the BMS algebra at null infinity with or without Virasoro generators~\cite{Bondi:1962px,Sachs:1962wk,Penrose:1962ij,Ashtekar:1978aa,Ashtekar:1990gc,Barnich:2009se,Barnich:2011mi,Strominger:2013jfa}. From these examples, we learn that the asymptotic symmetry algebra can be larger than the exact symmetry algebra of the background spacetime and it might in some cases contain an infinite number of generators. We also notice that several choices of boundary conditions, motivated from different physical considerations, might lead to different asymptotic symmetry algebras.

Let us now motivate boundary conditions for the near-horizon geometry of extremal black holes. There are two boundaries at $r = \infty$ and $r = -\infty$. It was proposed in~\cite{Guica:2008mu,Hartman:2008pb} to build boundary conditions on one boundary, for definiteness $r = \infty$, such that the asymptotic symmetry algebra contains one copy of the Virasoro algebra generated by 
\bea
\zeta_\eps &=& \eps(\phi)\p_\phi - r \eps^\prime (\phi)\p_r + (\text{subleading terms}),\label{Vir1}\\
\Lambda_\eps &=& -(f(\th)-\frac{e}{k})\eps(\phi)+ (\text{subleading terms})\, .\label{Vir2}
\eea
Part of the motivation behind this ansatz is the existence of a non-zero temperature $T_{\phi}$ associated with modes corotating with the black hole, as detailed in Section~\ref{sec:temp}. This temperature suggests the existence of excitations along $\p_\phi$. The ansatz for $\Lambda_\eps$ will be motivated in \eqref{motivL}. The subleading terms might be chosen such that the generator $\zeta_\eps$ is regular at the poles $\theta=0,\pi$. This ansatz has to be validated by checking if boundary conditions preserved by this algebra exist such that all charges are finite, well defined and conserved. We will discuss such boundary conditions below. Expanding in modes as\epubtkFootnote{The sign choice in this expansion is motivated by the fact that the central charge to be derived in Section~\ref{sec:c} will be positive with this choice. Also, the zero mode $\eps = -1$ is canonically associated with the angular momentum in our conventions.}
\be
\eps(\phi) = - e^{-i n \phi},
\ee
the generators $L_n \equiv (\zeta_n,\Lambda_n)$ obey the Witt algebra (Virasoro algebra without central extension)
\be
i \left[ L_m, L_n \right] = (m-n) L_{m+n}\, ,\label{Vir00}
\ee
where the bracket has been defined in \eqref{bragen}.

Finding consistent boundary conditions that admit finite, conserved and integrable Virasoro charges and that are preserved by the action of the Virasoro generators is a non-trivial task. The details of these boundary conditions depend on the specific theory at hand because the expression for the conserved charges depend on the theory. For the action~\eqref{generalaction}, the conserved charges defined using the Iyer-Wald formalim \cite{Lee:1990nz,Iyer:1994ys} can be found in~\cite{Compere:2009dp}.  In the simpler case of Einstein gravity, the surface charges associated with the asymptotic symmetries generated by $\zeta$ are defined as the charges $Q_\zeta[g] $ whose variations obey
\bea
\delta Q_\zeta[g] &=& \oint_S \boldsymbol k^{Einstein}_\zeta [\delta g ; g]. 
\eea
Here $S$ is a codimension 2 surface of integration and $\boldsymbol k_\zeta [\delta g ; g]$ is the surface charge form which reads for Einstein gravity as
\begin{align}
 & (16 \pi G) \boldsymbol k_\zeta^{Einstein}[\delta g ; g]
=
2 (d^{d-2}x)_{\mu \nu} 
\bigg\{
\zeta^\nu\nabla^\mu h
-\zeta^\nu\nabla_\sigma h^{\mu\sigma}
+\zeta_\sigma\nabla^{\nu}h^{\mu\sigma}
+\frac{1}{2}h\nabla^{\nu} \zeta^{\mu}
-h^{\rho\nu}\nabla_\rho\zeta^{\mu}\nonumber\\
&\hspace{5cm}
+ \frac{\alpha+1}{2}
h^{\sigma\nu}(\nabla^\mu\zeta_\sigma +
\nabla_\sigma\zeta^\mu)+\delta Y^{\mu\nu}[\delta_\zeta g ; g] - \delta_\zeta Y^{\mu\nu} [h;g ]
\bigg\} \ .
\label{kgrav}
\end{align}
Here we used the notation $(d^{d-2}x)_{\mu \nu} =\frac{1}{2 (d-2)!}\eps_{\mu\nu\alpha_1\dots \alpha_{d-2}}dx^{\alpha_1}\wedge \cdots \wedge dx^{\alpha_{d-2}}$, $h_{\mu\nu} = \delta g_{\mu\nu}$, $h = g^{\mu\nu} h_{\mu\nu}$ and indices are raised with $g^{\mu\nu}$. There is an ambiguity in the definition of the charge which is parametrized by the coefficient $\alpha$ and the codimension 2 form $Y^{\mu\nu} = Y^{[\mu\nu]}$. For $\alpha = -1$ and $Y^{\mu\nu} = 0$, this is the Iyer-Wald charge \cite{Lee:1990nz,Iyer:1994ys}. For $\alpha = 0$ and $Y^{\mu\nu} = 0$, this is the charge defined by Barnich-Brandt \cite{Barnich:2001jy} and Abbott-Deser \cite{Abbott:1981ff}. The terms $Y^{\mu\nu}[\delta g ; g]$ arise from a well-known ambiguity in the definition of the presymplectic structure of the theory \cite{Iyer:1994ys} and have been conjectured to be related to the counterterms necessary to regulate the action \cite{Compere:2008us,Compere:2011ve}. Such terms exactly vanish in the case of exact symmetries and for asymptotic symmetries where the linear theory applies \cite{Barnich:2001jy} but are relevant in general as illustrated e.g., in \cite{Compere:2014bia,Compere:2015bca}. In the case of the ansatz \eqref{Vir1}, it was assumed in~\cite{Guica:2008mu,Hartman:2008pb} that $\alpha = 0$ and $Y^{\mu\nu} = 0$. 

Let us first specialize to the case of the extremal Kerr black hole in Einstein gravity. The original approach followed in \cite{Guica:2008mu} (see also \cite{Amsel:2009pu}) was to postulate boundary conditions and perform check finiteness and integrability of the surface charges for perturbations around the background geometry. This analysis led to the fall-off conditions
\begin{align}
g_{tt}&= \mathcal{O}({r^2}),\qquad g_{t\phi}= k \Gamma(\th)\gamma(\th)^2 r + \mathcal{O}({1}),\nn\\
g_{t\theta} &= \mathcal{O}({1\over r}),\qquad g_{t r}= \mathcal{O}({1\over r^2}) ,\qquad g_{\phi\phi}= O(1), \nn \\
g_{\phi\theta}&= \mathcal{O}({1\over r}) ,\qquad g_{\phi r}= \mathcal{O}({1\over r}) ,\qquad g_{\theta r}= \mathcal{O}({1\over r^2}) , \label{strictbc} \\
g_{\theta\theta}&= \Gamma(\th)\alpha(\th)^2 + \mathcal{O}({1\over r}) ,\qquad 
g_{rr}= \frac{\Gamma(\th)}{r^2}+\mathcal{O}({1\over r^3}),\nn
\end{align}
together with the supplementary zero energy excitation condition 
\be
\delta \mathcal Q_{\p_t} = 0\, .\label{zeroee}
\ee
A non-trivial feature of the boundary conditions~\eqref{strictbc}\,--\,\eqref{zeroee} is that they are preserved precisely by the Virasoro algebra~\eqref{Vir1}, by $\p_t$ and the generator~\eqref{zeta0} (as pointed out in~\cite{Amsel:2009pu}) and subleading generators. (Note that these boundary conditions are not preserved by the action of the third $SL(2,\mathbb R)$ generator~\eqref{killing}.) It was shown in~\cite{Guica:2008mu} that the Virasoro generators are finite given the fall-off conditions and well defined around the background NHEK geometry. It was shown in~\cite{Amsel:2009pu} that the Virasoro generators are conserved and well defined around any asymptotic solution given that one additionally regularizes the charges using counter-term methods~\cite{Compere:2008us}, up to technical subteties. The explicit metric of the extremal Kerr black hole with left Virasoro descendants can be obtained \cite{Compere:2015bca}. Therefore, up to some technical details, it led to the claim that consistent boundary conditions admitting (at least) a Virasoro algebra as asymptotic symmetry algebra exist. The set of trivial asymptotic symmetries, i.e. asymptotic symmetries associated with vanishing charges, comprise two of the $SL(2,\mathbb R)$ generators. It is not clear if the boundary conditions could be enhanced in order to admit all $SL(2,\mathbb R)$ generators as trivial asymptotic symmetries.

Another approach was followed more recently \cite{Compere:2015mza,Compere:2015bca}. Since there is no local bulk dynamics in the near-horizon limit, a consistent phase space with a given asymptotic symmetry group can be generated by acting with finite diffeomorphisms which exponentiate the asymptotic symmetries. One can also consider varying the parameters of the near-horizon background geometry to generate a more complete phase space \cite{Hajian:2014twa}. Starting from a given ansatz for the asymptotic symmetry generator and choosing a particular gauge, one can then generate the corresponding phase space algorithmically and check for its consistency. According to the analysis of \cite{Compere:2015mza,Compere:2015bca}, the ansatz \eqref{Vir1} leads to a phase space where constant $t,r$ surfaces admit singularities which complicates the definition of surface charges, as also discussed in \cite{Amsel:2009pu}. Instead, the ansatz
\bea
\zeta_\eps &=& \eps(\phi)\p_\phi -  \eps^\prime (\phi)\left( r\p_r + \frac{1}{r} \p_t \right) + (\text{subleading terms}).\label{Vir1other}
\eea
was proposed in order to avoid these difficulties. The phase space can then be explicitly built and checked in detail. In particular, the conserved charges in the entire phase space can be explicitly constructed in terms of the stress-tensor of a Liouville field. Moreover, the asymptotic symmetries act everywhere in the bulk spacetime, which promotes them to symplectic symmetries similarly to $3d$ Einstein gravity \cite{Compere:2014cna,Compere:2015knw}. However in this case, a dynamical ambiguity in the definition of surface charges arose which could not be fixed uniquely \cite{Compere:2015bca}. 

Let us now generalize these arguments to the electrically-charged Kerr--Newman black hole in Einstein--Maxwell theory. First, the presence of the chemical potential $T_{e}$ suggests that the gauge field matters.  The associated conserved electric charge $\cQ_e$ can be shown to be canonically associated with the zero-mode generator $J_0 = (0,-1)$ with gauge parameter $\Lambda = -1$. It is then natural to define the current ansatz 
\be
J_n = (0,-e^{-i n \phi}),\label{Jn}
\ee
which obeys the commutation relations
\be
i\left[ L_m , J_n \right] = - n J_{m+n}, \qquad i\left[ J_m , J_n \right] = 0\, .\label{currentalg}
\ee
The non-trivial step consists in establishing the existence of boundary conditions such that the Virasoro and the current charges are well defined and conserved. One can simplify the problem of constructing boundary conditions by imposing the following additional constraints
\be
\delta \mathcal Q_{\p_t} = 0,\qquad \delta \mathcal Q_e = 0\, ,\label{extcond2}
\ee
which discard the current algebra. Such a simplification was used in~\cite{Hartman:2008pb} and the following boundary conditions were proposed (up to the term $e/k$, which was omitted in~\cite{Hartman:2008pb})
\bea
A_{t}&=& \mathcal{O}({r}), \qquad A_{\phi}=f(\th)-\frac{e}{k}+\mathcal{O}({1 \over r}), \nn \\
A_{\theta}&=& \mathcal{O}({1}), \qquad A_r= \mathcal{O}({1\over r^2}), \label{strictbc2}
\eea
which are preserved upon acting with the Virasoro generator~\eqref{Vir1}\,--\,\eqref{Vir2}. In particular, the choice of the compensating gauge transformation $\Lambda_\eps$ \eqref{Vir2} is made such that 
\be
\mathcal L_{\zeta_\eps}A_\phi + \p_\phi \Lambda_{\eps} = O(r^{-1}). \label{motivL}
\ee
It can be shown that the Virasoro generators are finite under these boundary conditions. 

In three dimensions, a Virasoro algebra can also be found in the near-horizon limit of the BTZ black hole~\cite{Balasubramanian:2009bg}. There it was shown that the  asymptotic symmetry group of the near-horizon geometry of the extremal BTZ black hole of angular momentum $J$ given in \eqref{selfdualorb} consists of one chiral Virasoro algebra extending the $U(1)$ symmetry along $\p_\phi$, while the charges associated with the $SL(2,\mathbb R)$ symmetry group are identically zero. These observations are consistent with the analysis of four-dimensional near-horizon geometries~\eqref{GenExt}, whose constant $\theta$ sections share similar qualitative features with the three-dimensional geometries~\eqref{selfdualorb}. 

Let us also discuss what happens in higher dimensions ($d > 4$). The presence of several independent planes of rotation allows for the construction of one Virasoro ansatz and an associated Frolov--Thorne temperature for each plane of rotation~\cite{Lu:2008jk,Isono:2008kx,Azeyanagi:2008kb,Nakayama:2008kg,Chow:2008dp}. More precisely, given $n$ compact commuting Killing vectors, one can consider an $SL(n,\mathbb Z)$ family of Virasoro ans\"atze by considering all modular transformations on the $U(1)^n$ torus~\cite{Loran:2009cr,Chen:2011wm}. However, no boundary condition is known that allows simultaneously two different Virasoro algebras in the asymptotic symmetry algebra~\cite{Azeyanagi:2008kb}. It was confirmed in the analysis of \cite{Compere:2015bca} that there are mutually-incompatible boundary conditions for each choice of Virasoro ansatz. However, there is an alternative boundary condition that exists in higher dimensions with an algebra which differs from the Virasoro algebra \cite{Compere:2015mza,Compere:2015bca}. These alternative boundary conditions will not be discussed here. 

Since two $U(1)$ circles form a torus invariant under $SL(2,\mathbb Z)$ modular transformations, one can then form an ansatz for a Virasoro algebra for any circle defined by a modular transformation of the $\phi_1$ and $\phi_2$-circles. More precisely, we define 
\be
\phi_1^\prime = p_1 \phi_1 + p_2 \phi_2, \qquad \phi_2^\prime = p_3 \phi_1 + p_4 \phi_2, \label{defp1p2}
\ee
where $p_1 p_4 - p_2 p_3 = 1$ and we consider the vector fields
\be
L^{(p_1,p_2)}_n = -e^{-i n \phi_1^\prime}\p_{\phi_1^\prime} - i r e^{-i n \phi_1^\prime}\p_r + (\text{subleading terms}) .\label{VirSL2Z}
\ee
The resulting boundary conditions have not been thoroughly constructed, but evidence points to their existence~\cite{Azeyanagi:2008kb,Loran:2009cr}.

The occurrence of multiple choices of boundary conditions in the presence of multiple $U(1)$ symmetries raises the question of whether or not the (AdS)--Reissner--N\"ordstrom black hole admits interesting boundary conditions where the $U(1)$ gauge symmetry (which is canonically associated to the conserved electric charge $Q$) plays the prominent role. One can also ask these questions for the general class of (AdS)--Kerr--Newman black holes.

It was argued in~\cite{Hartman:2008pb,Lu:2009gj} that such boundary conditions indeed exist when the $U(1)$ gauge field can be promoted to be a Kaluza--Klein direction of a higher-dimensional spacetime, or at least when such an effective description captures the physics. Denoting the additional direction by $\p_\chi$ with $\chi \sim \chi +2\pi R_\chi$, the problem amounts to constructing boundary conditions in five dimensions. As mentioned earlier, evidence points to the existence of such boundary conditions~\cite{Azeyanagi:2008kb,Loran:2009cr}. The Virasoro asymptotic-symmetry algebra is then defined using the ansatz 
\be
L^Q_n = -R_\chi e^{-\frac{i n \chi}{R_\chi}}\p_\chi - i r e^{-\frac{i n \chi}{R_\chi}}\p_r + (\text{subleading terms}) \label{Virchi}
\ee
along the gauge Kaluza--Klein direction. The same reasoning leading to the $SL(2,\mathbb Z)$ family of Virasoro generators~\eqref{VirSL2Z} would then apply as well. The existence of such a Virasoro symmetry around the Kerr--Newman black holes is corroborated by near-extremal scattering amplitudes as we will discuss in Section~\ref{sec:KerrCFT2}, and by the hidden conformal symmetry of probes, as we will discuss in Section~\ref{sec:hidden}.

\subsubsection{Absence of $SL(2,\mathbb R)$ asymptotic symmetries}
\label{sec:freez}

The boundary conditions discussed so far do not admit solutions with non-trivial charges under the $SL(2,\mathbb R)$ exact symmetry group of the background geometry generated by $\zeta_{0,\pm 1}$ \eqref{killing}. In fact, the boundary conditions are not even invariant under the action of the generator $\zeta_1$. One could ask whether such an enlargement of boundary conditions is possible, which would open the possibility of enlarging the asymptotic-symmetry group to include the $SL(2,\mathbb R)$ group and even a Virasoro extension thereof. We will now argue that such enlargement would result in trivial charges, which would not belong to the asymptotic symmetry algebra. 

First, we saw in Section~\ref{sec:nearext1} that there is a class of near-extremal solutions~\eqref{GenNearExt} obeying the boundary conditions~\eqref{strictbc}\,--\,\eqref{strictbc2} with near-horizon energy $\slash\hspace{-5pt}\delta \cQ_{\p_t} = T^{\mathrm{near-ext}}\delta \cS_{\ext}$. However, the charge $\slash\hspace{-5pt}\delta\cQ_{\p_t}$ is a heat term, which is not integrable when both $T^{\mathrm{near-ext}}$ and $\cS_{\ext}$ can be varied. Moreover, upon scaling the coordinates as $t \rightarrow t/\alpha$ and $r \rightarrow \alpha r$ using the $SL(2,\mathbb R)$ generator~\eqref{zeta0}, one obtains the same metric as \eqref{GenNearExt} with $T^{\mathrm{near-ext}} \rightarrow T^{\mathrm{near-ext}}/\alpha$. If one would allow the class of near-extremal solutions~\eqref{GenNearExt} and the presence of $SL(2,\mathbb R)$ symmetries in a consistent set of boundary conditions, one would be forced to fix the entropy $\cS_{\ext}$ to a constant, in order to define integrable charges. The resulting vanishing charges would then not belong to the asymptotic symmetry algebra. Since there is no other obvious candidate for a solution with non-zero near-horizon energy, we argued in Section~\ref{sec:nobulkdof} that there is no such solution at all. If that assumption is correct, the $SL(2,\mathbb R)$ algebra would always be associated with zero charges and would not belong to the asymptotic symmetry group. Hence, no additional non-vanishing Virasoro algebra could be derived in a consistent set of boundary conditions which contains the near-horizon geometries.\epubtkFootnote{For arguments in favor of $SL(2,\mathbb R)$ enhancement, see~\cite{Matsuo:2009sj,Matsuo:2009pg,Rasmussen:2009ix,Matsuo:2010ut}.} 

Second, as far as extremal geometries are concerned, there is no need for a non-trivial $SL(2,\mathbb R)$ or second Virasoro algebra. As we will see in Section~\ref{sec:microS}, the entropy of extremal black holes will be matched using a single copy of the Virasoro algebra, using the assumption that Cardy's formula applies. Matching the entropy of non-extremal black holes and justifying Cardy's formula requires two Virasoro algebras, as we will discuss in Section~\ref{sec:CFTmatch}. However, non-extremal black holes do not admit a near-horizon limit and, therefore, are not dynamical objects described by a consistent class of near-horizon boundary conditions. At most, one could construct the near horizon region of non-extremal black holes in perturbation theory as a large deformation of the extremal near-horizon geometry~\cite{Castro:2009jf}.

\subsection{Virasoro algebra and central charge} 
\label{sec:c}

Let us now assume in the context of the general theory~\eqref{generalaction} that a consistent set of boundary conditions exists that admits the Virasoro algebra generated by \eqref{Vir1}\,--\,\eqref{Vir2} as asymptotic-symmetry algebra. Current results are consistent with that assumption but, as emphasized earlier, boundary conditions have only been partially checked~\cite{Guica:2008mu,Amsel:2009pu,Azeyanagi:2008kb} and other ansatzes or boundary conditions exist \cite{Compere:2015mza,Compere:2015bca}. We will also assume the definition of the Barnich-Brandt surface charge and ignore the potential ambiguities $Y^{\mu\nu}$ in \eqref{kgrav}, see \cite{Compere:2015bca} for discussions.

Let us define the Dirac bracket between two charges as
\be
\{ \mathcal Q_{(\zeta_m,\Lambda_m)} , \mathcal Q_{(\zeta_n,\Lambda_n)} \} \equiv - \delta_{(\zeta_m,\Lambda_m)} \mathcal Q_{(\zeta_n,\Lambda_n)}\, .
\ee
Here, the operator $\delta_{(\zeta_m,\Lambda_m)} $ is a derivative in phase space that acts on the fields $g_{\mu\nu}$, $A^I_\mu$, $\chi^A$ appearing in the charge $\mathcal Q$ as~\eqref{actfields}. From general theorems in the theory of asymptotic symmetry algebras~\cite{Barnich:2001jy,Barnich:2007bf}, the Dirac bracket represents the asymptotic symmetry algebra \emph{up to a central term}, which commutes with each element of the algebra. Namely, one has 
\be
\{ \mathcal Q_{(\zeta_m,\Lambda_m)} , \mathcal Q_{(\zeta_n,\Lambda_n)} \} =\mathcal Q_{[(\zeta_m,\Lambda_m),(\zeta_n,\Lambda_n)]} + \mathcal K_{(\zeta_m,\Lambda_m),(\zeta_n,\Lambda_n)},\label{repr}
\ee
where the bracket between two generators has been defined in \eqref{bragen} and $\mathcal K$ is the central term, which is anti-symmetric in its arguments. Furthermore, using the correspondence principle in semi-classical quantization, Dirac brackets between generators translate into commutators of quantum operators as $\{ \dots \} \rightarrow -\frac{i}{\hbar } [ \dots ]$. Note that, according to this rule, the central terms in the algebra aquire a factor of $1/\hbar$ when operator eigenvalues are expressed in units of $\hbar$ (or equivalently, when one performs $\mathcal Q \rightarrow \hbar \mathcal Q$ and divide both sides of \eqref{repr} by $\hbar$.).

For the case of the Virasoro algebra~\eqref{Vir00}, it is well known that possible central extensions are classified by two numbers $c$ and $A$. The general result has the form 
\be
\,[ \mathcal L_m ,\mathcal L_n ] = (m-n) \mathcal L_{m+n} + \frac{c}{12} m(m^2 -A)\delta_{m,-n}\label{Virc}\, ,
\ee
where $A$ is a trivial central extension that can be set to 1 by shifting the background value of the charge $\mathcal L_0$. The non-trivial central extension $c$ is a number that is called the central charge of the Virasoro algebra. From the theorems~\cite{Barnich:2001jy,Barnich:2007bf}, the central term in \eqref{repr} can be expressed as a specific and known functional of the Lagrangian $\mathcal L$ (or equivalently of the Hamiltonian), the background solution $\bar \phi = (\bar g_{\mu\nu},\bar A_\mu^I,\bar \chi^A)$ (the near-horizon geometry in this case) and the Virasoro generator $(\zeta,\Lambda)$ around the background
\be
c = c(\mathcal L, \bar \phi, (\zeta,\Lambda) )\,. \label{csymb}
\ee
In particular, the central charge does \emph{not} depend on the choice of boundary conditions. The representation theorem leading to \eqref{Virc} only requires that such boundary conditions exist. The representation theorem for asymptotic Hamiltonian charges~\cite{Brown:1986ed} was famously first applied~\cite{Brown:1986nw} to Einstein's gravity in three dimensions around AdS, where the two copies of the Virasoro asymptotic-symmetry algebra were shown to be centrally extended with central charge $c= \frac{3 l }{2 G_N \hbar}$, where $l$ is the $\AdS$ radius and $G_N$ Newton's constant.

For the general near-horizon solution~\eqref{GenExt} of the Lagrangian~\eqref{generalaction} and the Virasoro ansatz~\eqref{Vir1}\,--\,\eqref{Vir2}, one can prove~\cite{Hartman:2008pb,Compere:2009dp} that the matter part of the Lagrangian (including the cosmological constant) does not contribute directly to the central charge, but only influences the value of the central charge through the functions $\Gamma(\th),\alpha(\th),\gamma(\th)$ and $k$, which solve the equations of motion. The central charge~\eqref{csymb} is then given as the $m^3$ factor of the following expression defined in terms of the fundamental charge formula of Einstein gravity as~\cite{Barnich:2001jy}
\be
c_J = 12 i \, \lim_{r\rightarrow \infty} \mathcal Q^{\text{Einstein}}_{L_m}[\mathcal L_{L_{-m}} \bar g ; \bar g] \Big|_{m^3},\label{cform}
\ee
where $\mathcal L_{L_{-m}} \bar g$ is the Lie derivative of the metric along $L_{-m}$ and 
\bea
\mathcal Q^{\text{Einstein}}_{L_m}[h ; g] &\equiv & \frac{1}{8 \pi G_N}\int_S (d^{d-2}x)_{\mu\nu}\Big( \xi^\nu D^\mu h +\xi^\mu D_\sigma h^{\sigma\nu}+\xi_\sigma D^\nu h^{\sigma\mu} +\frac{1}{2}h D^\nu \xi^\mu\nn \\
&&\qquad \qquad +\frac{1}{2}h^{\mu\sigma}D_\sigma \xi^\nu+\frac{1}{2}h^{\nu\sigma}D^\mu \xi_\sigma\Big).\label{QEin}
\eea
Here, the integrand is precisely the Abbott--Deser--Barnich--Brandt surface charge form \eqref{kgrav} (with $Y^{\mu\nu} = 0$, $\alpha = 0$) and $S$ is a surface at fixed time and radius $r$. Substituting the general near-horizon solution~\eqref{GenExt} and the Virasoro ansatz~\eqref{Vir1}\,--\,\eqref{Vir2}, one obtains 
\be
c_J = \frac{3 k }{G_N \hbar}\int_0^\pi d\theta \;\alpha(\th)\Gamma(\th)\gamma(\th)\, .\label{c1}
\ee
We will drop the factors of $G_N$ and $\hbar$ from now on. In the case of the NHEK geometry in Einstein gravity, substituting \eqref{NHEKvals}, one finds the simple result~\cite{Guica:2008mu}
\be
c_J = 12 J\, .\label{c2}
\ee
The central charge of the Virasoro ansatz~\eqref{Vir1}\,--\,\eqref{Vir2} around the Kerr--Newman black hole turns out to be identical to \eqref{c2}. We note in passing that the central charge $c_J$ of extremal Kerr or Kerr--Newman is a multiple of six, since the angular momentum is quantized as a half-integer multiple of $\hbar$. The central charge can be obtained for the Kerr--Newman--AdS solution as well~\cite{Hartman:2008pb} and the result is 
\be
c_J = \frac{12ar_+}{\Delta_0 },\label{ooc}
\ee
where $\Delta_0$ has been defined in \eqref{def2spAdSKN}.

When higher-derivative corrections are considered, the central charge can still be computed exactly, using as crucial ingredients the $SL(2,\mathbb R)\times U(1)$ symmetry and the $(t,\phi)$ reversal symmetry of the near-horizon solution. The result is given by~\cite{Azeyanagi:2009wf}
\be
c_J = -12 k \int_\Sigma \frac{\delta^{\text{cov}} L }{\delta R_{abcd}}\eps_{ab}\eps_{cd} vol(\Sigma)\, ,\label{c3}
\ee
where the covariant variational derivative $\delta^{\text{cov}}/\delta R_{abcd}$ has been defined in \eqref{Riemcov} in Section~\ref{sec:entropy}. One caveat should be noted. The result~\cite{Azeyanagi:2009wf} is obtained after auxiliary fields are introduced in order to rewrite the arbitrary diffeomorphism-invariant action in a form involving at most two derivatives of the fields. It was independently observed in~\cite{Krishnan:2009tj} that the formalism of~\cite{Barnich:2001jy,Barnich:2007bf} applied to the Gauss--Bonnet theory formulated using the metric only cannot reproduce the central charge~\eqref{c3} and, therefore, the black-hole entropy as will be developed in Section~\ref{sec:microS}. One consequence of these two computations is that the formalism of~\cite{Barnich:2001jy,Barnich:2007bf,Compere:2007az} is not invariant under field redefinitions. 
In view of the cohomological results of~\cite{Barnich:2001jy}, this ambiguity can appear only in the asymptotic context and when certain asymptotic linearity constraints are not obeyed. Nevertheless, it has been acknowledged that boundary terms in the action should be taken into account~\cite{Regge:1974zd,Hawking:1995fd}. Adding supplementary terms to a well-defined variational principle amount to deforming the boundary conditions~\cite{Breitenlohner:1982jf,Witten:2001ua,Marolf:2006nd} and modifying the symplectic structure of the theory through its coupling to the boundary dynamics~\cite{Compere:2008us}. Therefore, it remains to be checked if the prescription of~\cite{Compere:2008us} to include boundary effects would allow one to reconcile the work of~\cite{Krishnan:2009tj} with that of~\cite{Azeyanagi:2009wf}.

In five-dimensional Einstein gravity coupled to $U(1)$ gauge fields and scalars, the central charge associated with the Virasoro generators along the direction $\p_{\phi_i}$, $i=1,2$ can be obtained as a straightforward extension of \eqref{c1}~\cite{Hartman:2008pb,Compere:2009dp}. One has
\be
c_{\phi_i} = 6\pi k_i\int_0^\pi d\theta \;\alpha(\th)\Gamma(\th)\gamma(\th)\, ,\label{cphii}
\ee
where the extra factor of $2\pi$ with respect to \eqref{c1} originates from integration around the extra circle (see also~\cite{Goldstein:2011jh,Hayashi:2011uf} for some higher derivative corrections). Since the entropy~\eqref{entropy5d} is invariant under a $SL(2,\mathbb Z)$ change of basis of the torus coordinates $(\phi_1,\phi_2)$ as \eqref{defp1p2}, $c_{\phi_i}$ transforms under a modular transformation as $k_i$. Now, $k_i$ transforms in the same fashion as the coordinate $\phi_i$, as can be deduced from the form of the near-horizon geometry~\eqref{near5d}. Then, the central charge for the Virasoro ansatz~\eqref{VirSL2Z} is given by
\be
c_{(p_1,p_2)} = p_1 c_{\phi_1}+p_2 c_{\phi_2}\, .\label{c12}
\ee

Let us now discuss the central extension of the alternative Virasoro ansatz~\eqref{Virchi} for the extremal Reissner--Nordstr\"om black hole of electric charge $Q$ and mass $Q$. First, the central charge is inversely proportional to the scale $R_\chi$ set by the Kaluza--Klein direction that geometrizes the gauge field. One can see this as follows. The central charge is bilinear in the Virasoro generator and, therefore, it gets a factor of $(R_\chi)^2$. Also, the central charge consists of the $n^3$ term of the formula~\eqref{QEin}, it then contains terms admitting three derivatives along $\chi$ of $e^{-i n\chi/R}$ and, therefore, it contains a factor of $R_\chi^{-3}$. Also, the central charge is defined as an integration along $\chi$ and, therefore, it should contain one factor $R_\chi$ from the integration measure. Finally, the charge is inversely proportional to the five-dimensional Newton's constant $G_5 = (2\pi R_\chi) G_4$. Multiplying this complete set of scalings, one obtains that the central charge is inversely proportional to the scale $R_\chi$.

Using the simple embedding of the metric and the gauge fields in a higher-dimensional spacetime \eqref{KKlift}, as discussed in Section~\ref{KKup}, and using the Virasoro ansatz~\eqref{Virchi}, it was shown~\cite{Hartman:2008pb,Garousi:2009zx,Chen:2010yu} that the central charge formula~\eqref{cform} gives 
\be
c_Q = \frac{6 Q^3}{R_\chi}\,\label{cpsi} . 
\ee
One might object that \eqref{KKlift} is not a consistent higher-dimensional supergravity uplift. Indeed, as we discussed in Section~\ref{KKup}, one should supplement matter fields such as \eqref{KKliftA}. However, since matter fields such as scalars and gauge fields do not contribute to the central charge~\eqref{csymb}~\cite{Compere:2009dp}, the result~\eqref{cpsi} holds for any such consistent embedding. 

Similarly, we can uplift the Kerr--Newman black hole to five-dimensions, using the uplift~\eqref{KKlift}\,--\,\eqref{KKliftA} and the four-dimensional fields \eqref{GenExt}\,--\,\eqref{fctsKN}. Computing the central charge~\eqref{cphii} for the Virasoro ansatz~\eqref{Virchi}, we find again\epubtkFootnote{We thank Tom Hartman for helping deriving this central charge during a private communication.}
\be
c_Q = \frac{6 Q^3}{R_\chi}\,\label{cpsiagain} . 
\ee

Under the assumption that the $U(1)$ gauge field can be uplifted to a Kaluza--Klein direction, we can also formulate the Virasoro algebra~\eqref{VirSL2Z} and associated boundary conditions for any circle related by an $SL(2,\mathbb Z)$ transformation of the torus $U(1)^2$. Applying the relation~\eqref{c12} we obtain the central charge
\be
c_{(p_1,p_2)} = p_1 c_J + p_2 c_Q = 6 \left( p_1 (2J)+p_2 \frac{Q^3}{R_\chi} \right)\, .\label{c12KN}
\ee

Let us discuss the generalization to AdS black holes. As discussed in Section~\ref{KKup}, one cannot use the ansatz~\eqref{KKlift} to uplift the $U(1)$ 
 gauge field. Rather, one can uplift to eleven dimensions along a seven-sphere. One can then argue, as in~\cite{Lu:2009gj}, that the only contribution to the central charge comes from the gravitational action. Even though no formal proof is available, it is expected that it will be the case given the results for scalar and gauge fields in four and five dimensions~\cite{Compere:2009dp}. Applying the charge formula~\eqref{cform} accounting for the gravitational contribution of the complete higher-dimensional spacetime, one obtains the central charge for the Virasoro algebra~\eqref{Virchi} as~\cite{Lu:2009gj}
\be
c_Q = 6Q_e \frac{r_+^2-a^2}{\Xi \Delta_0 R_\chi}, \label{valcQAdSKN}
\ee
where parameters have been defined in Section~\ref{sec:defAdSKerrN} and $2\pi R_\chi$ is the length of the $U(1)$ circle in the seven-sphere.

The values of the central charges \eqref{c2}, \eqref{ooc}, \eqref{c3}, \eqref{cphii}, \eqref{c12}, \eqref{cpsiagain}, \eqref{c12KN}, \eqref{valcQAdSKN} are the main results of this section.

\subsection{Cardy matching of the entropy}
\label{sec:microS}

In Section~\ref{sec:c} we have discussed the existence of an asymptotic Virasoro algebra at the boundary $r = \infty$ of the near-horizon geometry. We also discussed that the $SL(2,\mathbb R)$ symmetry is associated with zero charges. Following semi-classical quantization rules, the operators that define quantum gravity with the boundary conditions~\eqref{strictbc}, \eqref{strictbc2}, \eqref{extcond2} form a representation of the Virasoro algebra and are in a ground state with respect to the representation of the $SL(2,\mathbb R)$ symmetry \cite{Strominger:1997eq,Guica:2008mu}. A consistent theory of quantum gravity in the near-horizon region, if it can be defined at all, can therefore be (i) a chiral CFT or (ii) a chiral half of a two-dimensional CFT or (iii) a chiral half of a two-dimensional deformed CFT with a Virasoro algebra in the IR and in the dual semi-classical gravity regime. A chiral CFT is defined as a holomorphically-factorized CFT with zero central charge in one sector, while a chiral half of a $2d$~CFT can be obtained, e.g., after a chiral limit of a $2d$~CFT, see Section~\ref{sec:DLCQ}. None of (i) and (ii) seem to apply for the description of the properties of near-extremal and non-extremal black holes as discussed in the Introduction~(\ref{sec:introduction}) and in the next sections. In the case (iii), the CFT can be deformed as long as the asymptotic growth of states is still captured by Cardy's formula. 

Before moving further on, let us step back and first review an analogous reasoning in AdS\sub{3}~\cite{Strominger:1997eq}. In the case of asymptotically AdS\sub{3} spacetimes, the asymptotic symmetry algebra contains two Virasoro algebras. Also, one can define a two-dimensional flat cylinder at the boundary of AdS\sub{3} using the Fefferman-Graham theorem~\cite{Fefferman:1985aa}. One is then led to identify quantum gravity in AdS\sub{3} spacetimes with a two-dimensional CFT defined on the cylinder. The known examples of AdS/CFT correspondences involving AdS\sub{3} factors can be understood as a correspondence between an ultraviolet completion of quantum gravity on AdS\sub{3} and a specific CFT. The vacuum AdS\sub{3} spacetime is more precisely identified with the $SL(2,\mathbb R)\times SL(2,\mathbb R)$ invariant vacuum of the CFT, which is separated with a mass gap of $-c/24$ from the zero-mass black holes. Extremal black holes with AdS\sub{3} asymptotics, the \emph{extremal BTZ black holes}~\cite{Banados:1992wn}, are thermal states in the dual CFT with one chiral sector excited and the other sector set to zero temperature. It was further understood in~\cite{Balasubramanian:2009bg} that taking the near-horizon limit of the extremal BTZ black hole corresponds to taking the DLCQ of the dual CFT (see Section~\ref{sec:DLCQ} for a review of the DLCQ procedure and~\cite{Balasubramanian:2010ys,Goldstein:2011jh} for further supportive studies). The resulting CFT is chiral and has a frozen $SL(2,\mathbb R)$ right sector. 

Given the close parallels between the near-horizon geometry of the extremal BTZ black hole~\eqref{selfdualorb} and the near-horizon geometries of four-dimensional extremal black holes~\eqref{GenExt}, it has been suggested in~\cite{Balasubramanian:2009bg} that extremal black holes are described by a chiral limit of two-dimensional CFT. This assumption nicely accounts for the fact that only one Virasoro algebra appears in the asymptotic symmetry algebra and it is consistent with the conjecture that no non-extremal excitations are allowed in the near-horizon limit as we discussed earlier. Moreover, the assumption that the chiral half of the CFT originates from a limiting DLCQ procedure is consistent with the fact that there is no natural $SL(2,\mathbb R)\times SL(2,\mathbb R)$ invariant geometry in the boundary conditions~\eqref{strictbc}, which would be dual to the vacuum state of the CFT. Indeed, even in the three-dimensional example, the geometric dual to the vacuum state (the AdS\sub{3} geometry) does not belong to the phase space defined in the near-horizon limit of extremal black holes. However, here there is no natural $SL(2,\mathbb R)\times SL(2,\mathbb R)$ invariant geometry which would be dual to the vacuum state due to the warping. This leads to considering a deformed $2d$ CFT which in the DLCQ limit would reproduce the growth of states.

We saw in Section~\ref{sec:temp} that scalar quantum fields in the analogue of the Frolov--Thorne vacuum restricted to extremal excitations have the temperature \eqref{TL}. Individual modes are co-rotating with the black hole along $\p_\phi$. Since we identify the left-sector of the deformed DLCQ CFT with excitations along $\p_\phi$ and the right $SL(2,\mathbb R)_R$ sector is frozen, the states are described by a thermal density matrix with temperatures
\be
T_L = T_{\phi},\qquad T_R = 0, \label{tempCFT}
\ee
where $T_{\phi}$ is given in \eqref{TL}. The other quantities $T_e$ and $T_m$ defined in \eqref{formTem} are then better interpreted as being proportional to auxiliary chemical potentials. One can indeed rewrite the Boltzman factor~\eqref{bol1} as 
\be
\exp\left( -\hbar\frac{m -q_e \,\mu_L^{J,e} - q_m \,\mu_L^{J,m}}{T_L} \right), \label{bol2}
\ee
where the left chemical potentials are defined as
\be
\mu^{J,e}_L \equiv - \frac{T_\phi}{T_e},\qquad \mu_L^{J,m} \equiv - \frac{T_\phi}{T_m}\, .\label{defmu}
\ee
It is remarkable that applying blindly Cardy's formula~\eqref{Cardy} using the central charge $c_L = c_J$ given in \eqref{c2} and using the temperatures \eqref{tempCFT}, one reproduces the extremal Bekenstein--Hawking black-hole entropy 
\be
\cS_{\text{CFT}} \overset{!}{=} \cS_{\ext}\, ,\label{match}
\ee
as first shown in~\cite{Guica:2008mu}. This matching is clearly not a numerical coincidence. For any spinning extremal black hole of the theory~\eqref{generalaction}, one can associate a left-moving Virasoro algebra of central charge $c_L = c_J$ given in \eqref{c1}. The black-hole entropy~\eqref{SEinstein} is then similarly reproduced by Cardy's formula~\eqref{match}.
 As remarkably, taking any higher curvature correction to the gravitational Lagrangian into account, one also reproduces the Iyer--Wald entropy~\eqref{entropy3} using Cardy's formula, while the central charge~\eqref{c3} is computed (apparently) completely independently from the entropy!

One can easily be puzzled by the incredible matching \eqref{match} valid for virtually any extremal black hole and outside the usual Cardy regime, as discussed in Section~\ref{sec:Cardy}. Indeed, there are no arguments for unitarity and modular invariance, since there is no clear definition of a dual deformed CFT, which we will refer to by the acronym $\CFT_J$. Moreover, the regime $T_R = 0$ lies outside the range of Cardy's formula even for exact CFTs with a sparse light spectrum. This might suggest the existence of a form of universality. Note also that the central charge depends on the black-hole parameters, such as the angular momentum or the electric charge. This is not too surprising since, in known AdS/CFT correspondences where the black hole contains an AdS\sub{3} factor in the near-horizon geometry, the Brown--Henneaux central charge $c=3l/2G_3$~\cite{Brown:1986nw} also depends on the parameters of the black hole because the AdS length $l$ is a function of the black hole charges~\cite{Maldacena:1997re}. 

Let us now discuss another matching valid when electromagnetic fields are present. Instead of assigning the left-moving temperature as \eqref{tempCFT}, one might instead emphasize that electrically-charged particles are immersed in a thermal bath with temperature $T_\chi = R_\chi T_e$, as derived in \eqref{defTchi} in Section~\ref{sec:temp}. Identifying the left sector of the dual field theory with a density matrix at temperature $T_\chi$ and assuming again no right excitations at extremality, we make the following assignment
\be
T_L = T_\chi = R_\chi T_e,\qquad T_R = 0.\label{TL2}
\ee
The other quantities $T_\phi$ and $T_m$ defined in \eqref{formTem} are then better interpreted as being proportional to auxiliary chemical potentials. One can indeed rewrite the Boltzman factor~\eqref{bol1} as 
\be
\exp\left( -\hbar\frac{q_\chi -m \, \mu_L^{Q,\phi} - q_m \, \mu_L^{Q,m}}{T_\chi} \right),\label{bol3}
\ee
where $q_\chi = R_\chi q_e$ is the probe electric charge in units of the Kaluza--Klein length and the left chemical potentials are defined as
\be
\mu_L^{Q,\phi} \equiv - \frac{T_\chi}{T_\phi},\qquad \mu_L^{Q,m} \equiv - \frac{T_\chi}{T_m}\, .\label{defmu2}
\ee
We argued above that in the near-horizon region, excitations along the gauge-field direction fall into representations of the Virasoro algebra defined in \eqref{Virchi}. As supported by non-extremal extensions of the correspondence discussed in Sections \ref{sec:KerrCFT2} and \ref{sec:hidden}, the left sector of the dual field theory can be argued to be the DLCQ of a deformed $2d$~CFT. Remarkably, Cardy's formula~\eqref{Cardy} with temperatures \eqref{TL2} and central charge~\eqref{cpsi} also reproduces the entropy of the Kerr--Newman black hole. When the angular momentum is identically zero, the black-hole entropy of the Reissner--Nordstr\"om black hole $\cS_{\ext}=\pi Q^2$ is then reproduced from Cardy's formula with left central charge $c_L = c_Q$ given in \eqref{cpsi} and left temperature $T_L=R_\chi/(2\pi Q)$ as originally obtained in~\cite{Hartman:2008pb}. As one can easily check, the entropy of the general Kerr--Newman--AdS black hole can be similarily reproduced, as shown in~\cite{Chen:2010bsa,Chen:2010jj,Chen:2010ywa,Chen:2011wm,Chen:2011gz}. We will refer to the class of conjectured dual deformed CFTs with Virasoro algebra~\eqref{Virchi} by the acronym $\CFT_Q$. Note that the entropy matching does not depend on the scale of the Kaluza--Klein dimension $R_\chi$, which is arbitrary in our analysis.

Finally, when two $U(1)$ symmetries are present, one can apply a modular transformation mixing the two $U(1)$ and one obtains a different description for each choice of $SL(2,\mathbb Z)$ element. Indeed, we argued that the set of generators~\eqref{VirSL2Z} obeys the Virasoro algebra with central charge~\eqref{c12KN}. After performing an $SL(2,\mathbb Z)$ change of basis in the Boltzman factor~\eqref{bol1}, we deduce the temperatures and Cardy's formula is similarly reproduced. We will denote the corresponding class of conjectured dual deformed CFTs by the acronym $\CFT_{(p_1,p_2,p_3)}$.

Several extensions of the construction of a Virasoro algebra which allows to reproduce the extremal black hole entropy via the chiral thermal Cardy formula exist. The same reasoning applies to magnetized black holes \cite{Astorino:2015lca,Astorino:2015naa,Siahaan:2015xia}, black holes in the  large $d$ limit \cite{Guo:2015swu}, superentropic black holes \cite{Sinamuli:2015drn}, black rings \cite{Sadeghian:2015hja} and black holes with acceleration \cite{Astorino:2016xiy}.

\newpage

\section{Scattering from Near-Extremal Black Holes}
\label{sec:KerrCFT2}

One can define near-extremal black holes as black holes with a Hawking temperature which is very small compared with the inverse mass
\be
M\, T_H \ll 1\, .\label{nearextregime0}
\ee
Closer and closer to extremality, a near-horizon region with enhanced $SL(2,\mathbb R)$ symmetry progressively develops and such a symmetry become relevant to describe physical processes occuring the vicinity of the horizon. The main idea of this section is that the infinite dimensional extension of conformal symmetry in both left and right sectors is relevant to describe the physical quantities in the near-horizon region \cite{Bredberg:2009pv,Hartman:2009nz}.

In practice, the existence of a nearly decoupled region allows the use of asymptotic matched expansions to solve otherwise complicated partial differential equations. The near region is described by the near-horizon geometries and the far region is described by the asymptotically flat near-extremal black hole geometry. Following \cite{Bredberg:2009pv,Hartman:2009nz} we will consider one such simple process: the scattering of a probe field on the near-extremal Kerr-Newman geometry (see also~\cite{Chen:2010bsa,Chen:2010ni,Chen:2010as,Chen:2010xu,Chen:2010yu,Agullo:2010hi}).\epubtkFootnote{Extensions to the Kerr--Newman--AdS black hole or other specific black holes in four and higher dimensions in gauged or ungauged supergravity can be found in~\cite{Bredberg:2009pv,Cvetic:2009jn,Chen:2010bh,Shao:2010cf,Birkandan:2011fr} (see also~\cite{Chen:2010jj,Chen:2010jc,Durkee:2010ea,Murata:2011my}). No general scattering theory around near-extremal black-hole solutions of \eqref{generalaction} has been proposed so far.} The near-horizon region is relevant only for probes with energy $\omega$ and angular momentum $m$ close to the superradiant bound $\omega \sim m \Omega^{\ext}_J +q_e \Phi^{\ext}_e$, 
\be
M(\omega - m \Omega^{\ext}_J -q_e \Phi^{\ext}_e) \ll 1\, . \label{nearextregime1}
\ee
This condition is equivalent to requiring probing the near-horizon region. In this approach, no explicit metric boundary conditions are needed. Moreover, since gravitational backreaction is a higher-order effect, it can be neglected. One simply computes the black-hole--scattering amplitudes on the black-hole background. In order to test the near-extremal black hole/CFT correspondence, one then has to determine whether or not the black hole reacts like a two-dimensional CFT to external perturbations originating from the asymptotic region far from the black hole. In order to simplify the notation, in this section we will drop all hats on quantities defined in the asymptotic region far from the black hole.

\subsection{Near-extremal Kerr--Newman black holes}
\label{sec:nearextgeo}

Near-extremal Kerr--Newman black holes are characterized by their mass $M$, angular momentum $J = M a$ and electric charge $Q$. (We take $a,Q \geq 0$ without loss of generality.) They contain near-extremal Kerr and Reissner--Nordstr\"om black holes as particular instances. The metric and thermodynamic quantities can be found in many references and will not be reproduced here. 

The near-extremality condition~\eqref{nearextregime0} is equivalent to the condition that the reduced Hawking temperature is small, 
\be
\tau_H \equiv \frac{r_+ - r_-}{r_+} \ll 1. \label{nearextregime0equiv}
\ee
Indeed, one has $\tau_H = M\, T_H [ 4\pi ((r_+/M)^2+(a/M)^2)/ (r_+/M)] $ and the term in between the brackets is of order one since $0 \leq a/M \leq 1$, $0 \leq Q/M \leq 1$ and $1 \leq r_+/M \leq 2$. Therefore, we can use interchangeably the conditions~\eqref{nearextregime0} and \eqref{nearextregime0equiv}. 

Since there is both angular momentum and electric charge, extremality can be reached both in the regime of vanishing angular momentum $J$ and vanishing electric charge $Q$. When angular momentum is present, we expect that the dynamics could be described by the $\CFT_J$ as defined in Section~\ref{sec:microS}, while when electric charge is present the dynamics could be described by the $\CFT_Q$. It is interesting to remark that the condition
\be
\frac{T_H}{\Omega_J} \ll 1, \label{extJ}
\ee
implies \eqref{nearextregime0}\,--\,\eqref{nearextregime0equiv} since $\tau_H = \frac{T_H}{\Omega_J} \left( 4\pi a/M \right) \ll 1$ but it also implies $a > 0$. Similarly, the condition 
\be
M\frac{T_H}{\Phi_e} \ll 1 \label{extQ}
\ee
implies \eqref{nearextregime0}\,--\,\eqref{nearextregime0equiv}, since $\tau_H = 4 \pi Q T_H/\Phi_e$, but it also implies $Q > 0$. In the following, we will need only the near-extremality condition~\eqref{nearextregime0}, and not the more stringent conditions~\eqref{extJ} or \eqref{extQ}. This is the first clue that the near-extremal scattering will be described by both the $\CFT_J$ and the $\CFT_Q$.


Near-extremal black holes are characterized by an approximative near-horizon geometry, which controls the behavior of probe fields in the window \eqref{nearextregime1}. Upon taking $T_H = O(\lambda)$ and taking the limit $\lambda \rightarrow 0 $ the near-horizon geometry decouples, as we saw in Section~\ref{sec:nearext1}. 

Probes will penetrate the near-horizon region close to the superradiant bound \eqref{nearextregime1}. When $T_H = O(\lambda)$ we need 
\be
\omega = m \Omega_J^{\ext} + q_e \Phi_e^{\ext}+O(\lambda).\label{nearextregime2}
\ee
Indeed, repeating the reasoning of Section~\ref{sec:temp}, we find that the Boltzman factor defined in the near-horizon vacuum (defined using the horizon generator) takes the following form 
\be
e^{\hbar \frac{\omega-m \Omega_J - q_e \Phi_e}{T_H} } = e^{-\hbar n} e^{- \hbar \frac{m}{T_\phi} - \hbar \frac{q_e}{T_e}},
\ee
where $\omega$, $m$ and $q_e$ are the quantum numbers defined in the exterior asymptotic region and 
\be
n \equiv \frac{ \omega - m \Omega_J^{\ext} - q_e \Phi_e^{\ext}}{T_H},\label{defn}
\ee
is finite upon choosing \eqref{nearextregime2}. 

The conclusion of this section is that the geometries~\eqref{GenNearExt} control the behavior of probes in the near-extremal regime \eqref{nearextregime0}\,--\,\eqref{nearextregime1}. We identified the quantity $n$ as a natural coefficient defined near extremality. It will have a role to play in later Sections~\ref{macroext} and \ref{sec:CFTmatch}. We will now turn our attention to how to solve the equations of motion of probes close to extremality.

\subsection{Macroscopic greybody factors}
\label{macrogen}

The problem of scattering of a general spin field from a Kerr black hole was solved in a series of classic papers by Starobinsky~\cite{Starobinsky:19aa}, Starobinsky and Churilov~\cite{Starobinsky:19ab} and Press and Teukolsky~\cite{Teukolsky:1972my,Teukolsky:1973ha,Press:1973zz,Teukolsky:1974yv} in the early 1970s (see also~\cite{Futterman:1988ni,Amsel:2009ev,Dias:2009ex}). The scattering of a spin 0 and 1/2 field from a Kerr--Newman black hole has also been solved~\cite{Teukolsky:1974yv}, while the scattering of spins 1 and 2 from the Kerr--Newman black hole cannot be solved analytically, except in special regimes \cite{Pani:2013ija,Pani:2013wsa}.

Let us review how to solve this classic scattering problem. First, one has to realize that the Kerr--Newman black hole enjoys a remarkable property: it admits a Killing--Yano tensor~\cite{Yano:1952aa,Penrose:1973ab,Floyd:1973aa}. (For a review and some surprising connections between Killing--Yano tensors and fermionic symmetries, see~\cite{Gibbons:1993ap}.) A Killing--Yano tensor is an anti-symmetric tensor $f_{\mu\nu} = - f_{\nu\mu}$, which obeys
\be
\nabla_{(\lambda}f_{\mu ) \nu} = 0.
\ee
This tensor can be used to construct a symmetric Killing tensor
\be
K_{\mu\nu} = f_{\mu}^{\; \, \lambda}f_{\lambda \nu}, \qquad \nabla_{(\lambda}K_{\mu\nu)} = 0,
\ee
which is a natural generalization of the concept of Killing vector $K_\mu$ (obeying $\nabla_{(\mu}K_{\nu)}=0$). This Killing tensor was first used by Carter in order to define an additional conserved charge for geodesics~\cite{Carter:1968rr}
\be
Q = K_{\mu\nu}\dot{x}^\mu \dot{x}^\nu\, ,
\ee
and thereby reduce the geodesic equations in Kerr to first-order equations. More importantly for our purposes, the Killing tensor allows one to construct a second-order differential operator $K^{\mu\nu}\nabla_\mu \nabla_\nu$, which commutes with the Laplacian $\nabla^2$. This allows one to separate the solutions of the scalar wave equation $\nabla^2 \Psi^{s=0} = 0$ as~\cite{Carter:1968rr} 
\be
\Psi^{s=0} = e^{- i \omega t + i m \phi} S_{\omega,A,m}(\th) R_{\omega,A,m}(r)\, ,
\ee
where $A$ is the real separation constant present in both equations for $S(\th)$ and $R(r)$. The underlying Killing--Yano tensor structure also leads to the separability of the Dirac equation for a probe fermionic field. For simplicity, we will not discuss further fermionic fields here and we refer the interested reader to the original reference~\cite{Hartman:2009nz} (see also~\cite{Becker:2012vd}). The equations for spin 1 and 2 probes in Kerr can also be shown to be separable after one has conveniently reduced the dynamics to a master equation for a master scalar $\Psi^s$, which governs the entire probe dynamics. As a result, one has 
\be
\Psi^{s} = e^{- i \omega t + i m \phi} S^s_{\omega,A,m}(\th) R^s_{\omega,A,m}(r)\, .
\ee
The master scalar is constructed from the field strength and from the Weyl tensor for spin 1 ($s=\pm 1$) and spin 2 ($s=\pm 2$) fields, respectively, using the Newman--Penrose formalism. For the Kerr--Newman black hole, all attempts to separate the equations for spin 1 and spin 2 probes have failed. Hence, there is no known analytic method to solve those equations (for details, see~\cite{Chandrasekhar:1983kt}). Going back to Kerr, given a solution to the master scalar field equation, one can then in principle reconstruct the gauge field and the metric from the Teukolsky functions. This non-trivial problem was solved right after Teukolsky's work~\cite{Cohen:1975aa,Chrzanowski:1975wv}; see Appendix~C of~\cite{Dias:2009ex} for a modern review (with further details and original typos corrected). 

In summary, for all separable cases, the dynamics of probes in the Kerr--Newman geometry can be reduced to a second-order equation for the angular part of the master scalar $S^s_{\omega,A,m}(\th)$ and a second-order equation for the radial part of the master scalar $R^s_{\omega,A,m}(r)$. Let us now discuss their solutions after imposing regularity as boundary conditions, which include ingoing boundary conditions at the horizon. We will limit our discussion to the non-negative integer spins $s=0,1,2$ in what follows. 

The angular functions $S^s_{\omega,A,m}(\th)$ obey the spin-weighted spheroidal harmonic equation
\bea
&&\bigg[ \frac{1}{\sin\theta}\p_\theta \left( \sin\theta \p_\theta \right) -a^2 (\omega^2 - \mu^2 \delta_{s,0}) \sin^2\theta - 2 a \omega s \cos\theta - \frac{(m+s \cos\theta)^2}{\sin^2\theta}\nn\\
&& \qquad \qquad + A \bigg] S^s_{\omega,A,m}(\th) = 0\, .\label{eqSs}
\eea
(The Kronecker $\delta_{s,0}$ is introduced so that the multiplicative term only appears for a massive scalar field of mass $\mu$.) All harmonics that are regular at the poles can be obtained numerically and can be classified by the usual integer number $l$ with $l \geq |m|$ and $l \geq |s|$. In general, the separation constant $A=A^s_{a\omega,l,m}$ depends on the product $a \omega$, on the integer $l$, on the angular momentum of the probe $m$ and on the spin $s$. At zero energy ($\omega = 0$), the equation reduces to the standard spin-weighted spherical-harmonic equation and one simply has $A^s_{0,l,m} = l(l+1)-s^2$. 
For a summary of analytic and numerical results, see~\cite{Berti:2005gp}. 

Let us now take the values $A^s_{a \omega,l,m}$ as granted and turn to the radial equation. The radial equation reduces to the following Sturm--Liouville equation
\be
\Big[ \Delta^{-s} \p_r \left( \Delta^{s+1}\p_r \right) - V^s(r) \Big] R^s(r)= 0,\label{eqRs}
\ee
where $\Delta(r) = (r-r_+)(r-r_-) = r^2 - 2M r +a^2 +Q^2$ in a potential $V^s(r)$. The form of the potential is pretty intricate. For a scalar field of mass $\mu$, the potential $V^0(r)$ is real and is given by
\be
V^0(r) = -\frac{H^2(r)}{\Delta(r)} - 2a m \omega + A^0_{a \omega,l,m} +\mu^2(r^2+a^2)\, ,
\ee
where $H(r) = \omega(r^2+a^2)-q_e Q r - am $. For a field of general spin on the Kerr geometry, the potential is, in general, complex and reads as 
\bea
V^s(r) &=& -\frac{H^2(r) - 2 i s(r-M)H(r)}{\Delta(r)}-4 i s \omega r - 2a m \omega + A^s_{a \omega,l,m}-s(s+1) ,
\eea
where $H(r) = \omega(r^2+a^2)- am $. This radial equation obeys the following physical boundary condition: we require that the radial wave has an ingoing group velocity -- or, in other words, is purely ingoing -- at the horizon. This is simply the physical requirement that the horizon cannot emit classical waves. This also follows from a regularity requirement. The solution is then unique up to an overall normalization. For generic parameters, the Sturm--Liouville equation~\eqref{eqRs} cannot be solved analytically and one has to use numerical methods. 

For each frequency $\omega$ and spheroidal harmonic $(l,m)$, the scalar field can be extended at infinity into an incoming wave and an outgoing wave. The absorption probability $\s_{\abs}$ or macroscopic greybody factor is then defined as the ratio between the absorbed flux of energy at the horizon and the incoming flux of energy from infinity,
\be
\s_{\abs}(\omega,l,m, s ; M,a,Q) = \frac{dE_{\abs} / dt}{dE_{\text{in}}/dt}\, .
\ee
An important feature is that in the superradiant range \eqref{superradrange} the absorption probability turns out to be negative, which results in stimulated as well as spontaneous emission of energy, as we reviewed in Section~\ref{propBH}. 


\subsection{Macroscopic greybody factors close to extremality}
\label{macroext}

The Sturm--Liouville problem~\eqref{eqRs} cannot be solved analytically. However, in the regime of near-extremal excitations~\eqref{nearextregime0}\,--\,\eqref{nearextregime1} an approximative solution can be obtained analytically using asymptotic matched expansions: the wave equation is solved in the near-horizon region and in the far asymptotically-flat region and then matched along their common overlap region.

For that purpose, it is useful to define the dimensionless horizon radius $x = (r-r_+)/r_+$ such that the outer horizon is at $x = 0$. The two other singular points of the radial equation~\eqref{eqRs} are the inner horizon $x = -\tau_H$ and spatial infinity $x = \infty$. One then simply partitions the radial axis into two regions with a large overlap as 
\begin{itemize}
\item Near-horizon region: $x \ll 1$,
\item Far region: $x \gg \tau_H$,
\item Overlap region: $\tau_H \ll x \ll 1$.
\end{itemize}
The overlap region is guaranteed to exist thanks to \eqref{nearextregime0equiv}.

In the near-extremal regime, the absorption probability $\s_{\abs}$ gets a contribution from each region as 
\bea
\s_{\abs} &=& \s_{\abs}^{\text{near}}\, \s_{\abs}^{\text{match}} ,\\
\s_{\abs}^{\text{near}} &=& \frac{dE_{\abs}/dt}{|\Psi(x=x_B)|^2},\label{defsabss} \\ 
\s_{\abs}^{\text{match}} &=& \frac{|\Psi(x=x_B)|^2}{dE_{\text{in}}/dt} ,
\eea
where $|\Psi(x=x_B)|^2$ is the norm of the scalar field in the overlap region with $\tau_H \ll x_B \ll 1$. One can conveniently normalize the scalar field such that it has unit incoming flux $dE_{\text{in}}/dt = 1$. The contribution $\s_{\abs}^{\text{match}}$ is then simply a normalization that depends on the coupling of the near-horizon region to the far region.

In the near-horizon region, the radial equation reduces to a much simpler hypergeometric equation. One can in fact directly 
obtain the same equation from solving for a probe in a near-extremal near-horizon geometry of the type~\eqref{GenNearExt}, which is, as detailed in Section~\ref{sec:spin}, a warped and twisted product of $AdS_2 \times S^2$. The presence of poles in the hypergeometric equation at $x=0$ and $x=-\tau_H$ requires one to choose the $\AdS_2$ base of the near-horizon geometry to be 
\be
ds_{(2)}^2 = -x(x+\tau_H) dt^2 + \frac{dx^2}{x(x+\tau_H)}\, .
\ee

One can consider the non-diagonal term $2 \Gamma(\th)\gamma(\th) k r\, dt d\phi$ appearing in the geometry~\eqref{GenNearExt} as a $U(1)$ electric field twisted along the fiber spanned by $d\phi$ over the $\AdS_2$ base space. It may then not be surprising that the dynamics of a probe scalar on that geometry can be expressed equivalently as a charged massive scalar on $\AdS_2$ with two electric fields: one coming from the $U(1)$ twist in the four-dimensional geometry, and one coming from the original $U(1)$ gauge field. By $SL(2,\mathbb R)$ invariance, these two gauge fields are given by 
\be
A_1 = \alpha_1 x\, dt,\qquad A_2 = \alpha_2 x\, dt.
\ee
The coupling between the gauge fields and the charged scalar is dictated by the covariant derivative
\be
\mathcal D = \nabla - i q_1 A_1 - i q_2 A_2 = \nabla - i q_{\mathrm{eff}} A,
\ee
where $\nabla$ is the Levi-Civita connection on $\AdS_2$ and $q_1$ and $q_2$ are the electric charge couplings. One can rewrite more simply the connection as $q_{\mathrm{eff}} A$, where $q_{\mathrm{eff}} = q_1 \alpha_1+ q_2 \alpha_2$ is the effective total charge coupling and $A=xdt$ is a canonically-normalized effective gauge field. The equation for a charged scalar field $\Phi(t,x)$ with mass $\mu_{\mathrm{eff}}$ is then 
\be
\mathcal D^2 \Phi - \mu_{\mathrm{eff}}^2 \Phi = 0.
\ee
Taking $\Phi(t,r) = e^{-i\omega_{\mathrm{eff}} \tau_H t}\Phi(x)$, we then obtain the following equation for $\Phi(x)$,
\be
\left[ \p_x \left( x(x+\tau_H)\p_x \right) + \frac{(\omega_{\mathrm{eff}}\tau_H+q_{\mathrm{eff}} x)^2}{x(x+\tau_H)} - \mu_{\mathrm{eff}}^2 \right] \Phi(x) = 0.\nn
\ee
Using the field redefinition 
\be
\Phi(x) = x^{s/2}(\frac{x}{\tau_H}+1)^{s/2}R^s(x),
\ee
we obtain the equivalent equation,
\be
x(x+\tau_H)\p_x^2 R^s+(1+s)(2x+\tau_H)\p_x R^s + V(x) R^s = 0,\label{eq45}
\ee
where the potential is 
\be
V(x) = \frac{(a x+b \tau_H)^2 - i s(2 x+\tau_H)(a x+b \tau_H)}{x(x+\tau_H)} - c\, .
\ee
Here, the parameters $a,b,c$ are related to $\mu_{\mathrm{eff}}$, $q_{\mathrm{eff}}$ and $\omega_{\mathrm{eff}}$ as\epubtkFootnote{There is a $\mathbb Z_2$ ambiguity in the definition of parameters since Eq.~\eqref{eq45} is invariant upon replacing $(a,b,c)$ by $(i s +2b -a,b,c+(2b-is)(is+2b-2a))$. We simply chose one of the two identifications.} 
\be
a = q_{\mathrm{eff}}+i s,\qquad b = \omega_{\mathrm{eff}}+\frac{i s}{2},\qquad c = \mu_{\mathrm{eff}}^2-s\, .\label{par45}
\ee
Finally, comparing Eq.~\eqref{eq45} with \eqref{eqRs}, where the potential $V^s(r)$ is approximated by the near-horizon potential, we obtain that these equations are identical, as previously announced, after identifying the parameters as
\bea
\omega_{\mathrm{eff}} &=& \frac{n}{4\pi}-\frac{i s}{2},\nn\\
q_{\mathrm{eff}} &=& 2 r_+ \omega - q_e Q - i s, \label{paramseff}\\
\mu^2_{\mathrm{eff}} &=& A^s_{a\omega,l,m}-2 a m \omega-s^2 + \mu^2(r_+^2+a^2)-2i m s.\nn
\eea
Moreover, using the expression of the frequency \eqref{nearextregime2} near extremality, one can write the effective charge in the convenient form 
\be
q_{\mathrm{eff}} = \frac{m}{2\pi T_\phi}+\frac{q_e}{2\pi T_e}-i s,\label{qeff2}
\ee
where the extremal Frolov--Thorne temperatures $T_e$ and $T_\phi$ are defined in \eqref{TKN}.

We can now understand that there are two qualitatively distinct solutions for the radial field $R^s(x)$. Uncharged fields in $\AdS_2$ below a critical mass are unstable or tachyonic, as shown by Breitenlohner and Freedman~\cite{Breitenlohner:1982bm}. Charged particles in an electric field on $\AdS_2$ have a modified Breitenlohner--Freedman bound 
\be
m_{\text{BF}}^2 = -\frac{1}{4}+ q_{\mathrm{eff}}^2,
\ee
in which the square mass is lifted up by the square charge. Below the critical mass, charged scalars will be unstable to Schwinger pair production~\cite{Pioline:2005pf,Kim:2008xv}. Let us define
\be
\beta^2 \equiv \mu_{\mathrm{eff}}^2 - m_{\text{BF}}^2\, .\label{defbeta}
\ee
Stable modes will be characterized by a real $\beta \geq 0$, while unstable modes will be characterized by an imaginary $\beta$. This distinction between modes is distinct from superradiant and non-superradiant modes. Indeed, from the definition of $n$ \eqref{defn}, superradiance happens at near-extremality when $n < 0$.

We can now solve the equation, impose the boundary conditions, compute the flux at the horizon and finally obtain the near-horizon absorption probability. The computation can be found in~\cite{Bredberg:2009pv,Cvetic:2009jn,Hartman:2009nz}. The net result is as follows. A massive, charge $e$, spin $s=0,\frac{1}{2}$ field with energy $\omega$ and angular momentum $m$ and real $\beta >0 $ scattered against a Kerr--Newman black hole with mass $M$ and charge $Q$ has near-region absorption probability
\bea
\sigma_{\abs}^{\text{near}} &\sim & {(T_H)^{2\beta} (e^{\frac{n}{2}}-(-1)^{2s}e^{-\frac{n}{2}}) \over \Gamma(2\beta)^2} |\Gamma\bigg(\half + \beta -s + i \, \text{Re}(q_{\mathrm{eff}})\bigg)|^2 \nn \\ && 
|\Gamma\bigg(\half + \beta + i\Big(\frac{n}{2\pi} -\text{Re}(q_{\mathrm{eff}})\Big)\bigg)|^2 \, .\label{gravsigma}
\eea
For a massless spin $s= 1,2$ field scattered against a Kerr black hole, exactly the same formula applies, but with $e=Q=0$. The absorption probability in the case where $\beta$ is imaginary can be found in the original papers~\cite{Press:1973zz,Teukolsky:1974yv}.

We will now show that the formulae~\eqref{gravsigma} are Fourier transforms of CFT correlation functions. We will not consider the scattering of unstable fields with $\beta$ imaginary in this review. We refer the reader to~\cite{Bredberg:2009pv} for arguments on how the scattering absorption probability of unstable spin 0 modes around the Kerr black hole matches with a dual CFT description as well.

\subsection{Microscopic greybody factors}
\label{sec:CFTmatch}

In this section we model the emission amplitudes from a microscopic point of view. We will first discuss near-extremal spinning black holes and we will extend our discussion to general charged and/or spinning black holes at the end of this section. The presentation mostly summarizes~\cite{Bredberg:2009pv,Cvetic:2009jn,Hartman:2009nz}. Relevant earlier work includes \cite{Maldacena:1997ih,Mathur:1997et,Gubser:1997qr}.

The working assumption of the microscopic model is that the near-horizon region of any near-extremal spinning black hole can be described and therefore effectively replaced by a dual two-dimensional CFT. This is a strong assumption since as we discussed earlier one would expect departure from a standard CFT in several respects. Assuming the existence of a $2d$ CFT with possible deformations, the near-horizon region is removed from the spacetime and replaced by the CFT glued along the boundary. Therefore, it is the near-horizon region contribution alone that we expect to be reproduced by the CFT. The normalization $\s_{\abs}^{\text{match}}$ defined in \eqref{defsabss} will then be dictated by the explicit coupling between the CFT and the asymptotically-flat region.

Remember from the asymptotic symmetry group analysis in Section~\ref{sec:BC} and \ref{sec:c} that boundary conditions were found where the exact symmetry of the near-horizon extremal geometry can be extended to a Virasoro algebra as 
\be
U(1)_L \times SL(2,\mathbb R)_R \rightarrow \text{Vir}_L \times SL(2,\mathbb R)_R\, .
\ee
The right sector was taken to be frozen at extremality. 

We will now assume that quantum gravity states form a representation of both a left and a right-moving Virasoro algebra with generators $L_n$ and $\bar L_n$. The value of the central charges are irrelevant for our present considerations. At near-extremality, the left sector is thermally excited at the extremal left-moving temperature \eqref{TL}. We take as an assumption that the right-moving temperature is on the order of the infinitesimal reduced Hawking temperature. As discussed in Sections~\ref{sec:nobulkdof} and \ref{sec:freez}, the presence of right-movers destabilize the near-horizon geometry. For the Kerr--Newman black hole, we have
\be
T_L = \frac{M^2+a^2}{4\pi J} ,\qquad T_R \sim \tau_H .\label{TLKN}
\ee

In order to match the bulk scattering amplitude for near-extremal Kerr--Newman black holes, the presence of an additional left-moving current algebra is required~\cite{Cvetic:2009jn,Hartman:2009nz}. This current algebra is expected from the thermodynamic analysis of charged rotating extremal black holes. We indeed obtained in Section~\ref{sec:temp} and in Section~\ref{sec:microS} that such black holes are characterized by the chemical potential $\mu_L^{J,e}$ defined in \eqref{defmu} associated with the $U(1)_e$ electric current. Using the expressions \eqref{TKN}, we find for the Kerr--Newman black hole the value
\be
\mu_L^e = - \frac{Q^3}{2J}.\label{mulLKN}
\ee

As done in~\cite{Bredberg:2009pv}, we also assume the presence of a right-moving $U(1)$ current algebra, whose zero eigenmode $\bar J_0$ is constrained by the level matching condition
\be
\bar J_0 = L_0.\label{match0}
\ee
The level matching condition is consistent with the fact that the excitations are labeled by three $(\omega,m,q_e)$ instead of four conserved quantities. The CFT state is then assumed to be at a fixed chemical potential $\mu_R$. This right-moving current algebra cannot be detected in the extremal near-horizon geometry in the same way that the right-moving Virasoro algebra cannot be detected, so its existence is conjectural (see, however,~\cite{Castro:2009jf}). This right-moving current algebra and the matching condition~\eqref{match0} will turn out to be adequate to match the gravitational result, as detailed below. 

Therefore, under these assumptions, the symmetry group of the CFT dual to the near-extremal Kerr--Newman black hole is given by the product of a $U(1)$ current and a Virasoro algebra in both sectors,
\be
(\text{Vir}_L \times \text{Curr}_L) \times (\text{Vir}_R \times \text{Curr}_R) . 
\ee

In the description where the near-horizon region of the black hole is replaced by a CFT, the emission of quanta is due to couplings 
\be
\Phi_{\text{bulk}} \mathcal O
\ee
between bulk modes $\Phi_{\text{bulk}}$ and operators ${\mathcal O}$ in the CFT. The structure of the scattering cross section depends on the conformal weights $(h_L,h_R)$ and charges $(q_L, q_R)$ of the operator. The normalization of the coupling is also important for the normalization of the cross section.

The conformal weight $h_R$ can be deduced from the transformation of the probe field under the scaling $\bar L_0 = t \p_t - r \p_r $ \eqref{zeta0} in the overlap region $\tau_H \ll x \ll 1$. The scalar field in the overlap region is $\Phi \sim \Phi_0(t,\theta,\phi) r^{-\frac{1}{2}+\beta}+\Phi_1(t,\theta,\phi) r^{-\frac{1}{2}-\beta}$. Using the rules of the AdS/CFT dictionary~\cite{Witten:1998qj}, this behavior is related to the conformal weight as $\Phi \sim r^{h_R-1},r^{-h_R}$. One then infers that~\cite{Hartman:2009nz}
\be
h_R = \frac{1}{2} + \beta\, .\label{valhR}
\ee
The values of the charges $(q_L,q_R)$ are simply the $U(1)$ charges of the probe, 
\be
q_L = q_e,\qquad q_R = m\, ,
\ee
where the charge $q_R = m$ follows from the matching condition~\eqref{match0}. We don't know any first-principle argument leading to the values of the right-moving chemical potential $\mu_R$, the right-moving temperature $T_R$ and the left-moving conformal weight $h_L$. We will deduce those values from matching the CFT absorption probability with the gravitational result. 

In general, the weight \eqref{valhR} will be complex and real weights will not be integers. However, a curious fact, described in~\cite{Durkee:2010ea,Murata:2011my}, is that for any axisymmetric perturbation ($m=0$) of any integer spin $s$ of the Kerr black hole, the conformal weight \eqref{valhR} is an integer
\be
h_R = 1 + l ,
\ee
where $l=0,1,\dots$. One can generalize this result to any axisymmetric perturbation of any vacuum five-dimensional near-horizon geometry~\cite{Murata:2011my}. Counter-examples exist in higher dimensions and for black holes in AdS~\cite{Durkee:2010ea}. There is no microscopic accounting of this feature at present.

Throwing the scalar $\Phi_{\text{bulk}}$ at the black hole is dual to exciting the CFT by acting with the operator $\mathcal O$. Reemission is represented by the action of the Hermitian conjugate operator. Therefore, the absorption probability is related to the thermal CFT two-point function~\cite{Maldacena:1997ih} 
\be\label{cftt}
G(t^+,t^-) = \langle \bigO^\dagger(t^+,t^-)\bigO(0)\rangle \, ,
\ee
where $t^\pm$ are the coordinates of the left and right moving sectors of the CFT. At left and right temperatures $(T_L, T_R)$ and at chemical potentials $(\mu_L, \mu_R)$ an operator with conformal dimensions $(h_L, h_R)$ and charges $(q_L, q_R)$ has the two-point function
\be\label{gzerotemp}
G \sim (-1)^{h_L+h_R}\left(\pi T_L\over \sinh(\pi T_L t^+)\right)^{2h_L} \left(\pi T_R\over \sinh(\pi T_R t^-)\right)^{2h_R}e^{iq_L \mu_L t^+ +iq_R\mu_Rt^-} \, ,
\ee
which is determined by conformal invariance. From Fermi's golden rule, the absorption cross section is~\cite{Bredberg:2009pv,Cvetic:2009jn,Hartman:2009nz}
\bea\label{cftform}
\sigma_{\abs}(\omega_L,\omega_R) &\sim & \int dt^+dt^- e^{-i\omega_R t^- - i\omega_Lt^+}\left[G(t^+-i\epsilon,t^--i\epsilon) \right. \nn\\
&& \left. - G(t^++i\epsilon,t^-+i\epsilon)\right] \, .
\eea
Performing the integral in (\ref{cftform}), we obtain\epubtkFootnote{The two-point function~(\ref{gzerotemp}) has a branch cut, and as a result, one must find a way to fix the choice of relative sign between the two exponentials in (\ref{cftsigma}). The sign is fixed by matching the gravitational computation to be $-(-1)^{2s}$, where $s$ is the spin of the corresponding field.}
\bea\label{cftsigma}
\sigma_{\abs}
&\sim& T_L^{2h_L-1}T_R^{2h_R-1} \left(e^{\pi \tilde{\omega}_L + \pi \tilde{\omega}_R} \pm e^{-\pi \tilde{\omega}_L - \pi \tilde{\omega}_R}\right) 
|\Gamma(h_L + i \tilde{\omega}_L) |^2 |\Gamma(h_R + i \tilde{\omega}_R) |^2 \, ,
\eea
where
\be\label{cftfreq}
\tilde{\omega}_L = {\omega_L - q_L \mu_L\over 2 \pi T_L} \, , \quad \ \tilde{\omega}_R = {\omega_R - q_R \mu_R\over 2\pi T_R} \, .
\ee

In order to compare the bulk computations to the CFT result~\eqref{cftsigma}, we must match the conformal weights and the reduced momenta $(\tilde \omega_L, \tilde \omega_R)$. The gravity result~\eqref{gravsigma} agrees with the CFT result~\eqref{cftsigma} if and only if we choose 
\bea
h_L &=& \frac{1}{2}+\beta - |s|,\qquad h_R = \frac{1}{2}+\beta ,\nn \\
\tilde \omega_L &=& \text{Re}(q_{\mathrm{eff}}),\qquad \tilde \omega_R = \frac{n}{2\pi}-\text{Re}(q_{\mathrm{eff}})\, .\label{match33}
\eea
The right conformal weight matches with \eqref{valhR}, consistent with $SL(2,\mathbb R)_R$ conformal invariance. The left conformal weight is natural for a spin $s$ field since $|h_L-h_R| = |s|$. The value for $\tilde \omega_L$ is consistent with the temperature \eqref{TLKN} and chemical potential \eqref{mulLKN}. Indeed, since the left-movers span the $\phi$ direction of the black hole, we have $\omega_L = m$. We then obtain
\be
\tilde \omega_L = \frac{2mJ+q_e Q^3}{r_+^2+a^2} = \text{Re}(q_{\mathrm{eff}}),\label{match34}
\ee
after using the value~\eqref{qeff2}. The value of $\tilde \omega_R$ is fixed by the matching. It determines one constraint between $\omega_R$, $\mu_R$ and $T_R$. However, there is a subtlety in the above matching procedure. The conformal weights $h_L$ and $h_R$ depend on $m$ through $\beta$. This $m$ dependence cannot originate from $\omega_L=m$ since $\omega_L$ is introduced after the Fourier transform~\eqref{cftform}, while $h_L,h_R$ are already defined in \eqref{gzerotemp}. One way to introduce this $m$ dependence is to assume that there is a right-moving current algebra and that the dual operator $\mathcal O$ has the zero-mode charge $q_R = m$, which amounts to imposing the condition~\eqref{match0}. (It is then also natural to assume that the chemical potential is $\mu_R \sim \Omega_J$, but the matching does not depend on any particular value for $\mu_R$~\cite{Bredberg:2009pv}.) This justifies why a right-moving current algebra was assumed in the CFT. The dependence of the conformal weights in $q_e$ is similarly made possible thanks to the existence of the left-moving current with $q_L = q_e$. The matching is finally complete. 

Now, let us notice that the matching conditions~\eqref{match33}\,--\,\eqref{match34} are ``democratic'' in that the roles of angular momentum and electric charge are put on an equal footing, as noted in~\cite{Chen:2010bsa,Chen:2010jj}. One can then also obtain the conformal weights and reduced left and right frequencies $\tilde \omega_L,\, \tilde\omega_R$ using alternative CFT descriptions such as the $\CFT_Q$ with Virasoro algebra along the gauge field direction, and the mixed $SL(2,\mathbb Z)$ family of CFTs. We can indeed rewrite \eqref{cftfreq} in the alternative form
\be
\tilde{\omega}_L = \frac{m T_e + q_e T_\phi}{2\pi T_\phi T_e} = \frac{q_\chi-\mu_L^{\phi,Q} m}{2\pi T_L^Q},
\ee
where $T_L^Q = R_\chi T_e$ is the left-moving temperature of the $\CFT_Q$, $\mu_L^{\phi,Q}$ is the chemical potential defined in \eqref{defmu2} and $q_\chi = R_\chi q_e$ is the probe electric charge in units of the Kaluza--Klein circle length. The identification of the right-moving sector is unchanged except that now $q_R = q_e$. One can trivially extend the matching with the $SL(2,\mathbb Z)$ family of CFTs conjectured to equally describe the \mbox{(near-)ext}remal Kerr--Newman black hole.

In summary, near-superradiant absorption probabilities of probes in the near-horizon region of near-extremal black holes are exactly reproduced by conformal field theory two-point functions. This shows the intriguing role of an underlying CFT description (or multiple CFT descriptions in the case where several $U(1)$ symmetries are present) of part of the dynamics of near-extremal black holes. We expect that a general scattering theory around any near-extremal black-hole solution of \eqref{generalaction} will also be consistent with a CFT description, as supported by all cases studied beyond the Kerr--Newman black hole~\cite{Cvetic:2009jn,Chen:2010bh,Shao:2010cf,Chen:2010jj,Chen:2010jc,Birkandan:2011fr}.

The conformal symmetries of the Kerr--Newman geometry close to extremality can be further investigated along several routes. First, one can attempt to match higher order correlation functions with CFT expectations. This line of thought has been partially developed for three-point functions~\cite{Becker:2010jj,Becker:2010dm,Becker:2014jla}. One can also match other gravitational observables in the near-extremal near-horizon region with CFT observables such as the gravitation emitted by a circular orbit or a plunge orbit at the vicinity of an extremal spinning black hole \cite{Porfyriadis:2014fja,Hadar:2014dpa,Hadar:2015xpa}. Conformal invariance in the near-horizon geometry also constraints electromagnetic radiation emitted close to the horizon \cite{Li:2014bta,Lupsasca:2014pfa,Zhang:2014pla,Lupsasca:2014hua,Compere:2015pja,Gralla:2016jfc}.

\subsection{Microscopic accounting of superradiance}

We mentioned in Section~\ref{propBH} that extremal spinning black holes that do not admit a globally-defined timelike Killing vector spontaneously emit quanta in the range of frequencies \eqref{superradrangeJ}. This quantum effect is related by detailed balance to the classical effect of superradiant wave emission, which occur in the same range of frequencies. 

It has been argued that the bound~\eqref{superradrangeJ} essentially follows from Fermi--Dirac statistics of the fermionic spin-carrying degrees of freedom in a dual two-dimensional CFT~\cite{Dias:2007nj} (see also~\cite{Emparan:2007en}). These arguments were made for specific black holes in string theory but one expects that they can be applied to generic extremal spinning black holes, at least qualitatively. Let us review these arguments here. 

One starts with the assumption that extremal spinning black holes are modeled by a $2d$~CFT, where the left and right sectors are coupled only very weakly. Therefore, the total energy and entropy are approximately the sum of the left and right energies and entropies. The state corresponding to an extremal spinning black hole is modeled as a filled Fermi sea on the right sector with zero entropy and a thermal state on the left sector, which accounts for the black-hole entropy. The right-moving fermions form a condensate of aligned spins $s=+1/2$, which accounts for the macroscopic angular momentum. It is expected from details of emission rates in several parametric regimes that fermions are only present on the right sector, while bosons are present in both sectors~\cite{Cvetic:1997ap,Cvetic:2009jn}. 

Superradiant spontaneous emission is then modeled as the emission of quanta resulting from interaction of a left and a right-moving mode. Using details of the model such as the fact that the Fermi energy should be proportional to the angular velocity $\Omega_J$, one can derive the bound~\eqref{superradrangeJ}. We refer the reader to~\cite{Emparan:2007en} for further details. It would be interesting to better compare these arguments to the present setup, and to see how these arguments could be generalized to the description of the bound~\eqref{superradrangeQ} for static extremal rotating black holes. 

\newpage

\section{Conformal Symmetry for Non-Extremal Black Holes}
\label{sec:hidden}

The analyses in Sections~\ref{sec:KerrCFT1} and~\ref{sec:KerrCFT2} strongly relied on the existence of a decoupled near-horizon region in the extremal limit with enhanced symmetry. More precisely, it was found that there is an exact $SL(2,\mathbb R)$ symmetry, an additional asymptotic Virasoro symmetry. Moreover, the full conformal group seems to be the symmetry controlling the formula for near-horizon scattering cross-sections. 
 Away from extremality, one cannot decouple the horizon from the surrounding geometry. Therefore, it is unclear why the previous considerations will be useful in describing any non-extremal physics. 

It might then come as a surprise that away from extremality, $SL(2,\mathbb R)\times SL(2,\mathbb R)$ invariance is present in the dynamics of probe scalar fields around the Kerr black hole in a specific regime (at low energy and close enough to the black hole as we will make more precise below)~\cite{Castro:2010fd} (see also~\cite{Maldacena:1997ih,Mathur:1997et,Gubser:1997qr,Cvetic:1997uw} for related earlier work). In that regime, the probe scalar field equation can be written as a $SL(2,\mathbb R)\times SL(2,\mathbb R)$ Casimir in a region close enough to the horizon. Such a local hidden symmetry is non-geometric but appears in the probe dynamics. The $2\pi$ periodic identification of the azimuthal angle $\phi$ breaks globally-conformal symmetry. Using the properties of this representation of conformal invariance, it was then argued that the Kerr black hole is described by a CFT with specific left and right-moving temperatures~\cite{Castro:2010fd}
\be
T_L = \frac{M^2}{2\pi J},\qquad T_R = \frac{\sqrt{M^4-J^2}}{2\pi J}.\label{TLTR}
\ee
It was also shown that the entropy of the Kerr black hole can be written as a thermal Cardy formula if one assumes that the conjectured CFT has left and right-moving central charges equal to the value $c_L=c_R=12J$, which matches with the value for the left-moving central charge~\eqref{c2} derived at extremality. Note however that there is no known derivation of two Virasoro algebras with central charges $c_L=c_R=12 J$ from the non-extremal Kerr geometry.\epubtkFootnote{Note that at extremality $J=M^2$, so the central charge at extremality~\eqref{c2} could as well be written as $c_L = 12M^2$. However, away from extremality, matching the black hole entropy requires that the central charge be expressed in terms of the quantized charge $c_L = 12 J$.}

In most of the regimes, probe scalar fields are not constrained by conformal invariance. There is therefore no $2d$ CFT dual to a non-extremal Kerr-Newman black hole, at best there might be a $2d$ CFTs with irrelevant deformations in both left and right sectors. This idea can be made more precise by considering deformations of the black hole which preserve its thermodynamics but transform the asymptotics to a geometry with an $\AdS_3$ geometry. Such deformations have been dubbed ``subtracted geometries''~\cite{Cvetic:2011hp,Cvetic:2011dn,Cvetic:2012tr}. Subtracted geometries are supported by additional matter fields, they can be uplifted in 5 dimensions to $\AdS_3 \times S^2$ and they can be formally generated using particular solution generating techniques named Harrison transformations \cite{Virmani:2012kw,Cvetic:2013cja}. In the asymptotically $\AdS_3$ geometry, the $\AdS_3/\CFT_2$ correspondence applies and one can obtain the operators necessary to deform the geometry to the asymptotically flat one \cite{Baggio:2012db}. Such operators correspond indeed to irrelevant deformations in both sectors. The main outcome of this analysis is that the IR cutoff set by the temperatures \eqref{TLTR} will in general be of the same order of magnitude as the UV cutoff set by the mass scale associated with irrelevant perturbations which make the CFT description of the Kerr-Newman geometry doubtful \cite{Baggio:2012db}. 

Yet, given the possible generalization of the extremal Kerr/CFT results to general classes of extremal spinning or charged black holes, it is natural to 
test the ideas proposed in~\cite{Castro:2010fd} to more general black holes than the Kerr geometry. First, hidden conformal symmetry can be found around the non-extremal Reissner--Nordstr\"om black hole~\cite{Chen:2010as,Chen:2010yu} under the assumption that the gauge field can be understood as a Kaluza--Klein gauge field, as done in the extremal case~\cite{Hartman:2009nz}. One can also generalize the analysis to the Kerr--Newman black hole~\cite{Wang:2010qv,Chen:2010xu,Chen:2010ywa}. In complete parallel with the existence of an $SL(2,\mathbb Z)$ family of CFT descriptions, there is a class of hidden $SL(2,\mathbb R)\times SL(2,\mathbb R)$ symmetries of the Kerr--Newman black hole related with $SL(2,\mathbb Z)$ transformations~\cite{Chen:2011kt}\epubtkFootnote{Therefore, one can expect that there will be also a  $SL(2,\mathbb Z)$ family of subtracted geometries for the Kerr-Newman black hole, which has not been constructed so far.}. As we will discuss in Section~\ref{sec:sl2r} each member of the $SL(2,\mathbb Z)$ family of CFT descriptions describes only probes with a fixed ratio of probe angular momentum to probe charge. Remarkably, for all cases where a hidden local conformal invariance can be described, the non-extremal black-hole entropy matches with Cardy's formula using the central charges $c_R=c_L$ and using the value $c_L$ in terms of the quantized conserved charges derived at extremality. Another natural question is whether or not this hidden conformal symmetry can be found for higher spin waves and in particular spin 2 waves around the Kerr black hole. The answer is affirmative \cite{Lowe:2013uea} and the same temperatures \eqref{TLTR} are found for any massless spin $s$ field around Kerr. Moreover it was found in \cite{Lowe:2013uea} that the angular part of the wave is described by a similar hidden $SU(2) \times SU(2)$ symmetry. Five-dimensional asymptotically-flat black holes were also discussed in~\cite{Krishnan:2010pv,Chen:2010jc}.

Another interesting property of non-extremal spinning or charged black holes is the remarkable product of area law which suggests an effective long string description of the microscopic degrees of freedom \cite{Larsen:1997ge,Cvetic:1997uw,Cvetic:1997xv}. The product of the inner horizon area and outer horizon area is independent of the mass and therefore only depends upon quantized charges. Even though the inner horizon is not physical because of classical instabilities, the thermodynamic quantities defined at the inner horizon have a role to play as illustrated in the scattering problem of probe fields around a black hole \cite{Cvetic:1997uw,Cvetic:2011hp}. A precise accounting of why inner horizon quantities appear in classical scattering problems has been given as follows \cite{Castro:2013kea,Castro:2013lba}. A scattering problem around an asymptotically flat black hole involves a radial second-order differential equation whose poles have as a location the black hole horizons and spatial infinity. Analytically continuing the solutions to the complex plane, it is natural to define the monodromies of two independent solutions around each of the poles. Since a circle encircling all the poles can be deformed to a point, its monodromy is trivial. On the other hand, it is equal to the sum of monodromies of all the poles. Then, when expressed in a common basis of solutions, the 2 by 2 monodromy matrices multiplie to the identity matrix, 
\bea
M_\infty M_+ M_- = \mathbb I. \label{mono}
\eea
Scattering amplitudes or greybody factors which require boundary conditions at the outer horizon and at infinity are therefore intrinsically also related to the inner horizon through the relation \eqref{mono}. Moreover, in the regime studied in \cite{Castro:2010fd} the irregular singular pole at infinity can be approximated by a regular pole and the wave equation reduces to an hypergeometric function similar to the one describing scattering around the BTZ black hole. This is another manifestation of the hidden conformal symmetry of these probes in the Kerr geometry. 

In attempting to generalize the hidden symmetry arguments to four-dimensional black holes in AdS one encounters an apparent obstruction, as we will discuss in Section~\ref{allwaves}. It is expected that hidden symmetries are present at least close to extremality, as illustrated by five-dimensional analogues~\cite{Birkandan:2011fr}. However, the structure of the wave equation is more intricate far from extremality because of the presence of complex poles (and associated additional monodromies), which might have a role to play in microscopic models~\cite{Cvetic:2010mn}.\epubtkFootnote{One single copy of hidden $SL(2,\mathbb R)$ symmetry can also be found around the Schwarzschild black hole~\cite{Bertini:2011ga} (see also~\cite{Lowe:2011aa}) but no analogue of the temperatures \eqref{TLTR} could be defined. Since there is no extremal limit of the Schwarzschild geometry, this approach cannot be supported to an asymptotic symmetry group derivation.}

In what follows, we first define various quantities at the inner horizon of black holes and review several of their puzzling features. We then review the equations of motion of scalar probing non-extremal black hole geometries and we study their separability properties. We then present a summary of the derivation of the hidden symmetries of the Kerr--Newman black hole and we discuss their possible CFT interpretation. We will mostly follow the approach of~\cite{Castro:2010fd} but we will generalize the discussion to the Kerr--Newman black hole and their generalization to supergravity theories, which contains several new interesting features. In particular, we will show that each member of the conjectured $SL(2,\mathbb Z)$ family of CFT descriptions of the Kerr-Newman black hole controls part of the dynamics of low energy, low charge and low mass probes. 

\subsection{Properties of inner horizons}

Inner horizons of non-extremal black holes are unstable to gravitational perturbations and therefore do not exist for realistic black holes. Nevertheless, since stationary and axisymmetric eternal black holes admit an inner horizon and realistic black holes are described in the first approximation by such ideal black holes, the properties of inner horizons are relevant even for realistic black holes, as illustrated e.g., by the monodromy argument \eqref{mono} presented earlier. Associating thermodynamic-like variables at the inner horizon of non-extremal black holes is an old idea  \cite{1979NCimB..51..262C,1979NCimB..52..165C,1992MNRAS.255..539O}. Two particularly interesting quantities are the inner horizon ``temperature'' $T_- = \frac{\kappa_-}{2\pi}$ and ``entropy'' $S_- = \frac{A_-}{4G}$ defined in terms of the area of the inner horizon $A_-$ and the surface gravity $\kappa_-$ corresponding to the null generator of the inner horizon. One can prove the first law and Smarr formula for asymptotically flat black holes following the same arguments as for the standard first law and Smarr formula. However, for all known 4-dimensional black holes, one can check that 
\bea
S_- T_- \leq 0,
\eea 
which obscures the physical interpretation of $S_-$ and $T_-$. 

Nevertheless, the role of quantities defined at the inner horizon in the microscopic description of black holes has been repeatedly emphasized \cite{Cvetic:1997xv,Cvetic:2011hp}. In the $\AdS_3/\CFT_2$ correspondence, the product of areas of the inner and outer horizon of the BTZ black hole is quantized as
\bea
\frac{A_+ A_-}{(8 \pi G_3)^2} = 
N_R - N_L  \label{ae1}
\eea 
where $N_{L,R}$ are the numbers of left/right-moving excitations in the dual CFT in the gravity approximation. Quite surprizingly, all known four and five dimensional asymptotically flat black holes enjoy the property that the product of inner and outer areas is independent of the mass. It only depends upon the angular momenta and electromagnetic charges and is therefore quantized. For example, the (non-extremal) Kerr-Newman black hole of electric and magnetic charges $Q$ and $P$ obeys\epubtkFootnote{In this section we follow the conventions of \cite{Cvetic:2010mn} for the normalizations of the charges.}
\bea
\frac{A_+ A_-}{(8 \pi G_4)^2} = J^2 + (Q^2 + P^2)^2.\label{ae2}
\eea
For other cases see \cite{Cvetic:2010mn,Castro:2012av,Chow:2014cca,Goldstein:2014gta} and references therein.

Asymptotically AdS black holes in 4 dimensions do not obey an area product law of the type \eqref{ae1}-\eqref{ae2}. Instead, one can analytically continue the black hole solution in the complex plane and consider the complex horizons (defined precisely later as the complex roots of \eqref{quart}) with areas $A_c$, $A_c^*$. The following quantization condition for the Kerr-Newman-AdS black hole then holds \cite{Cvetic:2010mn},
\bea
\frac{A_+ A_-}{(8 \pi G_4)^2} \frac{A_c A_c^*}{(4 \pi G_4 l^2)^2} = J^2 + (Q^2 + P^2)^2.
\eea

Four-dimensional asymptotically flat black holes also enjoy additional relations involving inner horizon quantities. The Kerr-Newman black hole obeys
\bea
T_+ S_+ + T_- S_- = 0, \\
\frac{\Omega_+}{T_+} + \frac{\Omega_-}{T_-} = 0, \label{kineq}\\
8 \pi^2 J = \Omega_+ S_+ \left( \frac{1}{T_+} + \frac{1}{T_-} \right) 
\eea
and these relationships extend as well as to any black hole solution of $\mathcal N = 8$ supergravity \cite{Chow:2014cca}. Black holes in theories of gravity with higher derivatives do not in general obey these relations \cite{Castro:2013pqa}. However, it is not clear to us whether or not black holes in string theory with $\alpha'$ corrections (with specific small higher derivative corrections) will disobey them.

\subsection{Scalar wave equation}
\label{allwaves}

Let us discuss general features of the massless Klein--Gordon equation in non-extremal black holes geometries. We restrict our discussions to four dimensions for simplicity. 
An essential property of all known black holes solutions is the existence of a conformal Killing--St\"ackel tensor which implies that the massless Klein-Gordon equation is separable. A rank-2 conformal Killing--St\"{a}ckel tensor is a symmetric tensor $Q_{\mu \nu} = Q_{(\mu \nu)}$ that satisfies $\nabla_{(\mu} Q_{\nu \rho)} = q_{(\mu} g_{\mu \nu)}$ for some $q_\mu$. The existence of such a tensor allows to build the operator $Q^{\mu\nu}\nabla_\mu \nabla_\nu$ which commutes with the Laplacian $\square \equiv \nabla^\mu \nabla_\mu$.

A general class of asymptotically flat or AdS metrics which admits a conformal Killing--St\"ackel tensor can be written in the following form \cite{Chow:2014cca} 
\begin{align}
\label{separablemetric}
\df s^2 & = - \fr{\Delta_r - \Delta_u}{W} \, \df t^2 - \fr{(L_u \Delta_r + L_r \Delta_u)}{\widehat{a} W} \, 2 \, \df t \, \df \phi  + \fr{(W_r^2 \Delta_u - W_u^2 \Delta_r)}{\widehat{a}^2 W} \, \df \phi^2 + W \bigg( \fr{\df r^2}{\Delta_r} + \fr{\df u^2}{\Delta_u} \bigg) ,
\end{align}
where
\be
\label{separableW}
W^2 = (\Delta_r - \Delta_u) \bigg( \frac{W_r^2}{\Delta_r} - \frac{W_u^2}{\Delta_u} \bigg) + \frac{(L_u \Delta_r + L_r \Delta_u)^2}{\Delta_r \Delta_u} .
\ee
The determinant is $\sr{-g} = W $. Here, $L_r,\Delta_r,W_r$ are functions of the radial coordinate $r$ while $L_u,\Delta_u,W_u$ are functions of the angular coordinate $u$. The parameter $\widehat{a}$ is chosen such that the identification $\phi \sim \phi +2 \pi$ leads to the standard asymptotic behavior at large radius.

The conformal Killing--St\"{a}ckel tensor is given by
\begin{align}
Q^{\mu \nu} \, \pd_\mu \, \pd_\nu & = \fr{1}{r^2 + u^2} \bigg[ \bigg( \fr{u^2 W_r^2}{\Delta_r} + \fr{r^2 W_u^2}{\Delta_u} \bigg) \, \pd_t^2 - \widehat{a} \bigg( \fr{u^2 L_r}{\Delta_r} + \fr{r^2 L_u}{\Delta_u} \bigg) \, 2 \, \pd_t \, \pd_\phi + \widehat{a}^2 \bigg( \fr{r^2}{\Delta_u} - \fr{u^2}{\Delta_r}\bigg) \, \pd_\phi^2 \nnr
& \qquad - u^2 \Delta_r \, \pd_r^2 + r^2 \Delta_u \, \pd_u^2 \bigg] .
\end{align}
It is generically irreducible, i.e., not a linear combination of the metric and products of Killing vectors. This conformal Killing--St\"{a}ckel tensor was identified for asymptotically flat black holes with four electric charges in \cite{Chow:2008fe}, for the asymptotically flat non-extremal rotating Kaluza--Klein black hole in \cite{Keeler:2012mq} and for two quite general classes of asymptotically AdS black holes in $SO(8)$ gauged supergravity in \cite{Chow:2013gba}.

The massless Klein--Gordon equation 
\be
\square \Phi = \fr{1}{\sr{-g}} \pd_\mu (\sr{-g} g^{\mu \nu} \pd_\nu \Phi) = 0
\ee
is separable using the ansatz
\be
\Phi = R (r) S(u) \expe{\im (m \phi - \omega t)} .\label{ans4}
\ee
The radial and angular equations read as
\begin{align}
& \fr{\df}{\df r} \bigg( \Delta_r \fr{\df R}{\df r} \bigg) + \bigg( \fr{\omega^2 W_r^2 - 2 \widehat{a} \omega m L_r + \widehat{a}^2 m^2}{\Delta_r} - A \bigg)R= 0 , \nnr
& \fr{\df}{\df u} \bigg( \Delta_u \fr{\df S}{\df u} \bigg) - \bigg( \fr{\omega^2 W_u^2 + 2 \widehat{a} \omega m L_u + \widehat{a}^2 m^2}{\Delta_u} -A \bigg) S = 0 ,\label{eqS}
\end{align}
where $A$ is a separation constant.  For asymptotically flat black hole solutions, one has
\bea
\Delta_r = (r-r_+)(r-r_-)
\eea
and the radial equation has regular singular points at the locations of the horizons, $r = r_\pm$, and an irregular singular point at infinity, similar to what happens for the Kerr solution. The solutions are Heun functions. 
For asymptotically $\AdS_4$ black hole solutions, one has
\bea
\Delta_r = (r-r_+)(r-r_-)\frac{(r-r_c)(r-r_c^*)}{l^2}\label{quart}
\eea
and the radial equation has regular singular points at the locations of the horizons, $r = r_\pm$, at the two complex roots $r_c$, $r_{c}^*$ of the function $\Delta_r$ and at infinity. The angular equation involving $u$ can be analyzed similarly.

For a large class of black hole solutions, one has $L_r = W_r$ and $L_u = W_u$, which implies $W = W_r + W_u$ (up to choosing a sign without loss of generality). The metric then takes the simpler form
\be
\df s^2 = - \fr{\Delta_r}{W_r + W_u} \bigg( \df t + \fr{W_u}{\widehat{a}} \, \df \phi \bigg)^2  + \fr{\Delta_u}{W_r + W_u} \bigg( \df t - \fr{W_r}{\widehat{a}} \, \df \phi \bigg) ^2 + (W_r + W_u) \bigg( \fr{\df r^2}{\Delta_r} + \fr{\df u^2}{\Delta_u} \bigg) .\label{sp}
\ee
This class of metrics has been studied in detail \cite{Chow:2013gba}, and has the property that metric possess one Killing--Yano tensor with torsion and one conformal Killing-Yano tensor with torsion. It implies that metric admits an exact Killing--St\"ackel tensor and the massive Klein-Gordon equation
\be
\square \Phi = \fr{1}{\sr{-g}} \pd_\mu (\sr{-g} g^{\mu \nu} \pd_\nu \Phi) = \mu^2 \Phi 
\ee
is separable. The resulting equations are 
\begin{align}
& \fr{\df}{\df r} \bigg( \Delta_r \fr{\df R}{\df r} \bigg) + \bigg( \fr{(\omega W_r - \widehat{a} m)^2}{\Delta_r} -\mu^2 W_r -A \bigg) R = 0 , \nnr
& \fr{\df}{\df u} \bigg( \Delta_u \fr{\df S}{\df u} \bigg) + \bigg( -\fr{(\omega W_u + \widehat{a} m)^2}{\Delta_u}- \mu^2 W_u +A \bigg) S = 0,
\end{align}
where $A$ is a separation constant. 

In particular, the Kerr--Newman--AdS black hole of mass $\frac{m}{\Xi^2}$, angular momentum $\frac{ma}{\Xi^2}$ and electromagnetic charges $\frac{Q_e}{\Xi}$, $\frac{Q_m}{\Xi}$ with  $\Xi = 1- \frac{a^2}{l^2}$ can be set in the form \eqref{sp} with $\widehat{a} = a \Xi$ and with structure functions (see e.g., \cite{Kostelecky:1995ei,Caldarelli:1999xj})
\bea
\Delta_r &=& (r^2 + a^2) (1+\frac{r^2}{l^2}) - 2 m r + Q^2_e + Q_m^2, \\
\Delta_u &=& (a^2-u^2) (1-\frac{u^2}{l^2}), \\
W_r &=& r^2 + a^2 ,\\
W_u &=& u^2 - a^2.
\eea
The standard polar angle $\theta$ is related to $u$ as $u = a \cos\theta$. The Kerr--Newman black hole is obtained in the limit $l \rightarrow \infty$. In the case of the Kerr--Newman black hole, the equations for the functions $S(\th)$ and $R(r)$ were also written down in~\eqref{eqSs} and \eqref{eqRs} (specialized for the spin 0 field) in Section~\ref{macrogen}. 

The charged massive Klein-Gordon equation with probe charge $q_e$, 
\bea
\mathcal D_\mu \mathcal D^\mu \Phi = \mu^2 \Phi,\qquad \mathcal D_\mu \equiv \nabla_\mu - i q_e A_\mu \label{we}
\eea
around the Kerr--Newman--AdS black hole is also separable. The extent to which the charged massive Klein--Gordon equation can be separated in general and the algebraic structures involving the gauge fields underlying separability remains to be totally clarified. 

In the case of the Kerr--Newman black hole, the charged analogue of Eqs.~\eqref{eqS} can be written conveniently as
\bea
&& \fr{\df}{\df r} \bigg( \Delta \fr{\df R(r)}{\df r} \bigg) + \bigg[ \frac{\alpha(r_+)^2}{(r-r_+)(r_+-r_-)}-\frac{\alpha(r_-)^2}{(r-r_-)(r_+-r_-)} -K_l+V(r)\bigg] R(r) = 0,\label{radeq}\\
&&\frac{1}{\sin\theta}\fr{\df}{\df \theta} \bigg( \sin\theta \fr{\df S(\theta)}{\df \theta} \bigg) + \bigg[- \frac{m^2}{\sin^2\theta}+a^2(\omega^2-\mu^2)\cos^2\theta + K_l \bigg]S(\th)=0,\label{angeq} 
\eea
where the separation constant is redefined as $A = K_l -2 a m \omega +a^2(\omega^2-\mu^2)$ and $\Delta=(r-r_+)(r-r_-)$. The function $\alpha(r)$ is defined as 
\be
\alpha(r) = (2Mr - Q^2)\omega-a m - Q r q_e\, ,\label{defalphar}
\ee
and is evaluated either at $r_+$ or $r_-$ and the potential $V(r)$ is given by
\bea
V(r) &=& (\omega^2-\mu^2)r^2+2\omega(M\omega-q_e Q)r -\omega^2 Q^2 +(2M\omega - q_eQ)^2.\label{defVr}
\eea
The radial equation is a Heun equation whose solutions can only be found numerically. 

In the case of the Kerr--Newman--AdS black hole, the charged analogue of Eqs.~\eqref{eqS} can be written as
\bea
&& \fr{\df}{\df r} \bigg( \Delta_r \fr{\df R(r)}{\df r} \bigg) + \Big[ \frac{[\omega(r^2+a^2)-m a \Xi - q_e Q_e r]^2}{\Delta_r} -\mu^2 r^2 -C_l \Big] R(r) = 0,\label{radeqAdS}\\
&&\frac{1}{\sin\theta}\fr{\df}{\df \theta} \bigg( \sin\theta \Delta_\theta \fr{\df S(\theta)}{\df \theta} \bigg) + \bigg[  - \frac{m^2\Xi^2}{\sin^2\theta \Delta_\theta}+\frac{2m a \Xi \omega - a^2\omega^2 \sin^2 \theta}{\Delta_\theta} -a^2 \mu^2 \cos^2\theta+ C_l \bigg]S(\th)=0,\nn
\eea
where $C_l$ is a separation constant and all parameters in the equations have been defined in Section~\ref{sec:defAdSKerrN}. In the flat limit, Eqs.~\eqref{angeq}\,--\,\eqref{radeq} are recovered with $C_l = K_l -2 a \omega m+a^2 \omega^2$.

The radial equation has a more involved form than the corresponding flat equation~\eqref{radeq} due to the fact that $\Delta_r$ is a quartic instead of a quadratic polynomial in $r$; see \eqref{defDeltar}\,--\,\eqref{quart}. The radial equation is a general Heun's equation due to the presence of two conjugate complex poles in \eqref{radeqAdS} in addition to the two real poles corresponding to the inner and outer horizons and the pole at infinity.

It has been suggested that all these poles have a role to play in the microscopic description of the AdS black hole~\cite{Cvetic:2010mn}. It is an open problem to unravel the structure of the hidden symmetries, if any, of the full non-extremal radial equation~\eqref{radeqAdS}. It has been shown that in the context of five-dimensional black holes, one can find hidden conformal symmetry in the near-horizon region close to extremality~\cite{Birkandan:2011fr}. It is expected that one could similarly neglect the two complex poles in the near-horizon region of near-extremal black holes, but this remains to be checked in detail.\epubtkFootnote{Alternatively, it was suggested in~\cite{Chen:2010bh,Chen:2010jj} that one can describe the dynamics of the scalar field in the near-horizon region using the truncated expansion of $\Delta_r(r)$ around $r_+$ at second order. However, the resulting function $\Delta_r^{\text{trunc}}$ has, in addition to the pole $r_+$, a fake pole $r_*$, which is not associated with any geometric or thermodynamic feature of the solution. Therefore, the physical meaning of this truncation is unclear.} Since hidden symmetries for AdS black holes are not understood, we will not discuss AdS black holes further.

\subsubsection{Near-region scalar-wave equation}

Let us go back to the scalar wave equation around the Kerr--Newman black hole. We will now study a particular range of parameters, where the wave equations simplify. We will assume that the wave has low energy and low mass as compared to the black hole mass and low electric charge as compared to the black hole charge, 
\be
\omega M = O(\eps),\qquad \mu M = O(\eps),\qquad q_e Q = O(\eps)\, ,\label{approxpp}
\ee
where $\eps \ll 1$. From these approximations, we deduce that $\omega a,\; \omega r_+,\; \omega Q$ and $\mu a = O(\eps)$ as well. 

We will only look at a specific region of the spacetime -- the ``near region'' -- defined by 
\be
\omega r = O(\eps),\qquad \mu r = O(\eps).\label{nearreg}
\ee
Note that the near region is a distinct concept from the near-horizon region $r-r_+ \ll M$. Indeed, for sufficiently small $\omega$ and $\mu$, the value of $r$ defined by the near region can be arbitrarily large.

Using the approximations~\eqref{approxpp}, the wave equation greatly simplifies. It can be solved both in the near region and in the far region $r \gg M$ in terms of special functions. A complete solution can then be obtained by matching near and far solutions together along a surface in the matching region $M \ll r \ll \omega^{-1}$. As noted in~\cite{Castro:2010fd}, conformal invariance results from the freedom to locally choose the radius of the matching surface within the matching region. 

More precisely, using \eqref{approxpp}, the angular equation~\eqref{angeq} reduces to the standard Laplacian on the two-sphere
\be
\frac{1}{\sin\theta}\fr{\df}{\df \theta} \bigg( \sin\theta  \fr{\df S(\theta)}{\df \theta} \bigg) + \left[- \frac{m^2}{\sin^2\theta}+ K_l \right]S(\th)=O(\eps^2).
\ee
The solutions $e^{i m \phi}S(\th)$ are spherical harmonics and the separation constants are 
\be
K_l = l(l+1)+O(\eps^2).\label{approx}
\ee

In the near region, the function $V(r)$ defined in \eqref{defVr} is very small, $V(r) = O(\eps^2)$. The near region scalar-wave equation can then be written as 
\bea
&& \fr{\df}{\df r} \bigg( \Delta \fr{\df R(r)}{\df r} \bigg) + \bigg[ \frac{\alpha(r_+)^2}{(r-r_+)(r_+-r_-)}-\frac{\alpha(r_-)^2}{(r-r_-)(r_+-r_-)}-l(l+1) \bigg] R(r) = 0,\label{eq:scJ}
\eea
where $\alpha(r)$ has been defined in \eqref{defalphar}.

\subsection{Hidden conformal symmetries}

\subsubsection{Local $SL(2,\mathbb R)\times SL(2,\mathbb R)$ symmetries}
\label{sec:sl2r}

We will now make explicit the local $SL(2,\mathbb R)\times SL(2,\mathbb R)$ symmetries of the near-horizon scalar field equations~\eqref{eq:scJ}. For this purpose it is convenient to define the ``conformal'' coordinates $(\omega^\pm,y)$ defined in terms of coordinates $(t,r,\phi^\prime)$ by (see~\cite{Castro:2010fd} and~\cite{Maldacena:1998bw} for earlier relevant work)
\bea
\omega^+ &=& \sqrt{\frac{r-r_+}{r-r_-}}e^{2\pi T_R (\phi^\prime - \Omega_R t) },\nn \\
\omega^- &=& \sqrt{\frac{r-r_+}{r-r_-}}e^{2\pi T_L (\phi^\prime - \Omega_L t)},\label{eq:95b}\\
y &=& \sqrt{\frac{r_+-r_-}{r-r_-}}e^{\pi T_L (\phi^\prime - \Omega_L t) +\pi T_R (\phi^\prime - \Omega_R t)}.\nn
\eea
The change of coordinates is locally invertible if $\Delta \Omega = \Omega_L - \Omega_R \neq 0$. We choose the chirality $\Delta \Omega > 0 $, as it will turn out to match the chirality convention in the description of extremal black holes in Section~\ref{sec:microS}. 

Several choices of coordinate $\phi^\prime \sim \phi^\prime+2\pi $ will lead to independent $SL(2,\mathbb R)\times SL(2,\mathbb R)$ symmetries. For the Kerr black hole, there is only one meaningful choice: $\phi^\prime=\phi$. For the Reissner--Nordstr\"om black hole, we identify $\phi^\prime=\chi/R_\chi$, where $\chi$ is the Kaluza--Klein coordinate that allows one to lift the gauge field to higher dimensions, as done in Section~\ref{sec:c}. For the Kerr--Newman black hole, we use, in general, a coordinate system $(\phi^\prime,\chi^\prime)\sim (\phi^\prime,\chi^\prime+2\pi)\sim (\phi^\prime+2\pi,\chi^\prime)$ parameterized by a $SL(2,\mathbb Z)$ transformation 
\bea
\phi^\prime &=& p_1 \phi + p_2 \chi/R_\chi ,\nn \\
\chi^\prime &=& p_3 \phi + p_4 \chi/R_\chi \, ,\label{coordsl2}
\eea
with $p_1 p_4 - p_2 p_3 = 1$ so that 
\bea
\p_{\phi^\prime} &=& p_4 \p_\phi -p_3 R_\chi \p_\chi ,\\
\p_{\chi^\prime} &=& -p_2 \p_\phi +p_1 R_\chi \p_\chi\, .
\eea

Let us define \emph{locally} the vector fields 
\bea
H_1 &=& i \p_+ ,\nn\\
H_0 &=& i (\omega^+ \p_+ + \frac{1}{2}y \p_y),\label{vec1} \\
H_{-1} &=& i(\omega^{+2}\p_+ + \omega^+ y \p_y - y^2 \p_-)\, ,\nn
\eea
and 
\bea
\bar H_1 &=& i \p_- ,\nn\\
\bar H_0 &=& i (\omega^- \p_- + \frac{1}{2}y \p_y), \label{vec2}\\
\bar H_{-1} &=& i(\omega^{-2}\p_- + \omega^- y \p_y - y^2 \p_+)\, .\nn
\eea
These vector fields obey the $SL(2,\mathbb R)$ Lie bracket algebra,
\be
[H_0,H_{\pm 1}] = \mp i H_{\pm 1},\qquad [H_{-1},H_1] = -2 i H_0,
\ee
and similarly for $(\bar H_0,\bar H_{\pm 1})$. Note that 
\be
T_L \bar H_0 + T_R H_0 = \frac{i}{2\pi}\p_{\phi^\prime}\, .\label{relphi}
\ee
The $SL(2,\mathbb R)$ quadratic Casimir is 
\bea
\mathcal H^2 &=& \bar{\mathcal H}^2 = -H_0^2+\frac{1}{2}(H_1 H_{-1}+H_{-1}H_1) \\
&=& \frac{1}{4} (y^2 \p_y^2 - y \p_y)+y^2 \p_+ \p_-\, .\label{Casimir}
\eea
In terms of the coordinates $(r,t,\phi^\prime)$, the Casimir becomes
\bea
 \mathcal H^2 &=& -\frac{r_+-r_-}{(r-r_+)(4\pi T_R)^2}\left(\p_{\phi^\prime} + \frac{T_L+T_R}{T_L \Delta \Omega }(\p_t+\Omega_R \p_{\phi^\prime}) \right)^2 \nn \\
&& +\frac{r_+-r_-}{(r-r_-)(4\pi T_R)^2}\left(\p_{\phi^\prime} + \frac{T_L-T_R}{T_L \Delta \Omega }(\p_t+\Omega_R \p_{\phi^\prime} ) \right)^2+\p_r \Delta \p_r\,, \nn
\eea
where $\Delta(r) = (r-r_+)(r-r_-)$. 

We will now match the radial wave equation around the Kerr--Newman black hole in the near region \eqref{eq:scJ} with the eigenvalue equation
\be
\mathcal H^2 \Phi = l(l+1) \Phi\, .\label{eigen}
\ee
The scalar field has the following eigenvalues $\p_t \Phi = - i \omega \Phi$ and $\p_\phi \Phi = i m \Phi$. In the case where an electromagnetic field is present, one can perform the uplift~\eqref{KKlift} and consider the five-dimensional gauge field \eqref{defphi5d}. In that case, the eigenvalue of the five-dimensional gauge field under $\p_\chi$ is the electric charge $\p_\chi \Phi = i q_e \Phi$. Let us denote the eigenvalue along $\p_{\phi^\prime}$ as $i m^\prime \equiv i(p_4 m -p_3 q_e R_\chi)$. Eqs.~\eqref{eq:scJ} and \eqref{eigen} will match if and only if the two following equations are obeyed
\be
\alpha(r_\pm) = \frac{r_+-r_-}{4\pi T_R}\left(-m^\prime+\frac{T_L \pm T_R}{T_L\Delta \Omega}(\omega - \Omega_R m^\prime )\right)\, ,
\ee
where $\alpha(r)$ has been defined in \eqref{defalphar}.

For simplicity, let us first discuss the case of zero probe charge $q_e=0$ and non-zero probe angular momentum $m \neq 0$. The matching equations then admit a unique solution 
\bea
\Omega_R &=& 0,\qquad \Omega_L = \frac{a}{2M^2-Q^2},\label{TLR1}\\
T_L &=& \frac{2M^2-Q^2}{4\pi J},\qquad T_R = \frac{M(r_+-r_-)}{4\pi J}\, ,\nn
\eea
upon choosing $\phi^\prime = \phi$ (and $\chi^\prime = \chi/R_\chi$). This shows in particular that the Kerr black hole has a hidden symmetry, as derived originally in~\cite{Castro:2010fd}. It is curious that $T_R$ can simply be expressed in terms of the Hawking temperature and angular velocity at the outer horizon as 
\bea
T_R = \frac{T_H}{\Omega_J}.
\eea

For probes with zero angular momentum $m=0$, but electric charge $q_e \neq 0$, there is also a unique solution,
\bea
\Omega_R &=& \frac{Q}{2M R_\chi},\qquad \Omega_L = \frac{M Q}{(2M^2-Q^2)R_\chi} ,\label{TLR2}\\
T_L &=& \frac{(2M^2-Q^2)R_\chi}{2\pi Q^3},\qquad T_R = \frac{M(r_+-r_-)R_\chi}{2\pi Q^3}\, ,\nn
\eea
upon choosing $\phi^\prime = \chi/R_\chi$ (and $\chi^\prime = -\phi$). This shows, in particular, that the Reissner--Nordstr\"om black hole admits a hidden symmetry, as pointed out in~\cite{Chen:2010as,Chen:2010yu}.

Finally, one can more generally solve the matching equation for any probe scalar field whose probe angular momentum and probe charge are related by 
\be
p_2 m - p_1 q_e R_\chi = 0\, .\label{zerooo}
\ee
In that case, one chooses the coordinate system~\eqref{coordsl2} and the unique solution is then
\bea
\Omega_R &=& \frac{p_2 Q}{2M R_\chi},\qquad \Omega_L = \frac{p_1 a+p_2 M Q/R_\chi}{2M^2-Q^2} ,\label{TLR3}\\
T_L &=& \frac{2M^2-Q^2}{2\pi (2p_1 J +p_2Q^3/R_\chi)},\qquad T_R = \frac{M(r_+-r_-)}{2\pi (2p_1 J +p_2Q^3/R_\chi)}\, .\nn
\eea
When $p_1 = 0$ and $Q\neq 0$ or $p_2 = 0$ and $J\neq 0$, one recovers the two previous particular cases. The condition~\eqref{zerooo} is equivalent to the fact that the scalar field has zero eigenvalue along $\p_{\chi^\prime}$. 

Let us now discuss the quantization of $q_e R_\chi$. In general, the scale $R_\chi$ of the Kaluza--Klein direction is constrained by matter field couplings. Let us illustrate this point using the simple uplift~\eqref{KKlift}. The wave equation \eqref{we} is reproduced from a five-dimensional scalar field $\phi_{(5d)}(x,\chi)$ probing the five-dimensional metric~\eqref{KKlift}, if one takes 
\be
\phi_{(5d)}(x,\chi) =\phi(x)e^{i q_e\chi},\label{defphi5d}
\ee
and if the five-dimensional mass is equal to $\mu^2_{(5d)}=\mu^2+q_e^2$. However, the five-dimensional scalar is multivalued on the circle $\chi$ unless 
\be
q_e R_\chi \in \mathbb N.\label{quantizR}
\ee
Since $m$ and $q_e R_\chi$ are quantized, as derived in \eqref{quantizR}, there is always (at least) one solution to~\eqref{zerooo} with integers $p_1$ and $p_2$. 

In conclusion, any low energy and low mass scalar probe in the near region \eqref{nearreg} of the Kerr black hole admits a local hidden $SL(2,\mathbb R)\times SL(2,\mathbb R)$ symmetry. Similarly, any low energy, low mass and low charge scalar probe in the near region \eqref{nearreg} of the Reissner--Nordstr\"om black hole admits a local hidden $SL(2,\mathbb R)\times SL(2,\mathbb R)$ symmetry. 
In the case of the Kerr--Newman black hole, we noticed that probes obeying \eqref{approxpp} also admit an $SL(2,\mathbb R)\times SL(2,\mathbb R)$ hidden symmetry, whose precise realization depends on the ratio between the angular momentum and the electric charge of the probe. For a given ratio~\eqref{zerooo}, hidden symmetries can be constructed using the coordinate $\phi^\prime = p_1 \phi+p_2 \chi/R_\chi$. Different choices of coordinate $\phi^\prime$ are relevant to describe different sectors of the low energy, low mass and low charge dynamics of scalar probes in the near region of the Kerr--Newman black hole. The union of these descriptions cover the entire dynamical phase space in the near region under the approximations~\eqref{approxpp}\,--\,\eqref{nearreg}.

\subsubsection{Symmetry breaking to $U(1)_L \times U(1)_R$ and Cardy entropy matching}

The vector fields that generate the $SL(2,\mathbb R)\times SL(2,\mathbb R)$ symmetries are not globally defined. They are not periodic under the angular identification
\be
\phi^\prime \sim \phi^\prime + 2\pi \, .\label{ident}
\ee
Therefore, the $SL(2,\mathbb R)$ symmetries cannot be used to generate new global solutions from old ones. In other words, solutions to the wave equation in the near region do not form $SL(2,\mathbb R)\times SL(2,\mathbb R)$ representations. In the $(\omega^+,\omega^-)$ plane defined in \eqref{eq:95b}, the identification~\eqref{ident} is generated by the $SL(2,\mathbb R)_L \times SL(2,\mathbb R)_R$ group element 
\be
e^{-i 4\pi^2 T_R H_0 - i 4\pi^2 T_L \bar H_0}\, ,\label{groupel}
\ee
as can be deduced from \eqref{relphi}. This can be interpreted as the statement that the $SL(2,\mathbb R)_L \times SL(2,\mathbb R)_R$ symmetry is spontaneously broken to the $U(1)_L \times U(1)_R$ symmetry generated by $(\bar H_0,H_0)$.

The situation is similar to the BTZ black hole in 2+1 gravity that has a $SL(2,\mathbb R)_L \times SL(2,\mathbb R)_R$ symmetry, which is spontaneously broken by the identification of the angular coordinate. This breaking of symmetry can be interpreted in that case as placing the dual CFT to the BTZ black hole in a density matrix with left and right-moving temperatures dictated by the $SL(2,\mathbb R)_L \times SL(2,\mathbb R)_R$ group element generating the $2\pi$ identification of the geometry~\cite{Maldacena:1998bw}. 

In the case of non-extremal black-hole geometries, one can similarly interpret the symmetry breaking using a CFT as follows~\cite{Castro:2010fd}. First, we need to assume that before the identification, the near region dynamics is described by a dual two-dimensional CFT, which possesses a ground state that is invariant under the full $SL(2,\mathbb R)_L \times SL(2,\mathbb R)_R$ symmetry. This is a very strong assumption since there are several obstacles to such a description as discussed in the Introduction~(\ref{sec:introduction}) and through the text, see in particular \cite{Amsel:2009ev,Dias:2009ex,Baggio:2012db,Castro:2013kea}. At best, if such a CFT description exists, it admits irrelevant deformations in both sectors and its range of applicability is very limited \cite{Baggio:2012db}. Nevertheless, assuming the existence of this vacuum state, the two conformal coordinates $(\omega^+,\omega^-)$ can be interpreted as the two null coordinates on the plane where the CFT vacuum state can be defined. At fixed $r$, the relation between conformal coordinates $(\omega^+,\omega^-)$ and Boyer--Lindquist $(\phi,t)$ coordinates is, up to an $r$-dependent scaling,
\be
\omega^\pm = e^{\pm t^\pm},
\ee
where
\bea
t^+ &=& 2\pi T_R (\phi^\prime - \Omega_R t),\\
t^- &=& -2\pi T_L (\phi^\prime - \Omega_L t).
\eea
This is precisely the relation between Minkowski $(\omega^\pm)$ and Rindler $(t^\pm)$ coordinates. The periodic identification~\eqref{ident} then requires that the Rindler domain be restricted to a fundamental domain under the identification
\be
t^+ \sim t^+ + 4\pi^2 T_R,\qquad t^- \sim t^- - 4\pi^2 T_L\, ,
\ee
generated by the group element~\eqref{groupel}. 

The quantum state describing this accelerating strip of Minkowski spacetime is obtained from the $SL(2,\mathbb R)_L \times SL(2,\mathbb R)_R$ invariant Minkowski vacuum by tracing over the quantum state in the region outside the strip. The result is a thermal density matrix at temperatures $(T_L,T_R)$. Hence, under the assumption of the existence of a CFT with a vacuum state, non-extremal black holes can be described as a finite temperature $(T_L,T_R)$ mixed state in a dual CFT. 

It is familiar from the three-dimensional BTZ black hole that the identifications required to obtain extremal black holes are different than the ones required to obtain non-extremal black holes~\cite{Banados:1992gq,Maldacena:1998bw}. Here as well, the vector fields \eqref{vec1}\,--\,\eqref{vec2} are not defined in the extremal limit because the change of coordinates~\eqref{eq:95b} breaks down. 
Nevertheless, the extremal limit of the temperatures $T_L$ and $T_R$ match with the temperatures defined at extremality in Section~\ref{sec:CFTmatch}. More precisely, the temperatures $T_L$ and $T_R$ defined in \eqref{TLR1}, \eqref{TLR2} and \eqref{TLR3} match with the temperatures defined at extremality $T_\phi,\, R_\chi T_e$ and $(p_1 T_\phi^{-1}+p_2 (R_\chi T_e)^{-1})^{-1}$, respectively, where $T_\phi$ and $T_e$ are defined in \eqref{TKN}. This is consistent with the interpretation that states corresponding to extremal black holes in the CFT can be defined as a limit of states corresponding to non-extremal black holes.

Still assuming a CFT description of non-extremal black holes, one can then show that the temperatures $T_L$ and $T_R$ obtained in Section~\ref{sec:hidden} combined with the analysis at extremality in Section~\ref{sec:KerrCFT1} lead to several Cardy matchings of the black hole entropy of the Kerr, Reissner--Nordstr\"om and Kerr--Newman black holes. The thermal version of Cardy's formula reads as
\be
\cS_{\text{CFT}} = \frac{\pi^2}{3} ( c_L T_L +c_R T_R ),\label{Cardy2}
\ee
which is valid when $T_L \gg 1$, $T_R \gg 1$. As explained in Section~\ref{sec:Cardy}, this range of applicability can be extended under certain conditions. In a CFT, the difference $c_R - c_L$ is proportional to the diffeomorphism anomaly of the CFT~\cite{Kraus:2005zm,Kraus:2006wn}. 
Assuming diffeomorphism invariance one could argue that the two left and right sectors should have the same value for the central charge,
\be
c_R = c_L\, .\label{cR}
\ee
The value $c_L$ was obtained at extremality in Section~\ref{sec:c} and it was checked that Cardy's formula reproduces the extremal black-hole entropy, for each choice of $U(1)$ circle which defines the corresponding Virasoro algebra. For the $\CFT_J$, we obtained $c_L = 12 J$. For $\CFT_Q$, we had $c_Q = 6Q^3/R_\chi$ and for the $\CFT_{(p_1,p_2,p_3)}$, we got $c_{(p_1,p_2)} = 6(p_1(2J)+p_2 Q^3/R_\chi)$. 

It turns out that in each case, the non-extremal black hole entropy matches Cardy's formula with the temperatures $T_L$, $T_R$ derived earlier and the central charges \eqref{cR} with $c_L$ computed at extremality. In particular, the central charge does not depend upon the mass $M$. This is a non-trivial feature of this Cardy matching which has no explanation so far.

\newpage

\section{Summary and Open Problems}
\label{ccl}

\subsection{Summary}
\label{sec:summary}

The Kerr/CFT correspondence is a set of relations between the classical physics of spinning or charged black holes and representation of conformal symmetry (which depending on the context is either $SL(2,\mathbb R)$ symmetry, Virasoro symmetry or the full $2d$ conformal group). In its strong original form, the Kerr/CFT correspondence is the statement that the microscopic degrees of freedom of black holes can be counted by an effective CFT model. Subsequent developments indicate that such putative dual theories would differ from standard CFTs in several respects (irrelevant deformations, warped deformations, complex conformal weights, $\dots$) but would still be constrained by conformal invariance to ensure that Cardy's formula applies. At present, no construction of such a dual theory has been achieved for an embedding of the extremal Kerr black hole in string theory.\footnote{There are however other extremal black holes in string theory which admit in addition to their near-horizon limit an intermediate  decoupling limit with a warped $AdS_3$ spacetime with finite energy excitations \cite{ElShowk:2011cm,Detournay:2012dz}. In those cases, one can construct of a dynamical phase space admitting two copies of the Virasoro algebra as asymptotic symmetry algebra, as long as no travelling waves are present, which point to the existence of a dual $2d$ CFT \cite{Compere:2014bia}. These toy models are encouraging but rely on the existence of an intermediate decoupling limit and on the absence of travelling wave instabilities, both hypotheses which are untrue in extremal Kerr.} One could therefore be skeptical on the validity of the strong Kerr/CFT correspondence. Nevertheless, several observations remain intruiging and not explained at present such as the occurrence and relevance of various symmetries and the incomprehensible universal match of Cardy's formula with the black hole entropy. 

We have reviewed that any extremal black hole containing a compact $U(1)$ axial symmetry admits  in its near-horizon geometry a Virasoro algebra with a non-trivial central charge. The black-hole entropy is reproduced by a chiral half of Cardy's formula. This result is robust for any diffeomorphism-invariant theory and holds even including scalar and gauge field couplings and higher-derivative corrections. Moreover, if a $U(1)$ gauge field can be geometrized into a Kaluza--Klein vector in a higher-dimensional spacetime, a Virasoro algebra can be defined along the Kaluza--Klein compact $U(1)$ direction and all the analysis goes through in a similar fashion as for the axial $U(1)$ symmetry. The deep similarity between the effects of rotation and electric charge can be understood from the fact that they are on a similar footing in the higher-dimensional geometry. When two $U(1)$ symmetries are present, one can mix up the compact directions using a modular transformation and the construction of Virasoro algebras can still be made. 

Independently of these constructions, the scattering probabilities of probes around the near-extremal Kerr--Newman black hole can be reproduced near the superradiant bound by manipulating near-chiral thermal two-point functions of a two-dimensional CFT. The result extends straightforwardly to other asymptotically-flat or AdS black holes in various gravity theories. Finally away from extremality, hidden $SL(2,\mathbb R) \times SL(2,\mathbb R)$ symmetries are present in some scalar probes around the Kerr--Newman black hole close enough to the horizon. We showed that several such hidden symmetries are required to account for the entire probe dynamics in the near region in the regime of small mass, small energy and small charge. This analysis does not extend straightforwardly to AdS black holes.

A fair concluding remark would be that several new and intriguing properties of the Kerr--Newman black hole and their generalizations in string theory have been uncovered over the last years, but there is still a long road ahead to comprehend what these results are really telling us about the nature of quantum black holes.

\subsection{Set of open problems}
\label{sec:open}

We close this review with a list of open problems. We hope that the interested reader will tackle them with the aim of shedding more light on the Kerr/CFT correspondence. We ordered the list of problems such that they range from concrete problem sets to much more involved questions. 

\begin{enumerate}

\item Near-horizon geometries of black-hole solutions of \eqref{generalaction} have been classified. Classify the four-dimensional near-horizon geometries of extremal black holes for gravity coupled to charged scalars, massive vectors, $p$-forms and non-abelian gauge fields. Are there new features?

\item Non-extremal asymptotically flat black holes admit universal features such as the product of area formula and relationships among inner and outer horizon quantities. Investigate whether or not all black holes in string theory (with higher curvature corrections) admit these features. Either try to formulate a proof or find a counterexample. Extend the analysis to AdS black holes.

\item A black hole in de Sitter spacetime can be extremal in the sense that its outer radius coincides with the cosmological horizon. The resulting geometry, called the rotating Narirai geometry, has many similarities with the near-horizon geometries of extremal black holes in flat spacetime or in AdS spacetime. The main difference is that the near-horizon geometry is a warped product of $dS_2$ with $S^2$ instead of $\AdS_2$ with $S^2$. Some arguments of the Kerr/CFT correspondence have been extended to this setting~\cite{Anninos:2009yc}. Extend the dictionary much further. 

\item Formulate a general scattering theory around near-extremal black-hole solutions of \eqref{generalaction}. Classify the geometries admitting a Killing--Yano tensor or other special algebraic tensors so that the wave equation could be separated. This would allow to check the matching with CFT two-point functions in much more generality.

\item In the analysis of near-extremal superradiant scattering for any spin, the modes that are below the Breitenlohner--Freedman bound were discarded. Such modes lead to non-conserved flux at the boundary, they lead to instabilities, and their interpretation by a CFT remains unclear. Clarify the match between these modes and CFT expectations for the Kerr--Newman black hole. Also, the match of near-extremal scattering waves with a CFT required to introduce a right-moving current algebra with the matching condition \eqref{match0}. Clarify why.

\item Understand how the extension of the Kerr/CFT correspondence to extremal AdS black holes fits within the AdS/CFT correspondence. As discussed in~\cite{Lu:2009gj}, the extremal AdS--Kerr/CFT correspondence suggests that one can identify a non-trivial Virasoro algebra acting on the low-energy states of strongly coupled large $N$ super-Yang--Mills theory in an extremal thermal ensemble. Try to make this picture more precise. 

\item Boundary conditions alternative to the Kerr/CFT boundary conditions have been proposed for higher dimensional extremal vacuum black holes \cite{Compere:2015mza}. These admit a generalized Virasoro algebra as asymptotic symmetry algebra with the black hole entropy as a central charge. Find criteria to assert which boundary condition is consistent with quantum gravity and in particular is relevant to describe the microscopic entropy. First find whether these boundary conditions can be extended with matter and higher derivative corrections. 

\item The hidden symmetry arguments for the non-extremal Kerr-Newman black hole rely on an choice of $U(1)$ circle on the two torus spanned by the azimuthal direction and the Kaluza-Klein direction obtained by the uplift of the gauge field. This leads to a $SL(2,\mathbb Z)$ family of $SL(2,\mathbb R) \times SL(2,\mathbb R)$ hidden symmetries. However only one subtracted geometry has been derived for the Kerr-Newman black hole. Does it exist a $SL(2,\mathbb Z)$ family?

\item Provide with a procedure to compute the central charges $c_L$ and $c_R$ away from extremality, or prove that it is not possible. 

\item Find the largest class of field theories and states such that Cardy's formula applies. 

\item Find astrophysical observables in the near-horizon region of near-extremal Kerr which are constrained by $SL(2,\mathbb R)$ symmetry and use the symmetry to analytically compute these observables. Find whether or not there exist signatures of Virasoro symmetry. 

\item Embed the extremal Kerr black hole in string theory and construct one exact holographic quantum field theory dual. Use this correspondence to precisely define the notion of a DLCQ of a warped deformation of a $2d$ CFT which is conjectured relevant to describe the microstates of the extremal Kerr black hole. In this model, compute the quantum corrections to the central charge $c_L$ and check that it reproduces the quantum-corrected entropy of extremal black holes derived in~\cite{Sen:2012cj}.

\end{enumerate}

\newpage

\section*{Acknowledgments}

This review originates from lectures given at Iberian Strings 2012 in Bilbao. I am very grateful to the organizers I.~Bandos, I.~Egusquiza, J.L.~Ma\~nes, M.A.~Valle and C.~Meliveo for the invitation to lecture in this outstanding and agreeable conference. I gratefully thank V.~Balasubramanian, J.~de Boer, B.~Chen, C.-M.~Chen, B.~Chowdury, A.~Castro, S.~Detournay, J.~Jottar, F.~Larsen, S.~Markoff, K.~Murata, M.~Rangamani, H.~Reall, S.~Sheikh-Jabbari, K.~Skenderis, A.~Strominger, A.~Virmani and especially M.~Guica and T.~Hartman for interesting exchanges during the writing of this review. I also thank the organizers D.~Berman, J.~Conlon, N.~Lambert, S.~Mukhi and F.~Quevedo of the program ``Mathematics and Applications of Branes in String and M-theory'' at the Isaac Newton Institute, Cambridge for support and hospitality during the final stages of the first version of this work. I finally thank K.~Hajian, S.S.~Jabbari. J.~Lucietti and A.~Seraj for precisions and anonymous referees for suggestions during the update to version 2. This work has been financially supported by the Nederlandse Organisatie voor Wetenschappelijk Onderzoek (NWO) via an NWO Vici grant. It is also currently supported by the FNRS, Belgium and the ERC Starting Grant 335146 ``HoloBHC''.

\newpage

\end{document}